\documentclass[8pt]{article}

\usepackage{amsmath,amsxtra,amsthm,amssymb,xr,mathtools}
\usepackage{graphicx}
\usepackage[latin1]{inputenc}
\usepackage{hyperref}
\usepackage[all]{xy}
\usepackage{xcolor}
\usepackage{stmaryrd}
\usepackage{rotating}
\usepackage{dsfont}
\usepackage{subcaption}
\usepackage{float}
\usepackage{diagbox}
\usepackage{adjustbox}

\newcommand{\bit}{\begin{itemize}}
\newcommand{\eit}{\end{itemize}}
\newcommand{\bd}{\begin{description}}
\newcommand{\ed}{\end{description}}

\newcommand{\bc}{\begin{center}}
\newcommand{\ec}{\end{center}}
\newcommand{\Ref}[1]{(\ref{#1})}

\newcommand{\C}{{\mathbb C}}

\newcommand{\R}{{\mathbb R}}
\newcommand{\Z}{{\mathbb Z}}

\newcommand{\SU}{\mathrm{SU}}

\newcommand{\SL}{\mathrm{SL}}
\newcommand{\SO}{\mathrm{SO}}
\newcommand{\U}{\mathrm{U}}

\newcommand{\cJ}{{\cal J}}
\newcommand{\be}{\begin{equation}}
\newcommand{\ee}{\end{equation}}
\newcommand{\bea}{\begin{eqnarray}}
\newcommand{\eea}{\end{eqnarray}}
\newcommand{\bs}{\begin{subequations}}
\newcommand{\es}{\end{subequations}}
\newcommand{\nn}{\nonumber}

\newcommand{\w}{\wedge}

\newcommand{\f}{\frac}

\newcommand{\Id}{\mathds{1}}

\newcommand{\scr}{\scriptscriptstyle\rm}

\newcommand{\ra}{\rangle}
\newcommand{\la}{\langle}
\newcommand{\bra}[1]{\langle {#1}|}
\newcommand{\ket}[1]{|{#1}\rangle}

\newcommand{\vet}[2]{\left( \begin{array}{cc}{#1}\\{#2}\end{array} \right)}
\newcommand{\mat}[4]{\left( \begin{array}{cc}{#1}&{#2}\\{#3}&{#4}\end{array} \right)}

\renewcommand{\a}{\alpha} \renewcommand{\b}{\beta} \newcommand{\g}{\gamma}
\renewcommand{\d}{\delta}  \newcommand{\eps}{\epsilon}  \newcommand{\z}{\zeta}
 \renewcommand{\th}{\theta}      \renewcommand{\l}{\lambda}
\let\m=\mu    \let\r=\rho \newcommand{\s}{\sigma}  \renewcommand{\t}{\tau}     
\let\G=\Gamma \let\D=\Delta  \let\Th=\Theta \let\L=\Lambda  \let\Om=\Omega

\newcommand{\norm}[1]{|\!|#1|\!|}

\newcommand{\Wthree}[6]{\left(\begin{array}{ccc} #1 & #2 & #3 \\ #4 & #5 & #6 \end{array}\right)}
\newcommand{\Wfour}[9]{\left(\begin{array}{cccc} #1 & #2 & #3 & #4 \\ #5 & #6 & #7 & #8 \end{array}\right)^{(#9)}}

\newcommand{\vaccachespacca}{\texttt{sl2cfoam}}

\usepackage[normalem]{ulem}

\begin{document}

\title{\bf Numerical study of the Lorentzian Engle-Pereira-Rovelli-Livine \\ spin foam amplitude}

\author{\Large{Pietro Don\`a$^a$, Marco Fanizza$^{b}$, Giorgio Sarno$^{c}$ and Simone Speziale$^c$}
\smallskip \\
\small{$^a$ IGC \& Physics Department, Penn State, University Park, PA 16802, USA}\\
\small{$^b$ NEST, Scuola Normale Superiore and Istituto Nanoscienze-CNR, I-56126 Pisa, Italy}\\
\small{$^c$ CPT, Aix Marseille Univ., Univ. de Toulon, CNRS, Marseille, France}
}
\date{\today}

\maketitle

%----------------------------------------------------------------------------
\begin{abstract}
\noindent 
The Lorentzian Engle-Pereira-Rovelli-Livine spin foam amplitude for loop quantum gravity is a multidimensional noncompact integral of highly oscillating functions. Using a method based on the decomposition of Clebsch-Gordan coefficients for the unitary infinite-dimensional representations of SL(2,C) in terms of those of SU(2), we are able to provide for the first time numerical evaluations of the vertex amplitude. 
The values obtained support the asymptotic formula obtained by Barrett and collaborators with a saddle point approximation, showing, in particular, a power law decay and oscillations related to the Regge action. 
The comparison offers a test of the efficiency of the method. Truncating the decomposition to the first few terms provides a qualitative matching of the power law decay and oscillations. For vector and Euclidean Regge boundary data, a qualitative matching is obtained with just the first term, which corresponds to the simplified EPRL model.
We comment on future developments for the numerics and extension to higher vertices.
We complete our work with some analytic results:
We provide an algorithm and explicit configurations for the different geometries that can arise as boundary data, and  explain the geometric consequences of the decomposition used.
\end{abstract}
%----------------------------------------------------------------------------

\tableofcontents

%----------------------------------------------------------------------------
\section{Introduction}
%----------------------------------------------------------------------------

We report on numerical studies of the Lorentzian Engle-Pereira-Rovelli-Livine (EPRL) vertex amplitude for spin foam quantum gravity \cite{EPRL,FK} (for reviews, see  \cite{PerezLR,RovelliVidotto}). These are aimed in particular at testing its large spin asymptotic behavior, as derived by Barrett et al. \cite{BarrettLorAsymp}. The asymptotics can be interpreted as a semiclassical limit for the vertex amplitude and shows the presence of the Regge action, for a suitable choice of boundary data. It establishes an important relation to (discrete) general relativity that is at the heart of most of the physical applications of the EPRL model, thereby the interest in numerical verifications.

Our numerical work is based on the code \vaccachespacca\ developed in \cite{Dona:2018nev} and publicly available at \cite{code-sl2cfoam}, optimized here for the specifics of the vertex amplitude. The code implements the method of \cite{Boosting} to compute Lorentzian Clebsch-Gordan coefficients and spin foam amplitudes. Our goal is twofold: on the one hand, to provide the first numerical test of Barrett's Lorentzian asymptotics results \cite{BarrettLorAsymp}, thus verifying the saddle point approximation underpinning it, as well as estimating its validity. On the other hand, to examine the robustness of \vaccachespacca\ and test its performances. This means being able to compute Lorentzian spin foam amplitudes in the deep quantum regime, a necessary step for the spin foam formalism.

The results we found are encouraging, albeit not completely satisfactory. The numerics confirm Barrett's formula for boundary data corresponding to vector geometries or to Euclidean 4-simplices. In these cases, the asymptotics is reached fast, with an error below $10\%$ at spins of order 10. This confirms the validity of the saddle point approximation already at small spins, a situation similar to the much more extensively studied case of asymptotics of SU(2) invariants, see e.g. \cite{PonzanoRegge,IoSU2asympt}. 
In parallel, it shows the accuracy and robustness of \vaccachespacca. To give an idea of the heavy computational load required to achieve these results, the plot used to confirm the Euclidean asymptotics at spins of order 10 needed the evaluation of $10^{11}$ configurations, each of which is determined by up to $10^4$ one-dimensional unbounded integrals of oscillating functions. 
We used a server provided by the CPT with 32 Intel\textregistered\ Xeon\textregistered\ CPUs E5-2687W v2 at 3.40GHz with 264 gigabytes of RAM. 
The evaluation times depend on the boundary data, as we will explain in the next sections. For an Euclidean 4-simplex, the main plot of section \ref{sec:euclidean} took approximately 3 weeks. For vector geometries the two plots in section \ref{sec:vector} took about 30h each. Lorentzian 4-simplices are even more demanding, and a bottleneck in our computational precision restricted the number of data points accessible to us. These are shown in section \ref{sec:lorentzian}, which took about 2 days.
These were enough to provide evidence of a critical behavior with the predicted power law falloff, as well as nonmonotonic behavior depending on the Immirzi parameter. But a more significative agreement with the asymptotic formula, in particular seeing the predicted oscillations, was not obtained, and will require more computational work.
Our results also show that the simplified model introduced in \cite{Boosting} captures the key features of the full EPRL asymptotics for vector and Euclidean Regge geometries, but fails to do so, and instead decays exponentially, for Lorentzian geometries.

The increased difficulty of Lorentzian 4-simplices has partially to do with the fact that one cannot resort to an equispin configuration. The best we found was a certain isosceles configuration, but even then one has to deal with individual spins that are much higher than the value of the rescaling parameter controlling the asymptotics. Furthermore, the internal sums introduced by the method of \cite{Boosting} turn out to converge more slowly than for a Euclidean 4-simplex. This fact actually has a geometric origin, which explains how the Lorentzian 4-simplex is reconstructed from many Euclidean 4-simplices compatible with the SU(2) $\{15j\}$ symbol, and whose tetrahedra are suitably transformed thanks to the half-edge booster functions. This discussion also clarifies the critical behavior observed for the simplified model.

To obtain the numerical results presented, it proved necessary to develop an efficient algorithm to construct the boundary data. This is a spin-off of our work, and we devote a section of the paper to explain how this problem can be solved and provide in the same online repository \cite{code-sl2cfoam} the relevant pieces of code.
For the vector geometries, we used the explicit parametrization worked out in \cite{IoSU2asympt}. For the Regge geometries, we used a ``deconstruction'' algorithm, whereby we start from a geometric 4-simplex identified by its vertices, and derive the 3D data associated with it by the EPRL model's $Y$ map. This procedure exposes some useful properties of the 3D data and was helpful for us to better understand the details of the saddle point analysis of \cite{BarrettLorAsymp}. We provide some more explicit formulas for the Hessian at the saddle point, needed for the numerical comparison of the analytic formula with the data points.

Barring the increased computational power to confirm precisely the analytic formula for Lorentzian 4-simplices, our work confirms the validity of this crucial result for the semiclassical limit of the EPRL model, but more importantly supports the robustness of the method \cite{Boosting} and code \vaccachespacca\ \cite{Dona:2018nev}, which can now be used to perform more ambitious calculations. To that end, we point out the heavy cost of using coherent intertwiners. These are needed for the asymptotic formula, but many physical applications of the EPRL spin foam model can be done in the much cheaper orthonormal basis of intertwiners. Some of us presented calculations with two nonsimplicial vertices in \cite{noiGen}. The present code is much more performing than the one there used, and we hope to apply it in future work to calculations with a few simplicial vertices and orthonormal intertwiners.

With the intent to make the paper concise and its technical results easier to appreciate, we refer to the existing literature for the necessary background. In particular, to \cite{EPRL,LS2,FK} for the analytic background on the EPRL model, to \cite{BarrettLorAsymp} for its saddle point approximation (see also \cite{BarrettCraneLor,BarrettEPRasymp,ConradyFreidel2,BarrettSU2,HanZhangLor} for related work), to \cite{Boosting} for the method used for the exact evaluation, and to \cite{Dona:2018nev} for the numerical code, describing its details, functioning scheme, and generic performance tests. Section 2 contains the minimal information about the vertex amplitude for the paper to be self-contained and gives an overview of the approach used for the calculation. Section 3 describes our algorithm to identify boundary data with the required properties, and the explicit configurations used in the numerical calculations.
Section 4 discusses the central point of the convergences of the internal sums and presents a useful approximation that we use to shorten significantly the numerical time of some numerical evaluations. Section 5 contains our main results, a selection of data and their comparison with the analytic asymptotic behavior. Section 6 contains a discussion on the role of the booster functions in mapping  Lorentzian 4-simplices from the Euclidean ones associated with the SU(2) $\{15j\}$ symbol,  as well as a summary of the numerical situation and future developments. 
After our conclusions, we supplement the paper with two Appendices. In one we provide explicit formulas to fix our conventions. In the second we summarize the asymptotic analysis of the EPRL vertex \cite{BarrettLorAsymp} and report on the form of the Hessian at the critical point. 

%----------------------------------------------------------------------------
\section{EPRL vertex amplitude and its asymptotic limit \label{Sec2}}
%----------------------------------------------------------------------------

The (coherent) vertex amplitude for the Lorentzian EPRL spin foam model \cite{EPRL,FK} is an $\SL(2,\C)$ invariant associated to the boundary graph of a 4-simplex,
\be\label{A1}
A_v(j_{ab},\vec n_{ab}) := \int \prod_{a=2}^5 dh_a \prod_{a<b}  D^{(\g j_{ab},j_{ab})}_{j_{ab},-\vec n_{ab}, j_{ab}, \vec n_{ba}}(h_a^{-1}h_b),
\ee
where $h_a\in\SL(2,\C)$, and $D^{(\r,k)}(h)$ are infinite-dimensional unitary representations of the principal series, labeled by $\r\in\R$ and $k\in\Z/2$. 
The irreducible representations (irreps) are restricted to satisfy $\r_{ab}=\g k_{ab}$, where $\g$ is the Immirzi parameter, and expressed using Naimark's canonical basis with minimal spin eigenvalues $j_{ab}=k_{ab}$, and unit vectors $\vec n_{ab}$ used to build coherent intertwiners \cite{LS,LS2}. Conventions and explicit formulas are reported in  Appendix~\ref{AppA}. The minus sign in the vectors appearing as rows is for later convenience.
The indices $a,b=1,\ldots 5$ stand for the nodes of the graph, and notice that one redundant integration has been removed, which is necessary to ensure finiteness \cite{EnglePereiraFiniteness,Baez:2001fh} (see also the discussion in \cite{noiGen}).

\begin{figure}[H]
\centering
\includegraphics[width=10cm]{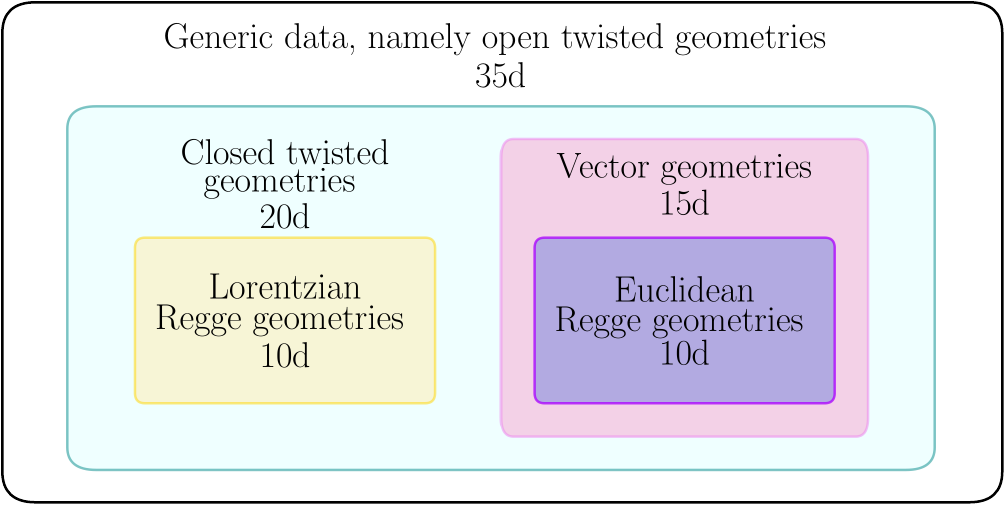}
\caption{\label{schemone} \small{\emph{Classification of the boundary data. It is interesting to compare the situation with SU(2) BF theory \cite{BarrettSU2,IoSU2asympt} and with the Euclidean EPRL model \cite{EPR,FK,LS2}. All three models have the same boundary states and thus some boundary data $(j_{ab},\vec n_{ab})$ in the coherent amplitudes. What changes are the group elements being integrated over [which belong to $\SL(2,\C)$, $\SU(2)$ and $\SO(4)$ respectively] and therefore the existence of critical points for a given set. For instance, Lorentzian Regge data only admit critical points for the $\SL(2,\C)$ model.}} } 
\end{figure}

The boundary data of the amplitude are ten spins $j_{ab}=j_{ba}$ and twenty unit vectors $\vec n_{ab}\neq \vec n_{ba}$. 
Four different subsets of data play an important role, see Fig.~\ref{schemone}: First, closed twisted geometries, defined by the following closure conditions, 
\be\label{clos}
\sum_{b\neq a} j_{ab} \vec n_{ab} = 0, \qquad \forall a=1.
\ee
This means that the spins and normals data around each node define a tetrahedron.\footnote{To be more precise, twisted geometries have additional `twist' angle variables, denoted $\xi_{ab}$ and canonically conjugated to the spins \cite{twigeo,twigeo2}. These angles do not show up in the coherent amplitude \Ref{A1}, defined by eigenvectors of the Casimir spin operator. They would appear in amplitudes where also the areas are semiclassical and not quantum, like in the graviton propagator calculations, see e.g. \cite{Bianchi:2006uf,IoASL}. The boundary data of \Ref{A1} are thus a subset of twisted geometries with $\xi_{ab}=0$.}

Within the space of closed twisted geometries, we can distinguish two disconnected subsets. One is vector geometries, if furthermore there exist five rotations $R_a\in\SO(3)$ such that the two normals associated to each face can be made opposite to one another:
\be\label{ori}
R_a \vec n_{ab}= - R_b\vec n_{ba}.
\ee  
These rotations are determined by $h_a$ at the corresponding critical point. Clearly, the simplest possible case is when the normals are pairwise antiparallel,
\be\label{btb}
\vec n_{ab}= - \vec n_{ba},
\ee  
which plays a useful role in choosing data. It is a remarkable fact that all Euclidean Regge geometries, namely all Euclidean 4-simplices in this context, are a subset of vector geometries. They are characterized by additional shape-matching constraints on top of Eq. \Ref{ori}, 
which can be described either using bivectors \cite{BarrettLorAsymp,Barrett:1997gw} or spherical cosine laws \cite{DittrichSpeziale,IoSU2asympt}. Their neat implication is to guarantee that the five tetrahedra determined by the closure conditions coincide with those determined as the boundary of a 4-simplex having the spins as areas.\footnote{That this is compatible with Eq. \Ref{ori} is not obvious, and one of the key results of the reconstruction.}
What happens in the simplest case \Ref{btb} is that there is always one critical point with all holonomies at the identity, and only when the shape-matching conditions are further satisfied does one find a second non trivial critical point that allows the reconstruction of the Euclidean 4-simplex.

The second disconnected subset of the closed twisted geometries are the Lorentzian Regge geometries, for which the boundary data coincide with those determined by a Lorentzian 4-simplex with all tetrahedra spacelike. The subset can again be characterized by shape-matching constraints in terms of bivectors \cite{BarrettLorAsymp,Barrett:1997gw} or Lorentzian spherical cosine laws \cite{IoFabio,toappear}.
Furthermore, while in this subset \Ref{ori} are not satisfied, it is still true that the 3D normals are related by five \emph{complex} transformation $H_a$ such that
\be\label{LorMatching}
H_{a} \vec{n}_{ab}= - H_{b} \vec{n}_{ba}.
\ee
These unimodular complex matrices are non unitary 3D representations of Lorentz boosts. They are again determined by the $h_a$ at the corresponding critical point.

How these conditions allow the embedding of the boundary data in a flat Euclidean or Lorentzian 4-simplex will be explained in detail below.
It is in the meantime useful to remark that the areas alone suffice to characterize a 4-simplex and distinguish whether it is Euclidean or  Lorentzian. Data with areas compatible with a Euclidean 4-simplex can be either a vector geometry or a Euclidean Regge geometry depending on the compatibility or lack thereof of the normals with the Euclidean 4-simplex. This requirement can be expressed in terms of the shape-matching conditions reducing a twisted geometry to a Regge geometry \cite{DittrichSpeziale,IoSU2asympt}. 

The coherent amplitude \Ref{A1} is a complex number, and it is important to make a remark about its phase before continuing.
The original formulation of the model \cite{EPRL} uses the standard basis of orthonormal intertwiners, which are labeled by a spin number $i_a$ per vertex, instead of the coherent intertwiners, labeled by unit vectors $\vec n_{ab}$. The resulting vertex amplitude $A_v(j_{ab},i_a)$  is a real number, provided a certain phase convention for the Lorentz irrep matrices is chosen \cite{Boosting}. The interest in using the complex coherent amplitude \Ref{A1} comes from the finer characterization of the boundary data and the clearer geometric interpretation that it permits. However, while the original amplitude is SU(2) invariant, Eq. \Ref{A1} is only SU(2) covariant: A rotation of the 4-normals on a node preserves the amplitude only up to a phase.\footnote{As for full Lorentz invariance, this is broken by the $Y$-map identifying $k=j$. Nonetheless, it can be shown that the vertex amplitude transforms covariantly under Lorentz transformations of the boundary data, and when various vertex amplitudes are glued together to form an extended spin foam, Lorentz invariance is restored in the bulk \cite{IoCarloCov}.}
This phase is irrelevant for geometric considerations because the coherent states themselves are only defined up to a phase, thus for any chosen boundary configuration, one can suitably tune the coherent state phase to have a real amplitude.
Accordingly, we can neglect this phase and factor out the rotational gauge freedom at each node in our geometric considerations. 
Up to rotations, the boundary data span a 35-dimensional space. The closed twisted geometries are a 20-dimensional subset, $10$ areas and $2\times5=10$ shape parameters, the vector geometries a 15-dimensional subset, ten areas and five angles \cite{IoSU2asympt}, and the Regge geometries a ten dimensional subset, ten lengths or ten areas (up to some discrete ambiguity in the Euclidean case).

Thanks to properties of the coherent states, integrands like Eq. \Ref{A1} can be written in exponential form with a linear dependence on the spins \cite{LS}. In particular, under a homogeneous rescaling  $j_{ab}\mapsto\l j_{ab}$ we can write the integrand as $\exp \l S(g_a;j_{ab},\vec n_{ab})$. This gives the possibility of evaluating the amplitude with a saddle point approximation in the homogeneous large spin limit, $\l\mapsto\infty$.
This approximation was studied in  \cite{BarrettLorAsymp},\footnote{\label{footphase}The definition of the coherent vertex amplitude used here differs from the one in  \cite{BarrettLorAsymp} by an overall phase, given for $\g>0$ by 
$$
\exp \{i \sum_{a<b} (2j_{ab}(\pi+\Phi_{ab}) + \arctan\g)\}.
$$ 
This is due to a phase difference between the Wigner matrices and the antilinear pairing of two unitary representations, as well as to our use of the antipodal map instead of the complex structure in associating coherent states to the target nodes.
See Appendix~\ref{AppA} for details.} 
and it was found that the leading order asymptotics of the coherent amplitude \Ref{A1} has the following dependence on the boundary data:
\begin{enumerate}
\item For generic open or closed twisted geometries: exponential decay in $\l$;
\item For vector geometries: power law decay $\l^{-12}$; 
\item For Regge geometries: power law decay $\l^{-12}$ with nonambiguous real oscillations, and frequency given by the Regge action. The Regge action is independent of $\g$ for Euclidean data and depends linearly on it for Lorentzian data.

\end{enumerate}

Our goal in this paper is to test numerically these saddle point approximations,  in particular the power law decays and the frequency of the oscillations. Such a numerical evaluation is not an easy task: The formula for the amplitude contains 12 unbounded integrations, and a complex and rapidly oscillating integrand built out of sums of products of hypergeometric functions. A direct approach using for instance adaptive Monte Carlo methods\footnote{As was done for the graviton propagator with the Euclidean Barrett-Crane model \cite{IoDan}.} appears daunting, and we are not aware of any results even for the much simpler Lorentzian Barrett-Crane integrals.
Facing these difficulties, the idea of \cite{Boosting} was to reduce as much as possible the unbounded integrations, taking advantage of the known factorization of $\SL(2,\C)$ Clebsch-Gordan coefficients into SU(2) ones. Applied to Eq. \Ref{A1}, the method in \cite{Boosting} splits the expression into a convolution between an SU(2) $\{15j\}$ symbol and one-dimensional boost integrals, called booster functions and denoted   $B^\g_4$:
\begin{align}
\label{Ac} 
A_v \left(j_{ab}, \, \vec n_{ab}\right) &= \sum_{l_{ab}\geq j_{ab}}\sum_{k_{a},i_{a}} \{15j\} 
\prod_{a=2}^{5} d_{k_{a}} B^\g_{4}(j_{ab},l_{ab};i_{a}, k_{a})  \prod_{a=1}^5 c_{i_a}(\vec n_{ab}) \\\label{AvG}
& = \sum_{l_{ab}, k_{a},i_{a}} \raisebox{-25mm}{ \includegraphics[width=8cm]{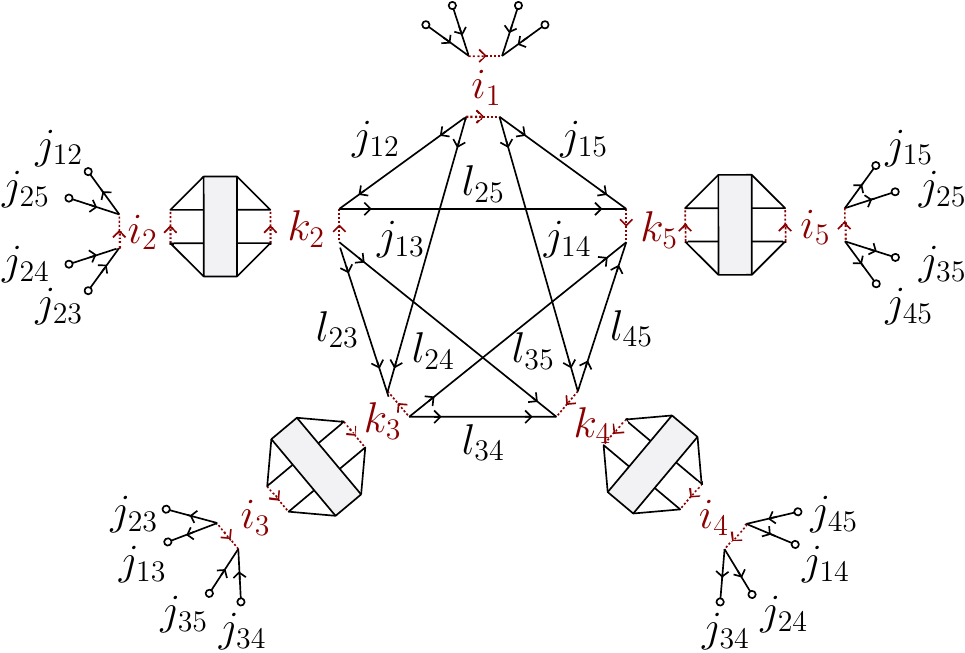}}.
\end{align}
Here $d_{j} := 2 j +1$. The $\{15j\}$ symbol is labeled by internal spins $l_{ab}$ and intertwiners $k_a$, except at the node $1$ without group integration, where it is labeled directly by the boundary spins and intertwiners. The unbounded sums over $l_{ab}$ go from $j_{ab}$ to infinity, whereas the sums over the intertwiner labels $k_a,i_a$ are bounded by the usual SU(2) triangle inequalities. We  define the simplified EPRL model consistently with \cite{Boosting,Dona:2018nev, noiGen} as the truncation of the EPRL amplitude \eqref{AvG} where only the first term in the summation $l_{ab} = j_{ab}$ is considered.

The booster functions depend both on the internal and the boundary data, and to obtain the coherent amplitude \Ref{A1}, they are contracted with the coefficients $c_i$ of the coherent intertwiners determined by the normals $\vec n_{ab}$. 
The graphical notation used in the second line helps to keep track of which links carry internal spins, as well as of the orientation of the normals.
The booster functions encode the details of the $Y$-map defining the EPRL model, and are defined as follows,
\begin{equation}
\label{boosters}
B^\g_4(j_f,l_f;i,k) = d_i d_k \sum_{m_f} \left(\begin{array}{c} j_f \\ m_f \end{array}\right)^{(i)} \left(\begin{array}{c} l_f \\ m_f \end{array}\right)^{(k)}\frac{1}{4\pi} \int_0^\infty  \mathrm{d} r \sinh^2r
 \prod_{f=1}^4 d^{(\gamma j_f,j_f)}_{j_f l_f m_f}(r) = \raisebox{-12.5mm}{ \includegraphics[width=2.2cm]{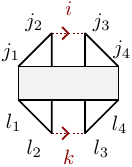}},
\end{equation}
in terms of generalized four-legged $(4jm)$ Wigner symbols and boost matrix elements  $d^{(\rho,k)}(r)$, see \cite{Boosting} for explicit formulas. To not be confused with $d_{i}$ and $d_k$ the dimension of the SU(2) irrep of spin $i$ and $k$, see Appendix \ref{AppA} for more details on our conventions. For all links outgoing, the coherent intertwiner coefficients are
\begin{equation}
\label{eq:coeffCS}
c_{i}(\vec n_f):= 
\sum_{m_f} \left(\begin{array}{c} j_f \\ m_f \end{array}\right)^{(i)} \prod_{f=1}^4D^{(j_f)}_{m_fj_f}(\vec{n}_f)
=  \raisebox{-5mm}{\includegraphics[width=2.8cm]{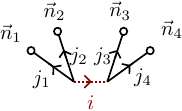}},
\end{equation}
where
\be
\label{eq:wignerM}
D^{(j)}(\vec n) = e^{-i\Phi J_z} e^{-i\Th J_y} e^{i\Phi J_z}
\ee
is a Wigner matrix with $\Th$ and $\Phi$ the zenithal and azimuthal angles of $\vec n$.
If one of the links is incoming one multiplies the above formula by $\eps^{(j)}_{mn}=(-1)^{j-m}\d_{m,-n}$.

In spite of its complexity, the vertex amplitude \Ref{Ac} is in a form suitable for numerical evaluations. The unbounded integrations have been limited to the booster functions, that are accessible numerically, but at the price of introducing infinite summations, the sums over the internal spins $l$, or ``internal sums'' for short. These may be just as hard to deal with numerically if they cannot be efficiently truncated. We will discuss this point below in detail. 
Let us first examine the different ingredients required in setting up the algorithm. 

The first is evaluating the $\{15j\}$. Optimized algorithms exist in the literature for the simpler $\{6j\}$ and $\{9j\}$ symbols, but the $\{15j\}$ can be significantly more complicated. In fact, the problem would be intractable if one had to compute it starting from its definition as ten $(3jm)$ symbols, contracted following the graph drawn in Eg. \Ref{AvG}. We can instead exploit the invariance of the sum over intertwiners under recoupling moves to trade the irreducible $\{15j\}$ appearing in Eq. \Ref{AvG} with a reducible one, a strategy successfully applied in \cite{IoSU2asympt} to compute numerically the semiclassical limit of the coherent $\{15j\}$ symbol. 
The simplest choice is a reducible $\{15j\}$ given by one sum of five $\{6j\}$ symbols. The $\{6j\}$ symbols can then be very efficiently computed thanks to the libraries \texttt{wigxjpf} developed in \cite{Johansson:2015cca}, based on smart storage of factorials used to compute the symbols. With this procedure, computing a single $\{15j\}$ for spins of order 10 takes about $10^{-6}$ s.
We refer to the original paper for a detailed discussion of the library's performances, accuracy, and memory management.
 
The second is the booster functions. Here and in \vaccachespacca\ we use the definition \Ref{boosters} with a polynomial expression for the boost matrix elements in $e^{-r}$ derived in \cite{Francois}, which is more efficient to implement numerically than the usual expression as hypergeometric functions.\footnote{The expression of \cite{Francois} is very similar to the one previously derived in \cite{RashidBoost}, but with the crucial difference that it only has finite sums. It is obtained starting from the integral representation, whereas the one of \cite{RashidBoost} starts from the infinite power series for the hypergeometric functions.} This expression is then amended with the phase conventions of \cite{Boosting} to make the booster functions and vertex amplitudes with orthonormal intertwiners real. 
 We compute the numerical integrals at fixed magnetic labels $m_f$, then resum over the $(4jm)$ coefficients. This gives an upper estimate of $4\l^3$ integrals, since only three magnetic labels are linearly independent. The results are then stored using a hash table. With this procedure, computing a booster for spins of order $10$ takes approximately $6$ minutes on a single core, and needs to be computed only once. 
The main limitation of this method is that numerical instabilities become important at spins of order $50$. These instabilities prevent the convergence of the boost integrals and are presumably due to the algebraic manipulations of very large numbers in the integrand. We worked with the GNU MPFR library for multiple-precision floating-point computations. See \cite{Dona:2018nev} for explicit formulas and more details. This makes the boosters the most delicate part of our code, and an obvious target for improvements, as we will discuss below in Sec.~\ref{SecOverview}.

The third is the coherent intertwiner coefficients \eqref{eq:coeffCS}. This is the simplest part of the code, obtained contracting the Wigner matrices \eqref{eq:wignerM} with $(4jm)$ symbols computed with \texttt{wigxjpf}. The evaluation time of a single coherent state coefficient in \vaccachespacca\ , 
is negligible with respect to the other computation times. 

Finally one has to tabulate all these quantities and combine them to compute the amplitude. This is a delicate task that is performed using the hash-table algorithm described in detail in \cite{Dona:2018nev}.
The resulting code is available in \cite{code-sl2cfoam}, and can be downloaded and used to independently rederive the results presented here, or push them forward using more powerful computers. %For this paper, we used a server with 32 cores at 3.4GhZ.

%----------------------------------------------------------------------------
\section{Constructing the boundary data}
%----------------------------------------------------------------------------
\label{SecBD}
Our code allows us to compute the vertex amplitude for any boundary data, in principle. However, some choices are more efficient than others: for instance, symmetric configurations maximize the number of data points with a given maximal value of the rescaling parameter $\lambda$, thus optimizing the numerical effort. This is simply a consequence of the discreteness of the areas, and the fact that the numerical costs are strongly determined by the highest spin in the amplitude. For twisted, vector and Euclidean Regge geometries, one has access to the most symmetric configuration with all equal spins $j_{ab} \equiv j$, and this is the one we focused on.

For Lorentzian Regge geometries, it is not possible to take all spins as equal since an equilateral Lorentzian 4-simplex does not exist. We then devised a Mathematica code to span Lorentzian 4-simplices with all tetrahedra spacelike, and we scanned the space of configurations with integer or half-integer areas. We found that the case with minimal highest spin is a configuration with six spins equal to 2, and four spins equal to 5, which can then be rescaled by an integer multiple (but not a half-integer). 
The four highest spins can all be associated to the same tetrahedron, and it is then best to choose to gauge fix the $\SL(2,\C)$ integration on the same tetrahedron, so to avoid having to compute the heaviest booster function.

In all cases, we define the homogeneous rescaling parameter $\l$ as an integer multiplying the smallest allowed configuration, hence $j_{ab}=\l/2$ for equispin data, and $j_{ab}=(2\l,5\l)$ for Lorentzian boundary data. The two spin configurations are summarized in Table~\ref{T1}.
From a purely numerical perspective, these configurations are rather generic, in the sense that they test generic structures of the numerical code.
\bgroup
\def\arraystretch{1.3}
\begin{table}[H]
\centering\framebox{\begin{tabular}{l|l|cl}
Twisted, vector, Euclidean data & All spins: & $\l/2$ & $=\tfrac12, 1,\tfrac32, 2,\ldots$ \\\hline
Lorentzian data & Four spins in gauge-fixed tetrahedron: & $5 \l$ & $=5,10,\ldots$  \\
& Remaining six spins: & $2 \l$ & $=2,4,\ldots$
\end{tabular}}
\caption{\label{T1} \small{\emph{The two configurations used for boundary Regge data throughout the paper. 
The equispin data used for the Euclidean Regge geometries can also be used for twisted and vector geometry data: It is the choice of normals that distinguish the three cases.}} }
\end{table}
\egroup

The information just given suffices to follow the numerical analysis in the paper.
On the other hand, the actual construction of the boundary data turned out to be a nontrivial part of our analysis, and for the interested reader, we explain in the rest of this section how it can be done, and provide more details on the specific configurations we chose. The online repository \cite{code-sl2cfoam} contains the Mathematica notebooks that we used to construct the boundary data we used in our asymptotic analysis. The notebook can be used to generate the boundary data for any spin foam vertex amplitude. 

\bigskip

The first thing to keep in mind is that the amplitude's boundary data $(j_{ab}, \vec n_{ab})$ are purely 3-dimensional: $j_{ab}$ label irreps of the canonical SU(2) matrix subgroup corresponding to rotations preserving the time direction $t^I:=(1,0,0,0)$, and $\vec n_{ab}$ are unit vectors in $\R^3$. 
We then have both an algebraic and a geometric mapping of the data into four-dimensional structures. The algebraic mapping is straightforward: The stabilizer of $t^I$ is also the canonical SU(2) subgroup of $\SL(2,\C)$ used in Naimark's basis of the unitary representation of the principal series $(\r,k)$ used in the EPRL model. Then the mapping is given by the map $Y:j\mapsto (\g j,j)$, and $\ket{j,\vec n}\mapsto\ket{\g j,j;j,\vec n}$. See \cite{EPRL,BarrettLorAsymp,PierreReview} for more details.
The geometric mapping is based on looking at the 3D geometry described by the data, and studying when and how it can be taken as the boundary of a flat 4-simplex. This mapping is finer and allows the classification of the data reported in Fig.~\ref{schemone}.
Given a set of data $(j_{ab},\vec n_{ab})$, it can be directly tested whether it describes one of the special geometric classes. For our purposes though, we are interested in the reverse, ``deconstructing'' problem: start from a geometric object of interest, like a Euclidean or Lorentzian 4-simplex, and deduce the 3D data to be used as boundary states.

%----------------------------------------------------------------------------
\subsection{A closed twisted geometry}
\label{SecCTG}
\begin{figure}[H]
    \begin{subfigure}[b]{0.49\textwidth}
		\centering        
        \includegraphics[width=3.5cm]{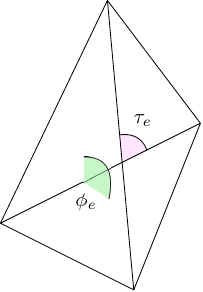}
    \end{subfigure}
    \begin{subfigure}[b]{0.49\textwidth}
    	\centering
       \raisebox{10mm}{\includegraphics[width=3.5cm]{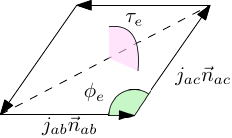}}
    \end{subfigure}
      \caption{\label{fig:twist}  \small{Left panel: \emph{Parametrization of the tetrahedron with given areas in terms of a dihedral angle $\phi_{e}$ and the corresponding twist angle $\tau_{e}$.} \\ Right panel: \emph{The dihedral angle $\phi_{e}=\phi^{(a)}_{bc}$ and the corresponding twist angle $\tau_{e}$ in the Kapovich-Millson polygon made with the scaled normals $j_{ab}\vec{n}_{ab}$ and $j_{ac}\vec{n}_{ac}$. The variable $\mu_{e}$ is the length of the dashed line.} } }
\end{figure}

%----------------------------------------------------------------------------
To generate the data of a closed twisted geometry it is convenient to start with five tetrahedra with areas matching along the graph of the 4-simplex. 
Each one can be characterized by the four areas and two independent 3D dihedral angles, $\phi_e\equiv \phi^{(a)}_{bc}$ at the edge $e$ of the tetrahedron $a$ identified as the one shared with the tetrahedra $b$ and $c$ in the graph. An alternative parametrization consists in using one dihedral angle and the twist angle $\t_e$ between the edge $e$ and the opposite edge, see Fig.~\ref{fig:twist}. The latter parametrization is closely related to the Kapovich-Millson \cite{Kapovich,IoPoly} conjugate variables $\left(\tau_{e}, \mu_{e} \right)$, where $\mu_{e}=({j_{ab}^2 + j_{ac}^2 - 2 j_{ab} j_{ac} \cos\phi^{(a)}_{bc}})^{1/2}$.
In terms of these variables, we considered the following set of equiarea tetrahedra:
\bgroup
\def\arraystretch{1.5}
\setlength{\arraycolsep}{2em}
\begin{equation*}
\begin{array}{ccccc}
\phi^{(1)}_{23}={\pi}/{2} & \phi^{(2)}_{12}={\phantom{2}\pi}/{3} & \phi^{(3)}_{12}={4\pi}/{5} & \phi^{(4)}_{12}={5\pi}/{6} & \phi^{(5)}_{12}={2\pi}/{3} \\ 
\tau^{(1)}_{23}={\pi}/{2} & \tau^{(2)}_{12}={2\pi}/{3} & \tau^{(3)}_{12}={5\pi}/{8} & \tau^{(4)}_{12}={3\pi}/{8} & \tau^{(5)}_{12}={5\pi}/{7}
\end{array} 
\end{equation*}
\egroup
These data correspond to five tetrahedra that, although all have the same areas, cannot be assembled to form a 4-simplex. As a result, we also have a certain freedom in picking up how to orient them in space. To keep things as simple as possible, we choose to orient all tetrahedra aligning a normal to a face with the $x$ axis. We further fix the rotational gauge by assuming that a normal to a second face lies in the $xy$ plane. The resulting normals are summarized in Table~\ref{Ttg}, and it can be explicitly checked that it is not possible to find local rotations so that all pairs at a given face satisfy the orientation equations \eqref{ori}.

\begin{table}[H]
\begin{center}
\begin{tabular}{c|ccccc} 
\parbox[c]{3em}{\includegraphics[scale=0.5]{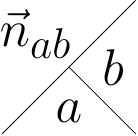}}  & $1$ & $2$ & $3$ & $4$ & $5$ \\ \hline
$1$ & & $\left(1, 0, 0 \right)$ & $\left(0 , 1 , 0 \right)$ & $\left(-0.5 , -0.5 ,  0.70 \right)$ & $\left(-0.5 , -0.5 ,  -0.70 \right)$  \\  
$2$ & $\left(1, 0, 0 \right)$ &  & $\left(0.5, 0.87, 0\right)$ & $\left(-0.62, -0.65, 0.43\right)$ & $\left(-0.88, -0.22, -0.43\right)$ \\ 
$3$ & $\left(1, 0, 0 \right)$ & $\left( -0.81, 0.59, 0 \right)$ &  & $\left( 0.25, -0.41, 
  0.88 \right)$ & $\left( -0.44, -0.18, -0.88 \right)$ \\ 
$4$ & $\left(1, 0, 0 \right)$ & $\left(-0.87, 0.5, 0 \right)$  & $\left(-0.42, -0.15, 
  0.89\right)$  &  & $\left(0.29, -0.35, -0.89\right)$  \\ 
$5$ & $\left(1, 0, 0 \right)$ & $\left(-0.5, 0.87, 0 \right)$  & $\left(0.22, -0.70, 
  0.68 \right)$  & $\left(-0.72, -0.17, -0.68 \right)$  &  
\end{tabular}
\end{center}
\caption{\label{Ttg} \small{\emph{Choice of normals for the closed twisted geometry described in the text. The gauge is fixed so that for every tetrahedron there is one normal in the $x$ direction, and a second normal in the $xy$ plane. The additional alignment $\vec n_{23}=(0,1,0)$ is a consequence of the straight angle $\phi_{23}=\pi/2$ among the shape parameters.}} }
\end{table}

%----------------------------------------------------------------------------
\subsection{A vector geometry}
%----------------------------------------------------------------------------
\label{SecVG}
Vector geometries were originally characterized in \cite{BCasympt2}, and an explicit parametrization was recently provided in \cite{IoSU2asympt}: One gives the ten areas, one dihedral angle $\phi^{(a)}_{bc}$ in four different tetrahedra, plus one additional angle between two normals belonging to different tetrahedra; the latter angle is not gauge invariant and has to satisfy an inequality. It is convenient to choose the gauge \Ref{btb} with all normals pairwise antiparallel.
A simple configuration following this construction, with all areas  equal and not giving a Regge geometry, is the following:
\begin{equation}\label{anglevg}
\phi^{(2)}_{34} = \frac{6}{15} \pi, \qquad \phi^{(3)}_{25} = \frac{8}{15} \pi, \qquad \phi^{(4)}_{25} = \frac{7}{15} \pi,
\qquad  \phi^{(5)}_{34} = \frac{8}{15}\pi,\qquad \phi_{53,42} = \frac{9}{15} \pi,
\end{equation}
where $\phi_{53,42}:=\arccos \left(\vec{n}_{53}\cdot \vec{n}_{42}\right)$ is the gauge-dependent angle. The resulting normals are summarized in Table~\ref{Tvg}.

\begin{table}[H]
\begin{adjustbox}{width=\columnwidth}
\begin{tabular}{c|ccccc} 
\parbox[c]{3em}{\includegraphics[scale=0.5]{_img/test.pdf}}  & $1$ & $2$ & $3$ & $4$ & $5$ \\ \hline
$1$ & & $\left(0.82, -0.45, -0.35 \right)$ & $\left(-0.96, 0.05, 0.28 \right)$ & $\left(-0.76, 0.54, -0.34 \right)$ & $\left(0.9, -0.15, 0.41 \right)$  \\  
$2$ & $\left(-0.82, 0.45, 0.35 \right)$ &  & $\left(0.31, 0.21, -0.93\right)$ & $\left(1, 0, 0\right)$ & $\left(-0.49, -0.66, 0.57\right)$ \\ 
$3$ & $\left(0.96, -0.05, -0.28 \right)$ & $\left(-0.31, -0.21, 0.93 \right)$ &  & $\left( -0.34, -0.69, -0.64 \right)$ & $\left( -0.31, 0.95, 0 \right)$ \\ 
$4$ & $\left(0.76, -0.54, 0.34 \right)$ & $\left(-1, 0, 0 \right)$  & $\left(0.34, 0.69, 0.64\right)$  &  & $\left(-0.1, -0.14, -0.98\right)$  \\ 
$5$ & $\left(-0.9, 0.15, -0.41 \right)$ & $\left(0.49, 0.66, -0.57 \right)$  & $\left(0.31, -0.95, 0\right)$  & $\left(0.1, 0.14, 0.98 \right)$  &  
\end{tabular}
\end{adjustbox}
\caption{\label{Tvg} \small{\emph{Choice of normals for the vector geometry \Ref{anglevg} described in the text. The gauge is fixed so that all normals are pairwise antiparallel, $\vec{n}_{ab}=-\vec{n}_{ba}$.}} }
\end{table}

%----------------------------------------------------------------------------
\subsection{A Euclidean Regge geometry}
\label{SecER}
%----------------------------------------------------------------------------
By Euclidean Regge geometry we refer to boundary data that can be embedded in $\R^4$ to form a Euclidean flat 4-simplex. To construct such data, we start from a flat Euclidean 4-simplex, as the convex envelope of its five vertices. 
Using the freedom to do SO(4) transformations, we can always put a vertex in the origin of $\R^4$, one on the $x$ axis, a third one in the $xy$ plane, and a fourth one in the $xyz$ hyperplane. This exhausts the rotational freedom, and the position of the last vertex is free. 
From the vertices we compute the edge vectors $\ell_{e}^I$, with $I=0,1,2,3$ a Cartesian coordinate index. Their lengths $\ell_e:=|\!| \ell_e^I |\!| := \sqrt{\ell_e^I \ell_e^J \d_{IJ}}$ provide ten numbers that characterize uniquely the 4-simplex up to SO(4) transformations. 
From the  edge vectors we can also define a simple bivector to each triangle $t$ of the 4-simplex, given by $B_{t}^{IJ}:=2\ell_{e}^{[I} \ell_{e'}^{J]}$, where $ee'$ is any pair of edges in $t$. Then from 
$A^2:=|\!|B^{IJ}|\!|^2=B_{IJ}B^{IJ}/2$, or equivalently using Heron's formula from the edge lengths, we obtain the ten areas of the 4-simplex.\footnote{Note that the resulting function from vertices (or equivalently from lengths) to areas is always injective and generically bijective; it fails to be bijective in special cases, like precisely for all equal areas, that can correspond to either an equilateral 4-simplex, or a 4-simplex with seven equilateral triangles and three isosceles. The 3D normals then distinguish the two cases.} 

The amplitude's boundary data are areas and 3D normals in the time-gauge frame. Therefore the next step is to find the transformation that takes all tetrahedra of our 4-simplex to the time-gauge, then compute the resulting 3D normals to the faces. Only then we can identify the SO(4)-invariant geometric area with the SU(2) irrep label $j$. To do so, we take as reference frame the $t=0$ hyperplane to which the first tetrahedron belongs (the one determined by the first four vertices fixed earlier). Its 4D normal is already the canonical one for the time-gauge, $t^I=(1,0,0,0)$. Hence we can directly compute the areas and 3D normals of this tetrahedron from the edge vectors, and take them as the first set of boundary data. For each of the remaining four tetrahedra, the procedure is not as straightforward. First, we determine the 4D normal from the triple product of edges with a common vertex, $e=1,2,3$,
\begin{equation}\label{defN}
N_{aI}=  \frac{\epsilon_{IJKL} \ell_{a1}^J \ell_{a2}^K \ell_{a3}^L}{|\!| \epsilon_{IJKL} \ell_1^J \ell_2^K \ell_3^L |\!|},
\end{equation}
ordered so that it is outgoing.\footnote{This is determined computing the Euclidean scalar product between the 4D normal and the vector connecting the chosen vertex to the center of the 4-simplex. Throughout the paper, we always work with outgoing normals, both 3D and 4D, and accordingly all dihedral angles are external.}
The SO(4) transformations mapping these vectors to $t^I$ can be found with a formula similar to Rodrigues's 3D rotation formula,
\begin{align}
\label{so4transf}
& \Lambda_a{}^I{}_J =\delta_{J}^I - \frac{1}{1+N_a \cdot t}\left(N_a^I N_{aJ}+  t^I t_J + N_a^I t_{J} -(1+2N\cdot t) t^I N_{aJ} \right), \\\nn
&  \L_a{}^I{}_J N_a^J= t^I, \qquad \det \L_a{}^I{}_J=1. 
\end{align}
The 4D normals define the 4D dihedral angles 
$\theta_{ab}=\arccos\left(N_a \cdot N_b \right)$, and provide alternative expressions for the face bivectors as well. Using the Hodge dual $\star:=(1/2)\eps_{IJKL}$, we have in fact
\begin{equation}
\label{defB}
B_{ab} = |\!| B_{ab}|\!| \f{\star N_a\w N_b}{|\!|\star N_a\w N_b |\!|} = - B_{ba}.
\end{equation}
The simplicity of the bivectors is now manifest in that
\begin{equation}
N_{aI} B^{IJ}_{ab} = 0 = N_{bI} B^{IJ}_{ab},
\end{equation}
and sums on the triangles in the same tetrahedron close,
\begin{equation}
\label{closB}
\sum_{b\neq a} B_{ab} = 0 \qquad \forall a.
\end{equation}
Using Eq. \Ref{so4transf} we map each tetrahedron to the time-gauge and define the rotated bivectors in the frame of the reference tetrahedron,
\be\label{defBo}
\overset{\circ}{B}{}^{IJ}_{ab} :=  \L_a^I{}_K B^{KL}_{ab} \L_a^J{}_L.
\ee
This is simple in the time-gauge frame, i.e.
$
t_I \overset{\circ}{B}{}^{IJ}_{ab} = 0. 
$
Therefore its magnetic part vanishes, $\overset{\circ}{B}{}^{0i}_{ab} = 0$,
and we can identify a 3D vector with its electric part,
\be
\label{3Dnormals}
j_{ab} n^k_{ab}:=\epsilon^k_{\phantom{k}ij} \overset{\circ}{B}{}^{ij}_{ab},\qquad   |\!|\overset{\circ}B_{ab}|\!| =j_{ab}.
\ee
Equation \Ref{3Dnormals} defines the remaining boundary data for the tetrahedra 2 to 5. 
Notice that we can also give a covariant formula for all 3D normals, 
\begin{equation}\label{3DnormC}
n^I_{ab}:=\left(0,\vec{n}_{ab}\right)= \f{2}{ |\!| \star \!\overset{\circ}B_{ab}|\!| } t_J (\star \overset{\circ}{B}_{ab})^{IJ} =
\Lambda_{a}{}^I{}_J \frac{N_{b}^J-N_{a}^J\left(N_{a}\cdot N_{b}\right)}{\sqrt{1-\left(N_{a}\cdot N_{b}\right)^{2}}},
\end{equation}
where the case $a=1$ has $\L_1=\Id$ and $N_1=t$.
 The covariant equations are actually more convenient to implement in the numerical algorithm because they use only the 4D normals, and bypass the explicit reconstruction of the bivectors.

Pictorially, we can describe the above procedure as the 4-simplex being opened up 
by SO(4) rotations of four tetrahedra to the same $\R^3$ frame of the first tetrahedron. The resulting 3D object was called a \emph{spike} in \cite{IoSU2asympt}, to which we refer the reader for explicit figures. 
Since the data $(j_{ab},\vec n_{ab})$ defined by Eq. \eqref{3Dnormals} or Eq. \Ref{3DnormC} come from a Euclidean 4-simplex, we expect that they satisfy the saddle point conditions \Ref{clos} and \Ref{ori} of the vertex amplitude asymptotic analysis. Indeed, the closure conditions \Ref{clos} are satisfied since Eq. \Ref{closB} is $SO(4)$-invariant, thus it applies to $\overset{\circ}{B}_{ab}$ as well and in turn to the 3D normals. As for the orientation conditions \Ref{ori}, they follow from Eqs. \Ref{defB} and \Ref{defBo}. To see this, one notices that the 3D vector \Ref{3Dnormals} also coincides with the electric part of the self-dual part of $\overset{\circ}{B}_{ab}$, and that under Eq. \Ref{defBo} the self-dual part transforms like a rotation.

It is convenient to perform these rotations, so to have a configuration with pairwise-opposite normals \Ref{btb}. This configuration was called \emph{twisted spike} in \cite{IoSU2asympt}. It was there shown that the required rotations have directions $\vec{n}_{1a}$ and angles given precisely by the 4D dihedral angle $\theta_{1a}=\arccos\left(N_1 \cdot N_a \right)$. In other words, the twisted spike twists the axis of the tetrahedra by an amount corresponding to the original 4D dihedral angle, and in doing so one achieves the condition of pairwise-opposite normals \Ref{btb}. See \cite{IoSU2asympt} again for illuminating pictures.
The twisted spike has a computational advantage because all holonomies at one of the two critical points are the identity.

We implemented the algorithm described above in the Mathematica code \texttt{EuclideanBoundaryDataMaker} in \cite{code-sl2cfoam}, which can be used to generate data corresponding to arbitrary Euclidean 4-simplices.
For our numerics, we chose an equilateral 4-simplex. The main advantage is that having all areas equal, we can obtain a maximal number of data points for a fixed maximal numerical effort. The resulting normals corresponding to the twisted spike are summarized in Table~\ref{TER}.
\begin{table}[H]
\begin{adjustbox}{width=\columnwidth}
\begin{tabular}{c|ccccc} 
\parbox[c]{3em}{\includegraphics[scale=0.5]{_img/test.pdf}}  & $1$ & $2$ & $3$ & $4$ & $5$ \\ \hline
$1$ & & $\left(0, 0, 1 \right)$ & $\left(0.94, 0, -0.33 \right)$ &$\left(-0.47, 0.82, -0.33 \right)$& $\left(-0.47,- 0.82, -0.33 \right)$ \\
$2$ & $\left(0, 0, -1 \right)$ &  & $\left(-0.24, -0.91, 0.33\right)$ & $\left(0.91, 0.25, 0.33\right)$ & $\left(-0.67, 0.66, 0.33\right)$ \\ 
$3$ & $\left(-0.94, 0, 0.33 \right)$ & $\left(0.24, 0.91, -0.33 \right)$ &  & $\left(0.09, -0.66, -0.75 \right)$ & $\left(0.62, -0.25, 0.75\right)$ \\ 
$4$ & $\left(0.47, -0.82, 0.33 \right)$ & $\left(-0.91, -0.25, -0.33 \right)$  & $\left(-0.09, 0.66, 0.75\right)$  &  & $\left(0.53, 0.41, -0.75\right)$  \\ 
$5$ & $\left(0.47, 0.82, 0.33\right)$ & $\left(0.67, -0.66, -0.33\right)$  & $\left(-0.62, 0.25, -0.75\right)$  & $\left(-0.53, -0.41, 0.75\right)$  &  
\end{tabular}
\end{adjustbox}
\caption{\label{TER} \small{\emph{Choice of normals for the Euclidean Regge data, corresponding to the twisted spike. These are reconstructed from an equilateral 4-simplex using the algorithm described in the text.}} }
\end{table}

As a final comment it is useful to understand the relation between our ``deconstruction'' of the 4-simplex and the reconstruction that one does in the saddle point approximation, the critical point equations for the Euclidean spike admit the following four solutions (see Appendix~\ref{appendixhessian}), 
\begin{equation}\label{hER}
h^{(c)}_{a} = \pm\exp \left(\f i2 \th_{1a} \vec{n}_{1a} \vec{\sigma} \right), 
\qquad h^{({\scr P}c)}_{a} = \pm\exp \left(-\f i2 \th_{1a} \vec{n}_{1a} \vec{\sigma} \right) = h^{(c)}_a{}^\dagger,
\end{equation}
all with $h_a\in \SU(2)\subset\SL(2,\C)$.
These group elements are precisely the $\L_a$ appearing in our procedure, namely one can show that
\begin{equation}
\Lambda_a{}^I{}_J = \frac{1}{2} \mathrm{Tr} \left( \sigma^I h_a^{\scr (c)} \sigma_J h_a^{\scr (c)}{}^\dagger\right)\ ,
\end{equation}
where $\sigma^I = \left(\mathds{1},\vec{\sigma}\right)$, and $\s_I$ is lowered with the Euclidean metric.

%----------------------------------------------------------------------------
\subsection{A Lorentzian Regge geometry}
%----------------------------------------------------------------------------
\label{SecLD}
To obtain the Lorentzian boundary data, we follow the same strategy as the previous case, but some additional care will be needed to take into account the presence of future pointing and past pointing timelike vectors. We start from a Lorentzian 4-simplex with all tetrahedra spacelike\footnote{A tetrahedron on the boundary of a Lorentzian 4-simplex is spacelike if its four-dimensional normal $N_a$ is timelike, $N_a^I N_{aI} = -1$. Conversely, the tetrahedron is said to be timelike if its four normal is spacelike.}, and derive the corresponding 3D data in the time-gauge from it. We consider the five vertices in $\R^4$ with metric $\eta ={\rm diag}(-+++)$, and use the SO(1,3) freedom to place them as before, a vertex in the origin, one on the $x$ axis, a third one in the $xy$ plane, and a fourth one in the $xyz$ hyperplane, with the last one completely free. The tetrahedron described by the first four vertices will always be spacelike, whereas the nature of the other four tetrahedra  depends on the location of the fifth vertex. We derive as before the edge vectors $\ell_e^I$, bivectors $B_{ab}^{IJ}$, and 4D normals $N_a^I$. The norms and scalar products are now given by $\eta$, and for the epsilon symbol we take the convention $\eps_{0123}=1$. 

To find a convenient choice of data, notice first that a Lorentzian 4-simplex cannot be equilateral, just like a triangle in 2D. Just like in 2D, the most regular 4-simplex with all spacelike sub-simplices is isosceles, with one equilateral tetrahedron and four equal isosceles ones, so in the simplest configuration we have four areas of one value and six of a different one. To find suitable ones, we need both values to be integers or half-integers.
Recall also that the numerical algorithm becomes the more costly the higher the spins involved. Therefore to optimize the numerical calculations we want a configuration in which the ratio between the two areas is maximal, while at the same time minimizing the smallest integer or half integer realization of each area value. Using the algorithm \texttt{FromAreasToVericesLorentzian} in \cite{code-sl2cfoam}, we scanned the space of admissible configurations, and settled on the ratio $2/5$ anticipated above.\footnote{
Limiting ourselves to numerators and denominators smaller than 40, we identified the following 32 possible ratios, in decreasing order:
$$\frac{2}{5},\frac{15}{38},\frac{13}{33},\frac{11}{28},\frac{9}{23},\frac{7}{18},\frac{12}{31},\frac{5}{13},\frac{13}{34},\frac{8}{21},\frac{11}{29},\frac{14}{37},\frac{3}{8},\frac{13}{35},\frac{10}{27},\frac{7}{19},\frac{11}{30},\frac{4}{11},\frac{13}{36},\frac{9}{25},\frac{14}{39},\frac{5}{14},\frac{11}{31},\frac{6}{17},\frac{13}{37},\frac{7}{20},\frac{8}{23},\frac{9}{26},\frac{10}{29},\frac{11}{32},\frac{12}{35},\frac{13}{38}.$$ We found an upper bound at $1/\sqrt{6}\simeq .41> 2/5=.4$ for which the 4-simplex becomes degenerate, and a lower bound at $1/3$ for which the isosceles tetrahedra become null.}
This is a 4-1 configuration, namely four of the timelike normals are future pointing and one past pointing, or vice versa.

\begin{figure}[H]
 \begin{subfigure}[b]{0.49\textwidth}
		\centering        
        \includegraphics[width=6cm]{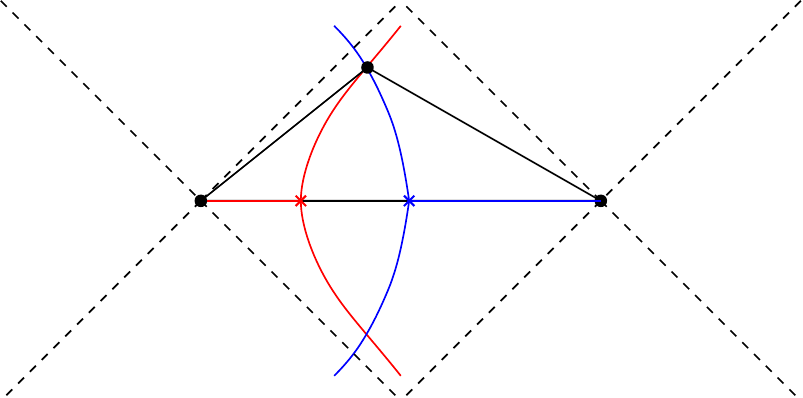}
    \end{subfigure}
    \begin{subfigure}[b]{0.49\textwidth}
    	\centering
       \includegraphics[width=6cm]{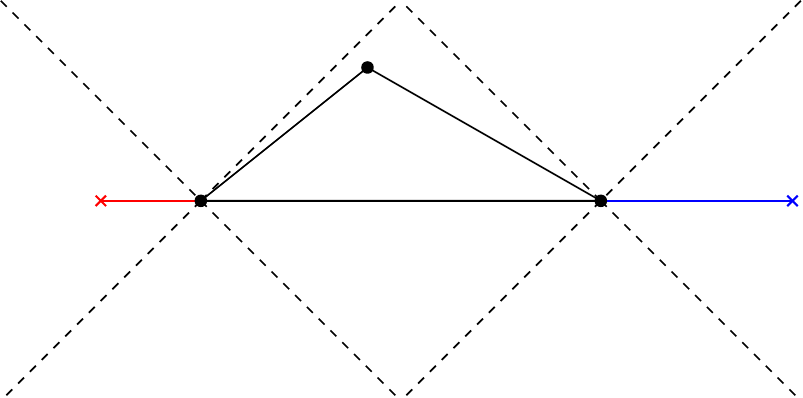}
    \end{subfigure}
      \caption{\label{antispike}  \small{Left panel:  \emph{$1+1$ spacelike Lorentzian simplex and its antispike configuration, obtained boosting two sides in the 1D frame of the base. Two normals are future pointing and one is past pointing. The two hyperbolae are the points you obtain boosting the two sides.}\\ Right panel: \emph{ The same simplex and the spike configuration, obtained acting with a $\hat T\hat P$ transformation after the antispike boosts. All normals are now pointing in the same direction.}}
      }
\end{figure}

Let us choose the gauge-fixed tetrahedron 1 to be past-pointing, i.e. $N_1=-t$. The remaining 4 normals are all future pointing, and can be computed from the edge vectors  as in \Ref{defN}.
To determine the 3D normals we proceed as in the Euclidean case, with $\L_a\in\SO(1,3)$ this time. To transform the future pointing 4-normals to $t$ we select the pure boost in the plane determined by $t$ and $N_a$, which can be computed to be
\begin{align}
\label{so31transf}
& \Lambda_a{}^I{}_J =\delta_{J}^I + \frac{1}{1- N_a \cdot t}\left(N_a^IN_{aJ} +t^It_J  + N_a^I t_J -(1-2N_a\cdot t) t^I N_{aJ}  \right), 
\qquad a\neq1, \\\nn
& \L_a{}^I{}_J N_a^J = t^I, \qquad \det \L_a{}^I{}_J=1.
\end{align}
We then have the expression \Ref{defB} for the bivectors and Eq. \Ref{defBo} for the time-gauge bivectors as before, and define again the 3D normals via Eq.  \Ref{3Dnormals}. The only difference is at the level of the covariant formula, which now reads
\begin{equation}
\label{3DnormalLOR}
n^I_{ab}:= \left(0,\vec{n}_{ab}\right)= \f{2}{ |\!| \star\! \overset{\circ}B_{ab}|\!| } t_J (\star \overset{\circ}{B}_{ab})^{IJ} =
- \Lambda_{a}{}^I{}_J \frac{N_{b}^J+N_{a}^J\left(N_{a}\cdot N_{b}\right)}{\sqrt{\left(N_{a}\cdot N_{b}\right)^{2}-1}}, \qquad a\neq1. 
\end{equation}
The 3D normals of the tetrahedron 1 are already in the chosen reference $\R^3$, and can then be directly computed from the vertices
(or equally from \Ref{3DnormalLOR} but with now $\L_1= - \mathds{1}$, due to the fact that the normal to this tetrahedron is past-pointing).
The resulting 3D object can be referred to as an antispike, because the tetrahedra boosted in the frame of the first will be ``sitting inside'' it, see Fig.~\ref{antispike} for a 1+1 example. And notice that the normals of the faces shared with the first tetrahedron will be aligned, and not antialigned. 
While this configuration is a good representation of a Lorentzian 4-simplex in 3D, it will \emph{not} feed a critical point to the coherent amplitude \Ref{As}, because of the minus signs in front of half of the vectors, which were chosen for convenience in dealing with the vector and Euclidean geometries.\footnote{The same would be true had we worked with the antipodal spinor conventions of \cite{BarrettLorAsymp} instead of the antipodal vectors.} 
This can be fixed if we turn around the reference tetrahedron applying an inversion, namely the composition $\hat T\hat P\not\in\SL(2,\C)$ of a time reversal and a parity transformation. Doing so we obtain again a picture with all tetrahedra outside, which we call the Lorentzian spike. Pictorially, it can be distinguished from a Euclidean spike because the tetrahedra spiking out are too ``short'' to allow the mapping to a closed Euclidean 4-simplex, see again Fig.~\ref{antispike} to get some intuition.

For the Lorentzian spike, all 4-normals are future pointing, and the 3D normals of the first tetrahedron are each antialigned with the corresponding one of the tetrahedron sharing that face, as in the Euclidean case. On the other hand, there is not now a set of five rotations so that Eq. \eqref{ori} holds, because the $\L_a$ are pure boosts for $a\neq 1$. This means that there is no twisted spike configuration, and no configuration with all critical holonomies at the identity. The 3D normals satisfy instead Eq. \eqref{LorMatching}, namely they can be antialigned by a set of five complex rotations.

The Mathematica code implementing this algorithm is called \texttt{LorentzianBoundaryDataMaker} in \cite{code-sl2cfoam}. 
The normals obtained for the 2/5 ratio 4-simplex are summarized in Table~\ref{TLR} below. 

\begin{table}[H]
\begin{adjustbox}{width=\columnwidth}
\begin{tabular}{c|ccccc} 
\parbox[c]{3em}{\includegraphics[scale=0.5]{_img/test.pdf}}  & $1$ & $2$ & $3$ & $4$ & $5$ \\ \hline
$1$ & & $\left(1, 0, 0 \right)$ & $\left(-0.33,0.94,0 \right)$ &$\left(-0.33,-0.47,0.82 \right)$& $\left(-0.33,-0.47,-0.82\right)$ \\
$2$ & $\left(-1, 0, 0 \right)$ &  & $\left(0.83,0.55,0\right)$ & $\left(0.83,-0.28,0.48\right)$ & $\left(0.83,-0.28,-0.48\right)$ \\ 
$3$ & $\left(0.33,-0.94,0 \right)$ & $\left(0.24,0.97,0 \right)$ &  & $\left(-0.54,0.69,0.48 \right)$ & $\left(-0.54,0.69,-0.48\right)$ \\ 
$4$ & $\left(0.33,0.47,-0.82 \right)$ & $\left(0.24,-0.48,0.84 \right)$  & $\left(-0.54,0.068,0.84\right)$  &  & $\left(-0.54,-0.76,0.36\right)$  \\ 
$5$ & $\left(0.33,0.47,0.82\right)$ & $\left(0.24,-0.48,-0.84\right)$  & $\left(-0.54,0.068,-0.84\right)$  & $\left(-0.54,-0.76,-0.36\right)$  &  
\end{tabular}
\end{adjustbox}
\caption{\label{TLR} \small{\emph{Choice of normals for the Lorentzian Regge data. These are reconstructed from the chosen isosceles 4-simplex with area ratio $2/5$, using the algorithm described in the text, and correspond to the spike configuration. }} }
\end{table}

In determining the analytic asymptotic formula to be compared with this configuration, let us recall the definition of the Lorentzian dihedral angles $\th^{\scr L}_{ab}\geq 0$. Following \cite{BarrettLorAsymp}, the case with both timelike normals inside the same causal patch is called thick wedge, and the case with the timelike normals inside opposite causal patches is called thin wedge. With signature $-+++$,  the scalar product of normals is negative in the first case, and positive in the second. Accordingly, we define
\begin{align}
& {\rm thick \ wedge} && N_a\cdot N_b =-\cosh\th^{\scr L}_{ab}, \\
& {\rm thin \ wedge} && N_a\cdot N_b =\cosh\th^{\scr L}_{ab}.
\end{align}
The 4-simplex of Table~\ref{TLR} is of type 4-1, and we have four thin angles $\th^{\scr L}_{1a}$ and six thick ones $\th^{\scr L}_{ab}$ for $a,b\neq1$. 

The solutions of the critical point equations for the Lorentzian spike are
\be\label{hLR}
h^{\scr (c)}_{a} = \pm\exp \left((\th^{\scr L}_{1a}+i\pi) \vec{n}_{1a}\cdot \frac{\vec{\sigma}}{2}\right), 
\qquad h^{\scr (Pc)}_{a} = \pm\exp \left((-\th^{\scr L}_{1a}+i\pi) \vec{n}_{1a} \cdot \frac{\vec{\sigma}}{2}\right) = h^{\scr (c)}_a{}^{-1}.
\ee
They are a combination of a boost given by the thin angle and an additional rotation by $\pi$ in the same direction. The origin of this rotation is the inversion performed to have the data describing a spike, so that the 3D normals in the reference tetrahedron are the opposite of the corresponding ones in the adjacent tetrahedra, and not parallel as in the antispike configuration with the reference tetrahedron being past pointing. To recover the pure boost \Ref{so31transf} we remove this additional rotation, and indeed one can check that
\begin{equation}
\Lambda_a{}^I{}_J = \frac{1}{2} \mathrm{Tr} \left( \sigma^I\hat h_a^{\scr (c)} \sigma_J \hat h_a^{\scr (c)}{^\dagger}\right),
\qquad \hat h^{\scr (c)}_{a} = \pm\exp \left(\th^{\scr L}_{1a} \vec{n}_{1a}\cdot \frac{\vec{\sigma}}{2}\right)\ ,
\end{equation}
where $\sigma^I = \left(\mathds{1},\vec{\sigma}\right)$, and where $\s_I$ is lowered with the Minkowski metric.

%----------------------------------------------------------------------------
\section{Methods: Shelled sums and a useful approximation}
%----------------------------------------------------------------------------

The finiteness of the vertex amplitude \cite{EnglePereiraFiniteness} guarantees that the internal sums in Eq. \eqref{AvG} converge. However, the speed of the convergence 
depends on the boundary data considered, since for different configurations, both the booster functions and the $\{15j\}$ symbol can decrease either exponentially or polynomially. This speed determines crucially the efficiency of our numerical method:
If the convergence is fast we can keep the lowest order terms only, but if the convergence is slow one has to add up more and more terms, and the numerical algorithm becomes slower and slower. 

To study the convergence, we introduce a homogeneous integer cutoff $\Delta$ on the range of the internal spins, 
\be\label{cutoff}
\sum_{l_{ab}=j_{ab}}^\infty \, \to \, \sum_{l_{ab}=j_{ab}}^{j_{ab}+\Delta}
%j_{ab}\leq l_{ab}\leq j_{ab}+\D,
\ee
and we compare the value of the amplitude for successive truncations.  
Notice that each of these truncated sums, or \emph{shelled} sums to use the terminology of \cite{Dona:2018nev}, can have a priori arbitrarily high internal spins $l$. It is only the difference with the boundary spins $j$ that is being truncated, namely the number of terms being summed over in Eq. \Ref{Ac}. The truncation $\Delta=0$ gives the simplified EPRL model defined in \cite{Boosting}.

\begin{figure}[H]
    \centering  \includegraphics[width=12cm]{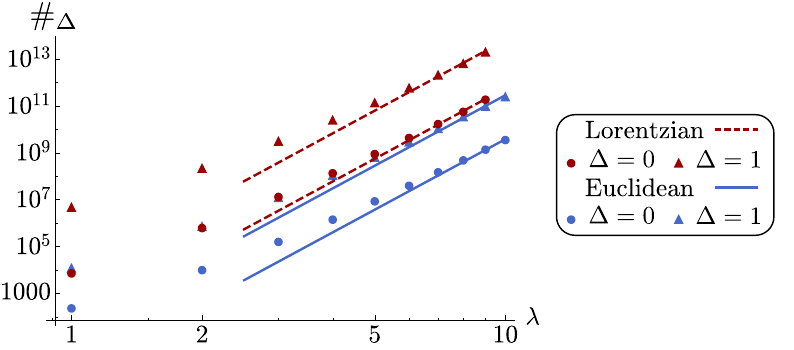}
      \caption{\label{fig:NumberTerms}  \small{ \emph{Total number of terms to be summed over at a given $\l$, for Euclidean (blue) and Lorentzian (red) boundary data, for the simplified model (circles) and the first shelled sum (triangles). The lines are obtained through a numerical fit with the $(\D+1)^6 \l^{10}$ estimate from \Ref{gio}. For Euclidean data, the fitted coefficients give $.38$ at $\D=0$ and $.45$ at $\D=1$. These numbers give an indication of the percentage of configurations removed by the triangle inequalities. For Lorentzian data, we cannot apply the counting leading to the estimate \Ref{gio} because the spin configurations are isosceles. Nonetheless, the numerical fit shown in the figure is also with slope $(\D+1)^6\l^{10}$, and it is in qualitative agreement (a numerical fit with free exponent gives $\l^{9.96}$ for $\D=0$ and $\l^{9.12}$ for $\D=1$). 
      This counting shows that for the same $\l$ (and thus same number of data points in the asymptotic plots), the Lorentzian configurations are numerically much heavier. Generating this plot took about 24h on a standard laptop.}}
      }
\end{figure}

Let us estimate the number of terms in a given truncation, for the equispin case $j_{ab}=j$. The $l$ configurations  
compatible with Eq. \Ref{cutoff} are 
$
(\D+1)^6.
$
For each $l$'s configuration, we have to sum over the internal intertwiners $k$. 
Notice from the graphical representation \Ref{AvG} that each internal intertwiner always has at least one external spin. Hence increasing $l$ has the effect of tightening the triangular inequalities for the intertwiners. Consequently, the external triangular inequalities from the $j$ spins provide an upper bound to the number of $k$ configurations, which is $(2j)^4$ when all $j$'s are equal. 
To complete the estimate we have to include the $2j$ terms coming from the sum defining the reducible $\{15j\}$ symbol used. Overall, this gives an upper bound of  
\be\label{gio}
(\D+1)^6 \times (2j)^5
\ee
configurations for the amplitude defined by boundary spins and intertwiners. For the coherent amplitude \eqref{Ac}, one also has to sum over the external intertwiners $i$, which gives an additional $(2j)^5$ terms. Recall that for this configuration $\l = 2j$, the final upper bound estimate is
\be\label{gioc}
\#_\D \leq (\D+1)^6 \times \l^{10}.
\ee
An exact numerical counting confirms this estimate, see the Euclidean data in Fig.~\ref{fig:NumberTerms}. A numerical fit reported there shows that the neat effect of the triangular inequalities is to remove a bit more than half of the configurations.

To give an idea of the complexity of the numerics involved, the amplitude for Euclidean data with $\l=16$ and $\D=2$ contains $10^{14}$ configurations, each of which requires the evaluation of $10^4$ integrals and five $\{6j\}$ symbols. The evaluation of this data point alone took three weeks on our server.

For Lorentzian boundary data, the situation is worse, since one cannot use equal spins. 
Therefore for the same value of the rescaling parameter $\l$, higher individual spins will enter, increasing the number of terms to be cycled over. 
For the $2/5$ ratio used as Lorentzian boundary data, we are limited at $\l=9$, which corresponds to highest spins $45$, beyond which we encounter numerical instabilities in the booster functions.
We counted numerically all the configurations allowed by triangular inequalities, and the results are shown in Fig.~\ref{fig:NumberTerms}.
The most expensive data point we bought for Lorentzian data has $\l=9$ and $\D=1$, and $10^{13}$ configurations to be evaluated.

These estimates  show the high price of increasing the cutoff $\D$ on the internal sums, making configurations for which the convergence is fast  much more affordable.
We also point out the $\l^5$ numerical cost coming from the use of coherent states in the boundary, which requires the additional summations over intertwiners. Hence our methods are much faster for calculations requiring orthogonal intertwiners and not coherent ones.

%------------------------------------------------------
\subsection{Convergence of the shells}
\begin{figure}[H]
    \centering
    \begin{subfigure}[b]{0.49\textwidth}
        \includegraphics[width=7.5cm]{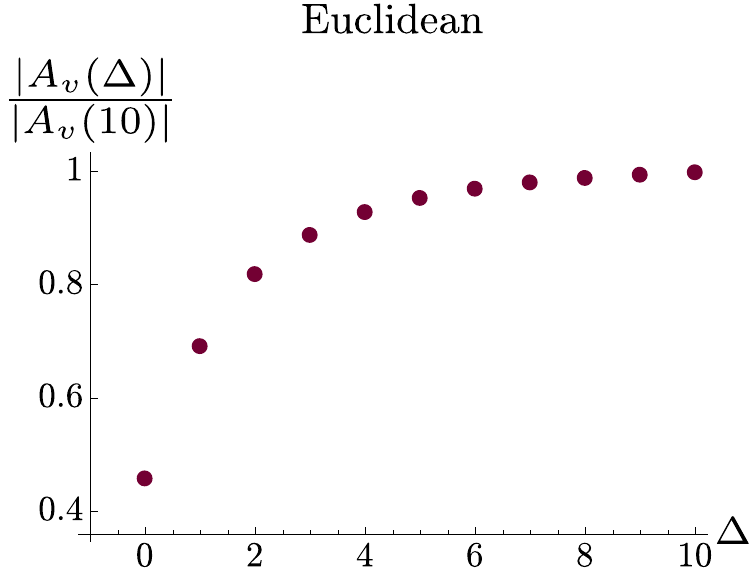}
    \end{subfigure}
    \begin{subfigure}[b]{0.49\textwidth}
        \includegraphics[width=7.5cm]{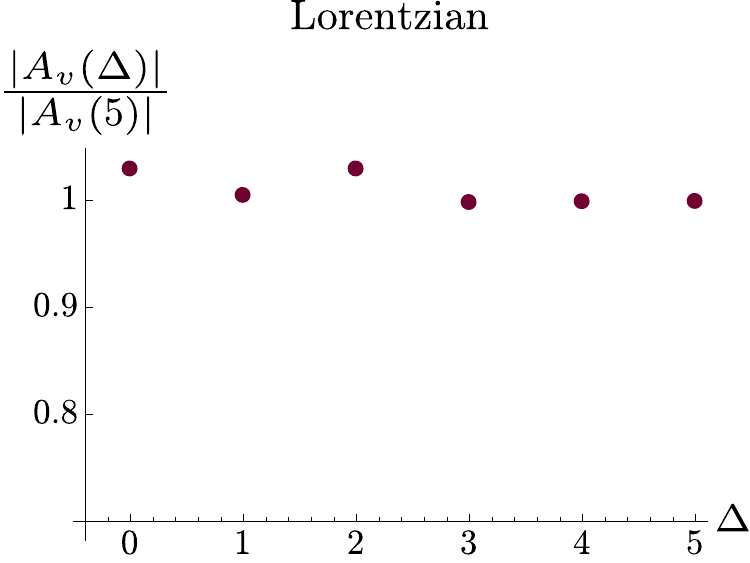}
    \end{subfigure}
      \caption{\label{FigConv}  \small{\emph{Convergence of the internal sums at $\l=1$ for Euclidean (left panel) and Lorentzian (right panel) data. 
      Closed twisted and vector geometries give plots undistinguishable from the one on the left panel, with numerical differences typically at 4 digits. 
      } } }
\end{figure}

%------------------------------------------------------
Having discussed the numerical costs of the evaluations, let us now present some investigations on the convergence of the internal sums. We restricted attention to the four types of boundary data presented earlier, which will be relevant to study the asymptotic scaling.
Figure~\ref{FigConv} shows examples of convergence for $\g=1.2$, and $\l=1$. At this $\l$ the convergence is quite fast for all four cases, with differences of $1\%$ or less when increasing the truncation from $\D=4$ to $\D=5$. 

\begin{figure}[H]
    \centering
    \begin{subfigure}[b]{0.49\textwidth}
        \includegraphics[width=7.5cm]{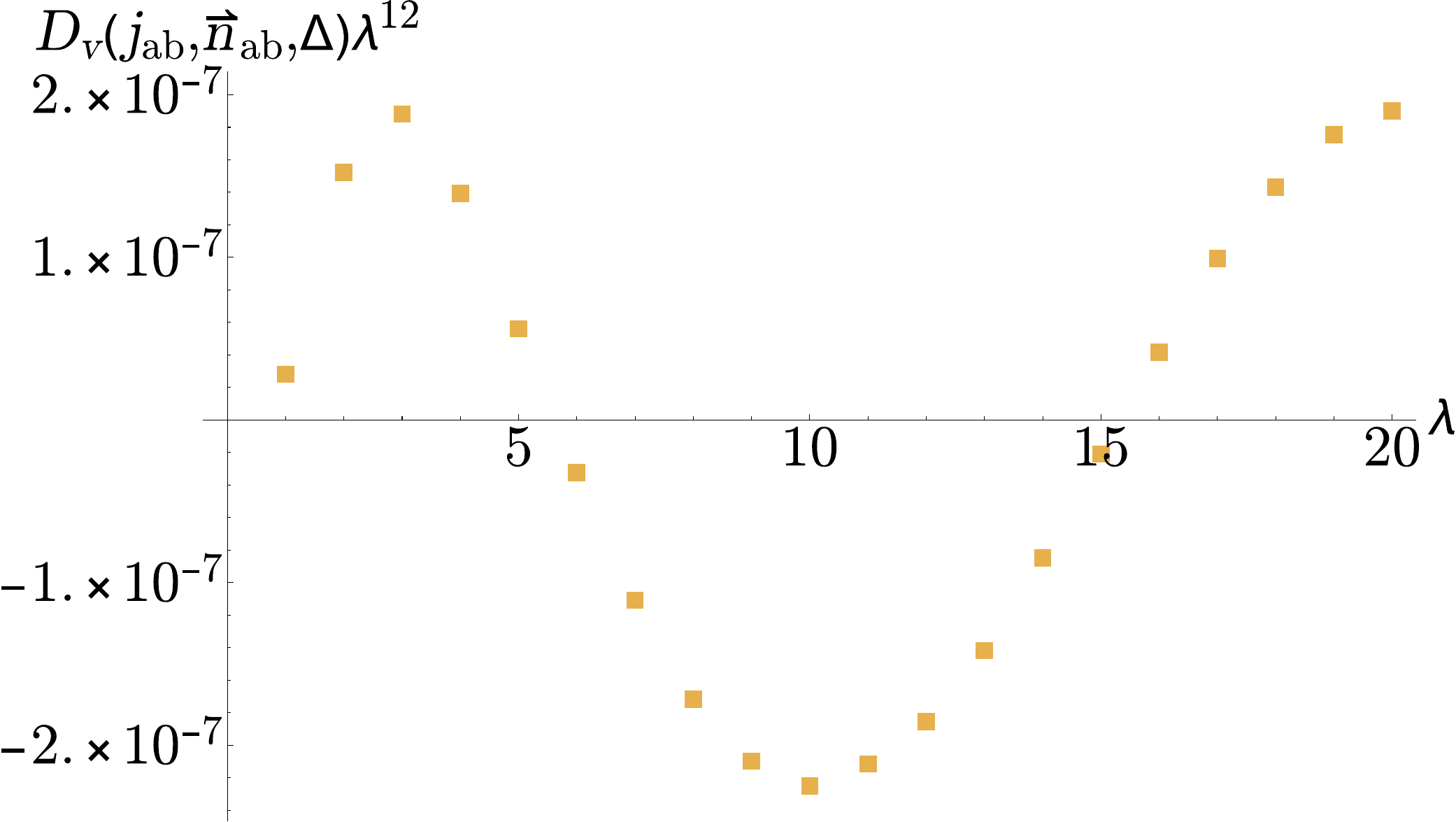}
    \end{subfigure}
    \begin{subfigure}[b]{0.49\textwidth}
        \includegraphics[width=7.5cm]{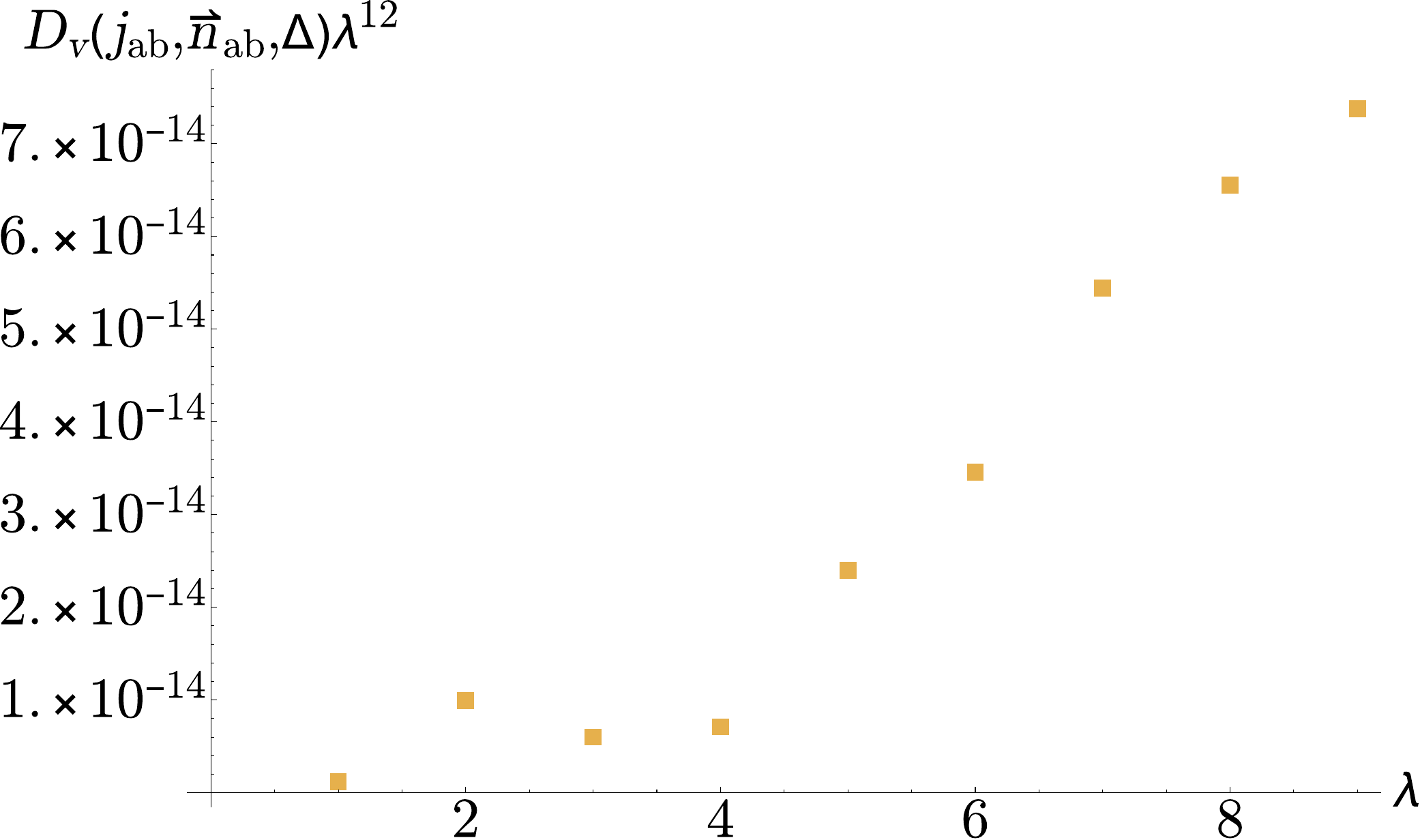}
    \end{subfigure}
      \caption{\label{fig:GapEvolution}  \small{\emph{ Behavior of the difference $D_v (j_{ab}, \vec{n}_{ab}, \Delta) = A_v (j_{ab}, \vec{n}_{ab}, \Delta = 1) - A_v (j_{ab},  \vec{n}_{ab}, \Delta = 0)$, for Euclidean (on the left) and Lorentzian (on the right) boundary data. The oscillating behavior in the Euclidean case is the Regge asymptotic behavior, as explained in the next Section. The growing behavior in the Lorentzian case is a clear indication that the simplified model is becoming subdominant with respect to the first shell as $\lambda$ increases. This will be clarified in the next Section, where it will be shown that  the simplified model misses the critical behavior and decays exponentially, while the first shell decays at power law.} } }
\end{figure}

The convergence can, however, change quite significantly for different configurations or different values of $\g$. More importantly for our scopes, it can also change as we increase $\l$. 
An analysis like Fig.~\ref{FigConv} for higher $\l$ is unfortunately  too costly and could not be performed. As a cheaper alternative, we looked at the gap $\D_1-\D_0$ between the first shelled sum and the simplified model. 
This is reported in Fig.~\ref{fig:GapEvolution}. 
We see that for Euclidean data the gap remains bounded at 30\% of the $\D=1$ amplitude. If the same happens for the other differences $\D_2-\D_1$, etc, it would mean that the speed of convergence is constant in $\l$. 
This possibility is supported by the full asymptotic study presented in the next section. 
In that context, we will also explain the oscillation seen in (the right panel of) Fig.~\ref{fig:GapEvolution} as the rephasing of the asymptotic formula performed by the shells.

For Lorentzian data, on the other hand, we observe a generic growth of the gap. This suggests that while the convergence was (very) fast at $\l=1$, it will likely slow down as $\l$ increases.
This slower convergence for Lorentzian data as $\l$ is increased will also be confirmed by the full asymptotic study of the next section, and suggests that the simplified model will miss important aspects of the full model for such boundary configurations.

Summarizing, our investigations indicate that convergence is fast at very small spins for all types of data considered, but that while it remains fast for twisted, vector and Euclidean Regge geometries, it slows down for Lorentzian Regge geometries.

%-----------------------------------------------------
\subsection{Approximating the coherent states}
\begin{figure}[H]
    \centering
    \begin{subfigure}[b]{0.49\textwidth}
        \includegraphics[width=7.5cm]{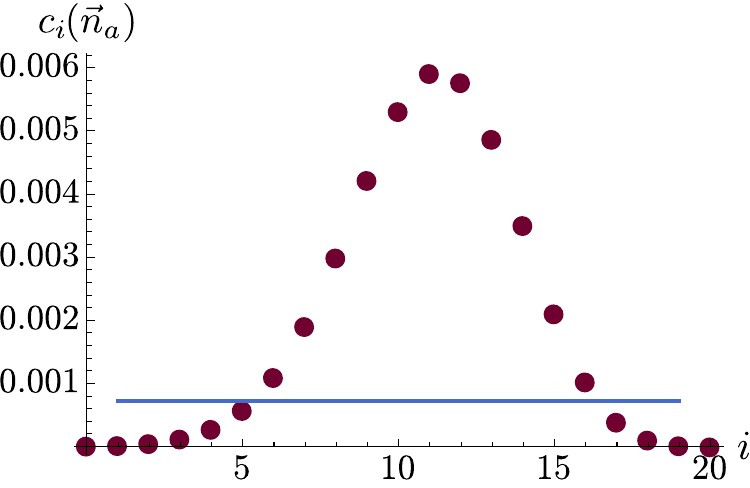}
    \end{subfigure}
    \begin{subfigure}[b]{0.49\textwidth}
        \includegraphics[width=7.5cm]{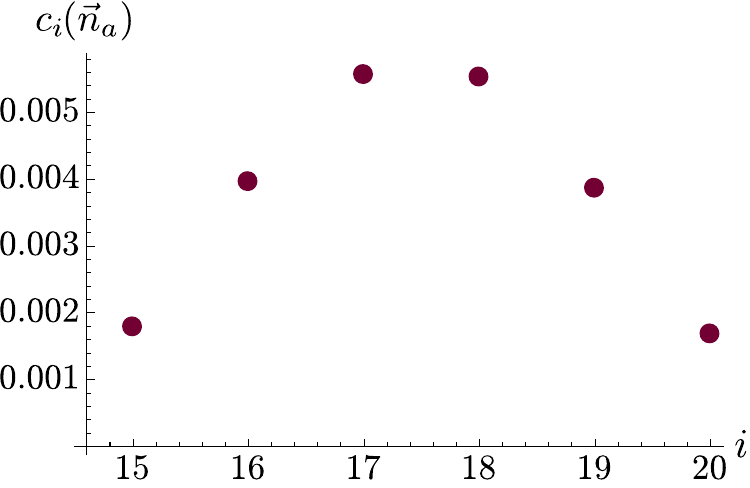}
    \end{subfigure}
      \caption{\label{fig:CoherentCut}  \small{Left panel: \emph{The distribution of a coherent state with  $j_{ab}= 10$ and normals corresponding to an equilateral tetrahedron, as a function of the intertwiner label $i=j_{12}$. 
      To estimate numerically the width we interpolate the data points with a Gaussian, and use the function \texttt{gsl\_stats\_sd} of the \texttt{GNU Scientific Library} \cite{contributors-gsl-gnu-2010}. Numerically it is then convenient to define a cutoff $\a$ excluding terms for which $c_{i}(\vec n_f)\leq \s(j,\vec n)/\alpha$. This means cutting the tails at $\sqrt{1+2\log\alpha}\, \s(j,\vec n)$. The horizontal line in the figure corresponds to $\alpha=3$, namely $1.79\s$ for which $93\%$ of the Gaussian is covered.} \\
      Right panel: \emph{The distribution of an isosceles coherent state with three spins equal to 10 and one equal to 25. For this configuration there are only 6 intertwiner spins allowed, and the spread is too broad to allow any reliable truncation. These are the tetrahedra entering the most numerically economical Lorentzian boundary data.}} }
\end{figure}

%-----------------------------------------------------
In an effort to reduce the numerical cost of the coherent amplitude, we studied the effect of truncating the sums over the boundary intertwiners $i$. 
In fact, $c_{i}(\vec n_f)$ is roughly a Gaussian distribution in $i$ \cite{LS,EteraHoloQT}, see e.g. Fig.~\ref{fig:CoherentCut}, the more accurate and the sharper for larger spins, and one can explore the impact of cutting the tails.
An analytic study of the Gaussian width shows that it depends on both the spins and the normals \cite{EteraHoloQT}. However the use of those formulas is not easy beyond the equilateral case. For our purposes it is more practical to estimate $\s(j_f,\vec n_f)$ numerically, as explained in the caption of Fig.~\ref{fig:CoherentCut}. 

\begin{figure}[H]
    \centering
    \begin{subfigure}[h]{0.49\textwidth}
        \includegraphics[width=7.5cm]{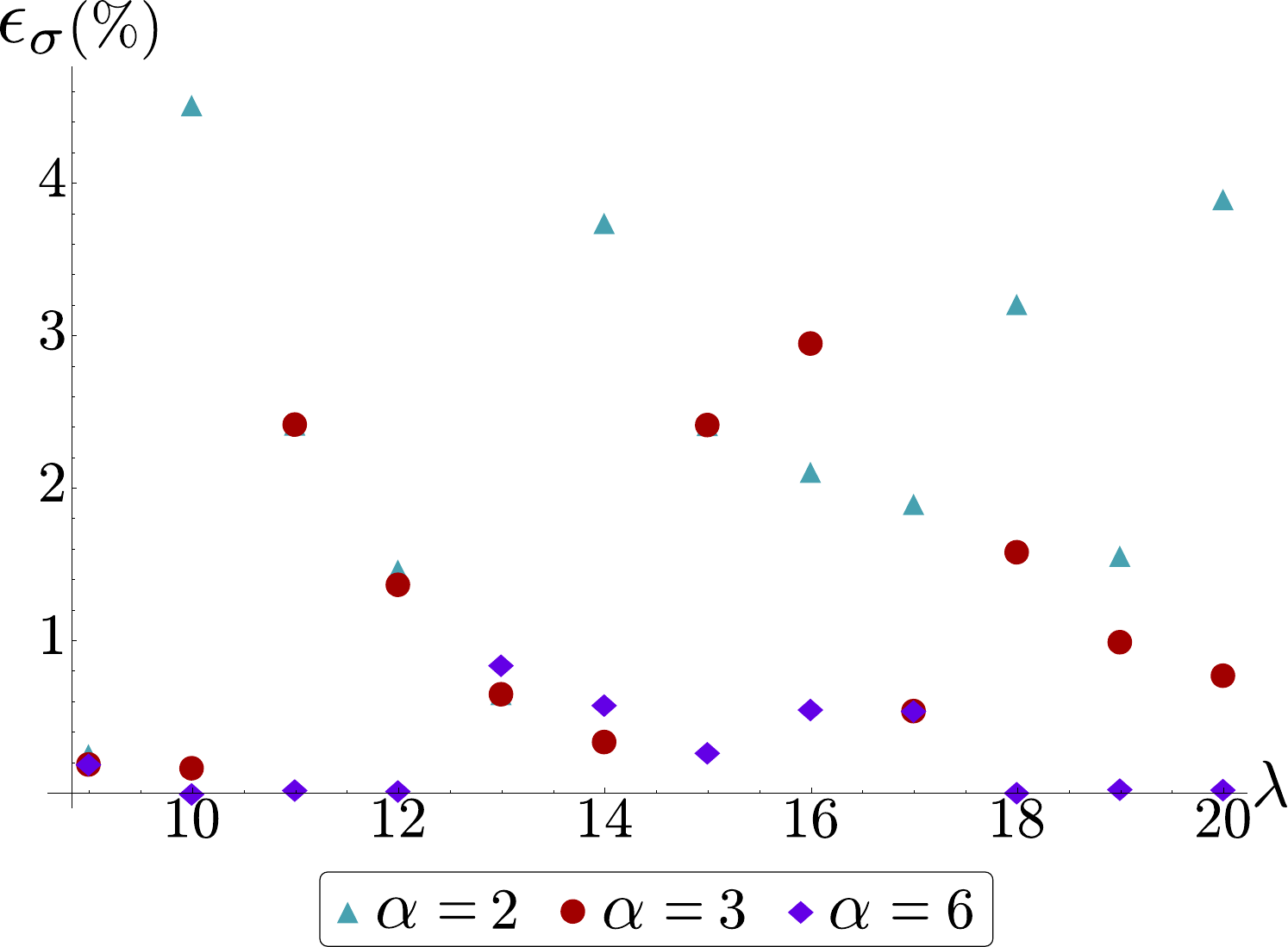}
    \end{subfigure}
    \begin{subfigure}[h]{0.49\textwidth}
    \vspace{0.35cm}
        \includegraphics[width=7.5cm]{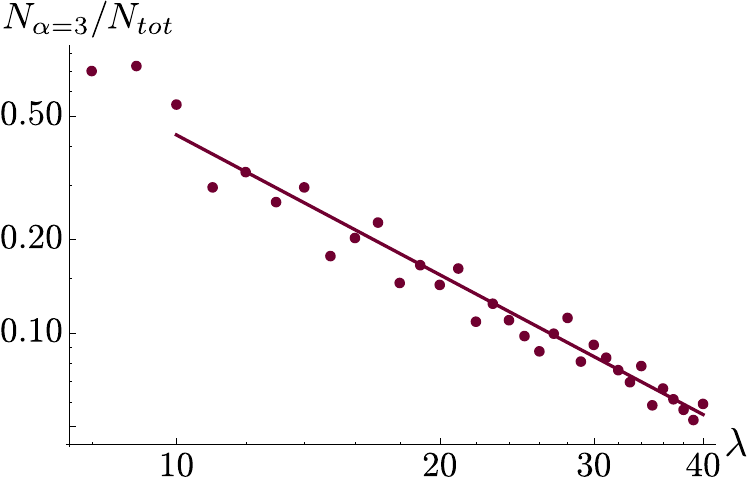}
    \end{subfigure}
      \caption{\label{fig:NumberTermsEucl}  \small{\emph{Study of the impact of truncating three of the five coherent states distributions, for the simplified model with Euclidean equilateral boundary data, and $\g=1.2$.} \\ Left panel,
      The error introduced:  \emph{We compute the relative error $\epsilon_\sigma$ between the exact amplitude and various truncations corresponding to different values of $\alpha$ -- defined in the previous figure, the smaller $\a$ the bigger the truncation -- as a function of the rescaling parameter $\lambda$. Setting ourselves an error tolerance of $3\%$, we truncate at $\a=3$.
      }\\
      Right panel, the improvement on the evaluation times: \emph{ 
      We compute the ratio between the number of terms $N_{\alpha=3}$ surviving in a truncation with ${\alpha=3}$ and the total number $N_{tot}$. The impact of the truncation grows roughly linearly in $\l$ (the solid line is $13.6\lambda^{-3/2}$), and we see that for instance at $\lambda = 20$ only 5\% of the terms survive.  }}}
\end{figure}

There is however no guarantee that truncating the tails of the Gaussians gives a valid approximation for the vertex amplitude because the intertwiner dependence of the $\{15j\}$ symbol can compensate for the Gaussian damping. This happens for instance in the equilateral configurations used for Euclidean boundary data, where the coherent state distribution is roughly peaked around the middle value of the intertwiner label,\footnote{More precisely slightly after, at $\bar\imath=(2/\sqrt{3}) j\simeq 1.15 j$ \cite{LS}.} 
whereas the $\{15j\}$ is peaked on extremal intertwiners 0 or $2j$. 
In this case, it turns out that cutting the tails even just a bit introduces large errors. For the simplified model and $\l$ around 10, one needs at least $5\s$ in the Gaussians to reduce the error on the full amplitude below $5\%$. This makes the approximation basically useless and pushed us to devise an alternative scheme. 
Instead of truncating all five Gaussians of the vertex amplitude, we kept two Gaussians exact and truncated only the other three (it does not matter which because of the symmetry of the problem).  
Using the simplified model as trial, we found that truncating three Gaussians at $1.79\s$ (namely covering $93\%$) 
we reduce drastically the number of terms included in the summation, see Fig.~\ref{fig:NumberTermsEucl}, while introducing a small error $\leq 3\%$.
This reduces the required computational time to roughly 1/20, for the simplified model and spins of order 10.

\begin{figure}[H]
    \centering
        \includegraphics[width=7.5cm]{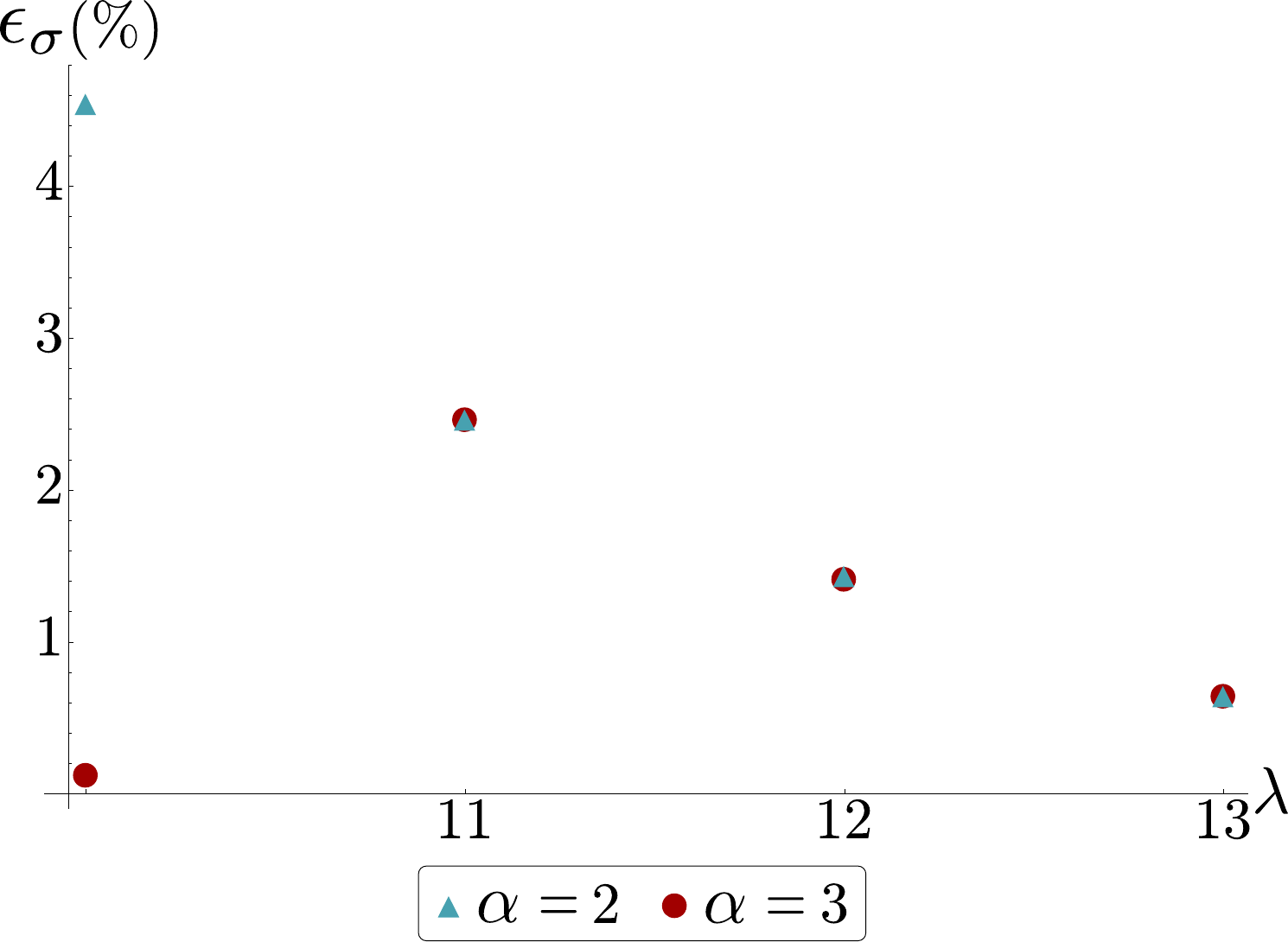}
      \caption{\label{fig:ErrRelLambdaShell}  \small{\emph{The truncation relative error in the $\Delta=1$ contribution of the shelled summation and Euclidean boundary data.} }}
\end{figure}

The advantage of such a truncation becomes even more important when adding the internal sums. To make sure that the error introduced is not significantly different than in the simplified model, we performed an analog study for $\D=1$ and $\l\sim 10$, reported in Fig.~\ref{fig:ErrRelLambdaShell}.
It confirms that $\a=3$ is a good compromise between cutting as many terms as possible and keeping the error small.

Having so chosen the truncation $\a=3$, we apply it to higher spins where the evaluation times are the longest. 
We have no quantitative estimate of the error made, but the coherence of the plots shown in the next section reassures us that we are not deviating significantly from the few percentage points established for $\l\sim 10$, and the evaluation times are very conveniently shortened.

\begin{figure}[H]
    \centering
    \begin{subfigure}[h]{0.49\textwidth}
        \vspace{0.35cm}
        \includegraphics[width=7.5cm]{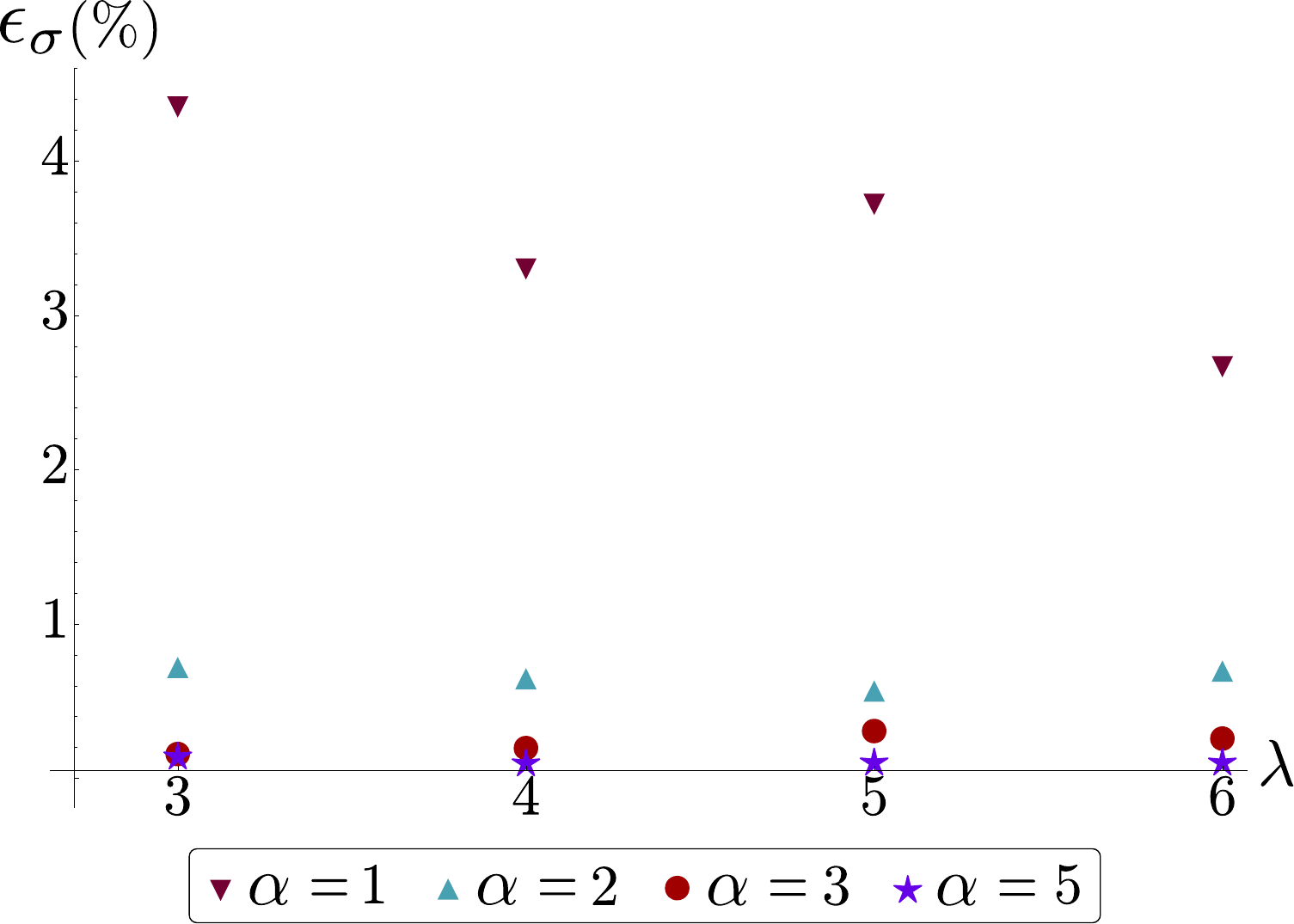}
    \end{subfigure}
    \begin{subfigure}[h]{0.49\textwidth}
        \includegraphics[width=7.5cm]{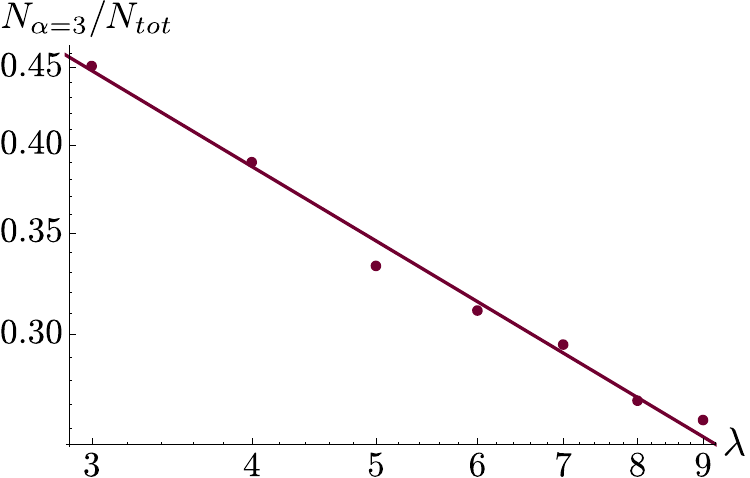}
    \end{subfigure}
      \caption{\label{fig:NumberTermsLor}  \small{\emph{ The improvement of the evaluation times obtained truncating one of the Gaussians. } Left panel: \emph{We perform the same test as shown in Fig. \ref{fig:NumberTermsEucl} for the Lorentzian configuration, truncating only on one coherent state (the equilateral one). We can cut only a single tetrahedron due to the fact that, for low rescaling parameters, the isosceles tetrahedra do not approximate a Gaussian, as shown in Fig. \ref{fig:CoherentCut}. In this case, the fit gives $0.8\lambda^{-1/2}$.}\\
       Right panel: \emph{The truncation relative error in the simplified model contribution for Lorentzian boundary data, used in Section \ref{sec:lorentzian}.}}}
\end{figure}

On the other hand, a similar truncation does not work for Lorentzian boundary data because of the isosceles tetrahedra there used. For the isosceles configuration there is a smaller range of intertwiners and the Gaussians are more spread out, see Fig.~\ref{fig:NumberTermsLor}. We found that any truncation in this case introduces large errors. Therefore for Lorentzian data we contented ourselves to truncate the only equilateral coherent intertwiner, again with $\a=3$. This gives an improvement of roughly 1/4 of the computational time for $\l=9$. On the other hand, we can introduce the truncation already at $\l=3$, something that with three Gaussians truncated as for Euclidean data was introducing a bigger error.

Summarizing, in the numerics we used the Gaussian truncation with $\a=3$ on three coherent intertwiners for vector and Euclidean Regge data, starting at $\l=8$;  and a similar truncation with $\a=3$ but on the unique equilateral coherent intertwiner for Lorentzian data, starting at $\l=3$. This approximation enters the plots shown in Fig.~6, etc. An estimated error bar of 3\% could be added but would not be visible.

%----------------------------------------------------------------------------
\section{Numerical analysis of the asymptotic formula}
\label{sec:numerical}
%----------------------------------------------------------------------------

In this section, we present a selection of our numerical data. We consider the four different types of boundary geometries constructed in Sec.~\ref{SecBD}, plus a completely random configuration not satisfying even closure, to test the different possible asymptotic behaviors derived in \cite{BarrettLorAsymp}.
\begin{enumerate}
\item For arbitrary areas and normals, or for the closed twisted geometry: the data confirm the exponential decay;
\item For the vector geometry: The data confirm a power law decay $\l^{-12}$ and no oscillations; 
\item For the Euclidean Regge geometry:  
The data confirm a power law decay $\l^{-12}$ and an oscillation with a frequency given by the Regge action;
 \item For the Lorentzian Regge geometry: The data indicate a power law decay $\l^{-12}$ and an oscillation with a frequency depending on $\g$, but more conclusive results will require additional computational power.
\end{enumerate}

%----------------------------------------------------------------------------
\subsection{Generic data, open and closed twisted geometry}
%----------------------------------------------------------------------------
\subsubsection{Open boundary data}
\begin{figure}[H]
\centering
\includegraphics[width=7.5cm]{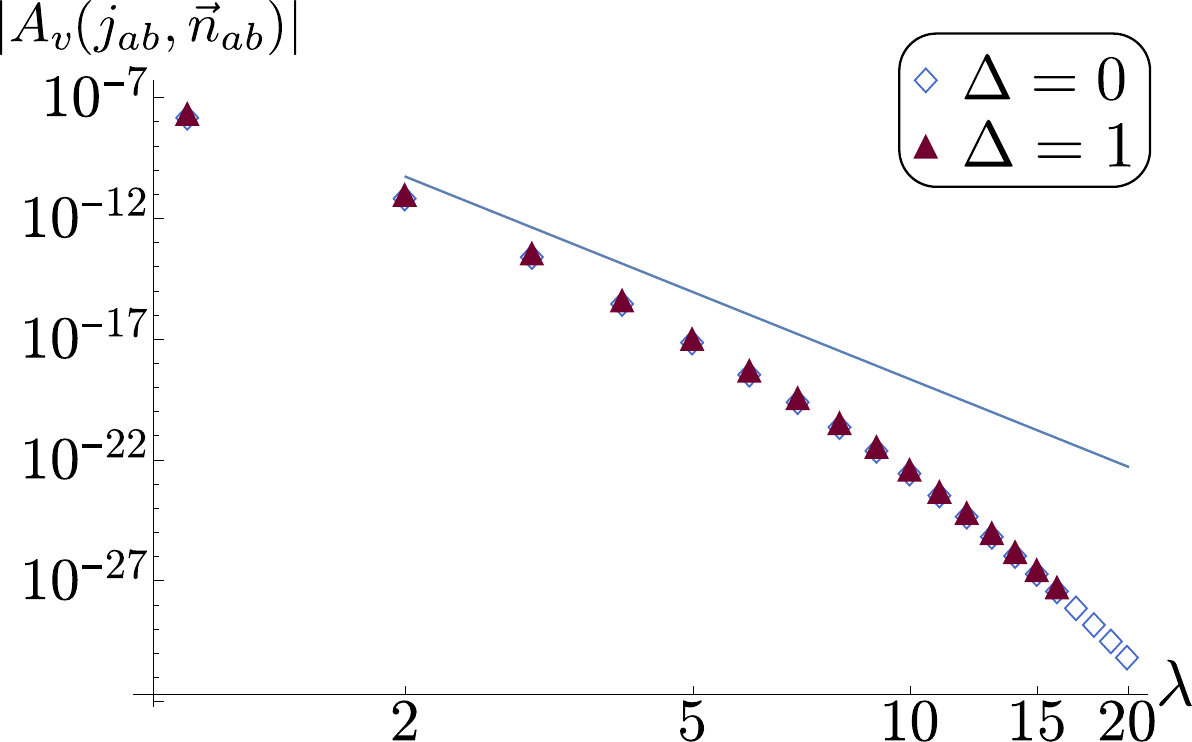}
\hspace{1cm}
\includegraphics[width=7.5cm]{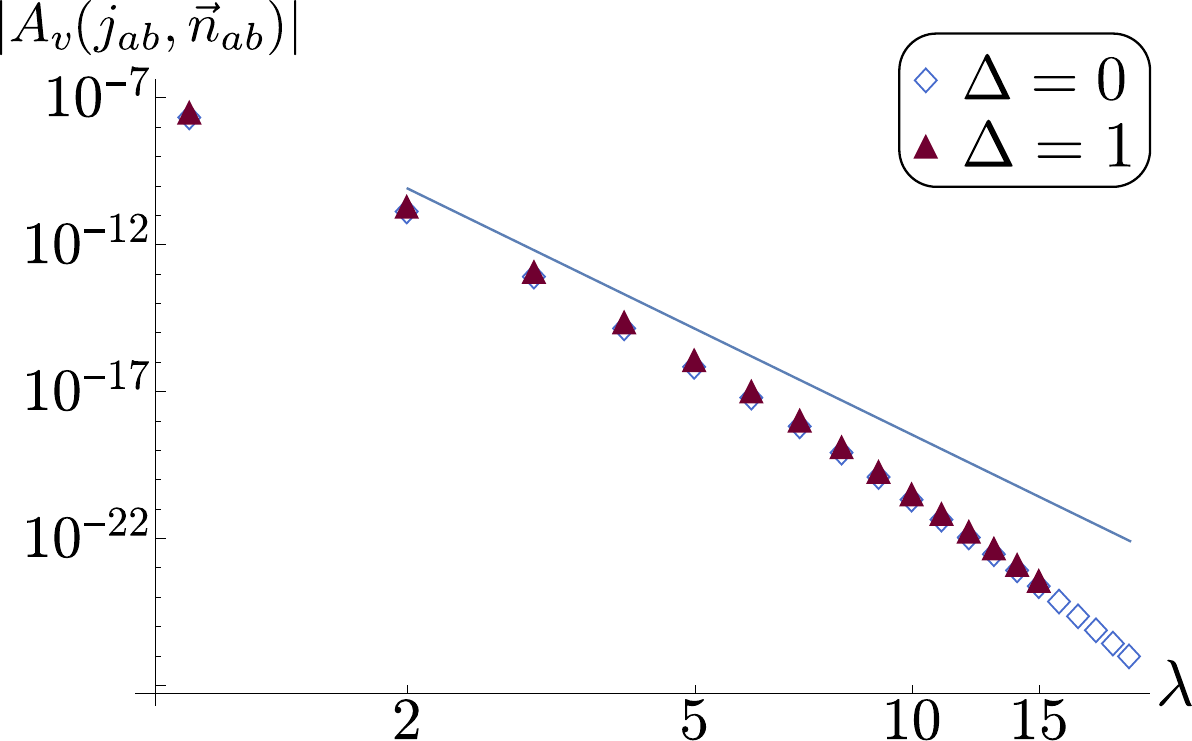}
\caption{\label{fig:Exponential1}  \small{\emph{The amplitude for generic boundary data not admitting critical points, showing an exponential falloff. Both log-log plots showing the absolute value of the amplitude for the simplified model ($\D=0$) and the first shelled sum ($\D=1$), with $\g=1.2$. 
Each plot took roughly 2h for the simplified model, and 36h for $\D=1$.} \\
Left panel: \emph{Open data, with spins equal and random normals not satisfying closure. The line is $2.1\times 10^{-7} \l^{-12} $, added to help the eye identifying the exponential falloff.} \\
Right panel: \emph{Closed twisted geometries data, with the normals satisfying closure but not the vector geometry conditions. The line is $3.2 \times 10^{-12}  \l^{-12}$, added to help the eye identifying the exponential falloff.} } } 
\end{figure}

For generic data we expect from \cite{BarrettLorAsymp} an exponential decay, namely
\be\label{LOexp}
A_v = o(\lambda^{-K}) \quad\quad\quad\quad \forall \text{ nonnegative integer } K.
\ee
To confirm this numerically, we took all spins to be equal, but the normals were randomly generated. We did not truncate the coherent states, since the Gaussian profile of Fig.~\ref{fig:CoherentCut} is lost for open configurations.\footnote{The reader may wonder why we can use open configurations in the boundary states, since these represent coherent intertwiners and closure is the classical counterpart of the quantum gauge invariance. The answer is that the normals are just classical labels, therefore one can have gauge invariant states with nonclosed normals. These states are on the other hand over-redundant, and can be effectively removed without losing coherence. The result is a family of coherent states depending only on the cross ratio parametrizing the intrinsic shape of the tetrahedron \cite{EteraHoloQT}. This family was used for numerical calculations of the volume operator in \cite{IoPoly}, but we refrain from using it here to keep the treatment as simple as possible.}
We evaluate the amplitude in the simplified model, $\D =0$, and the first shelled summation, $\D=1$. The vertex amplitude is exponentially suppressed, see Fig.~\ref{fig:Exponential1}.    

%---------------------------------------------------------------
\subsubsection{Closed twisted geometry}
%---------------------------------------------------------------
A closed configuration is enough to have a critical point for the norm of the coherent intertwiners \cite{LS}, but not for the full amplitude \Ref{Ac}, and \cite{BarrettLorAsymp} predicts again an exponential falloff \eqref{LOexp}.
We computed numerically the amplitude using the configuration constructed in Sec.~\ref{SecCTG}, and verified the exponential suppression for the simplified model and the first shelled sum, see Fig.~\ref{fig:Exponential1}.
In generating these data, we did not use any truncation on the coherent states. 
The coherent state distributions are in fact too broadly spread for this configuration, and it is thus not possible to efficiently truncate them for the values of spins considered.

%----------------------------------------------------------------------------
\subsection{Vector geometry}
\label{sec:vector}
%----------------------------------------------------------------------------
For vector geometries, namely data satisfying both closure \eqref{clos} and orientation \eqref{ori} constraints, there is one critical point (doubly degenerate), and one finds the following power law asymptotic behavior \cite{BarrettLorAsymp}
\begin{equation}
\label{LOvec}
A_v = \f1{\l^{12}} N_c e^{i \l \Phi_c} + O(\l^{-13}).
\end{equation}
Here $N_c$ contains the (inverse square root of the) Hessian determinant, the integration measure evaluated at the critical point, and the factors of 2 and $\pi$ coming from the critical point degeneracies and the Gaussian integrations. The action at the critical point is purely imaginary, hence the phase $\Phi_c$ in Eq. \Ref{LOvec}. See \cite{BarrettLorAsymp} and our appendixes for details and explicit formulas.
With the vector geometry constructed in Sec.~\ref{SecVG} one finds
\begin{equation}
\label{Nvg}
N_c= 7.86 \times 10^{-7}. 
\end{equation}
The value of the $\Phi_c$ is irrelevant because of the global phase ambiguity discussed earlier, and we restrict attention to the absolute value of the amplitude.

For the numerical evaluation of the amplitude, we fixed $\g=1.2$, and evaluated first the simplified model, $\D=0$, and then the first shelled sum, $\D=1$. In both cases we included a truncation on three Gaussians, for reference, on tetrahedra 2, 3 and 4. Even if the configurations are not equilateral, equal area is enough to have sufficiently peaked Gaussians. 

\begin{figure}[H]
\centering
\includegraphics[width=7.5cm]{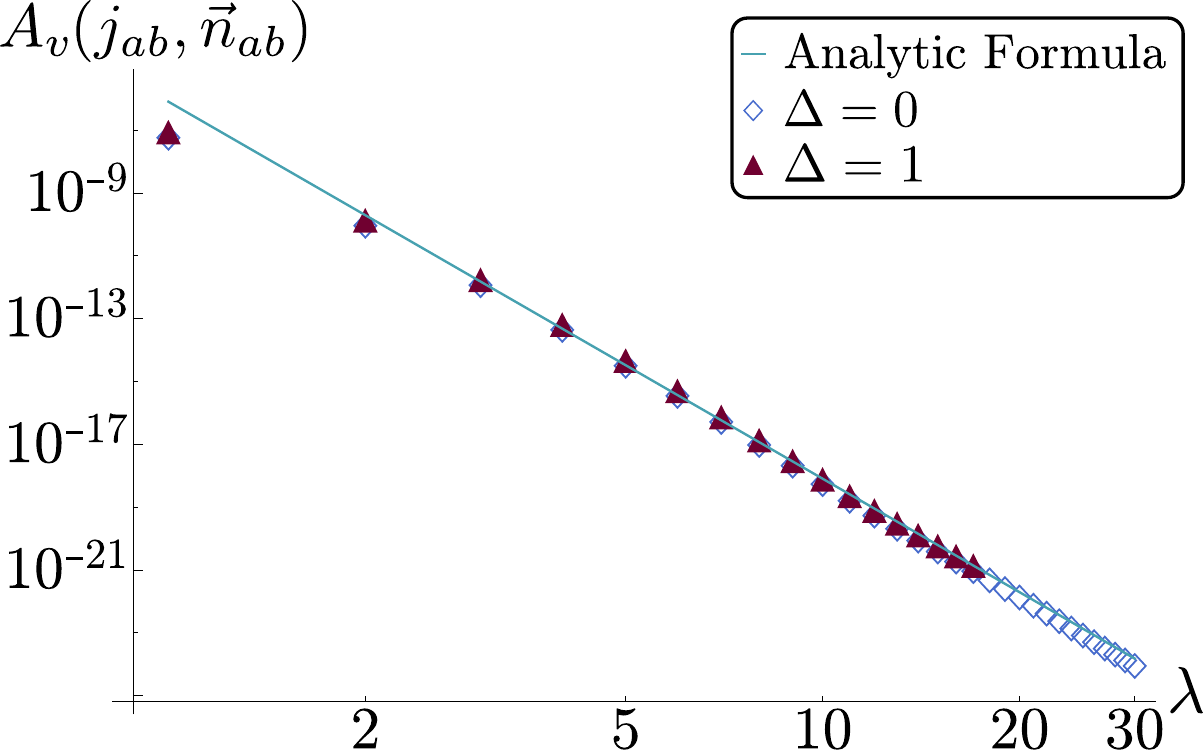} \hspace{0.3cm} \includegraphics[width=7.5cm]{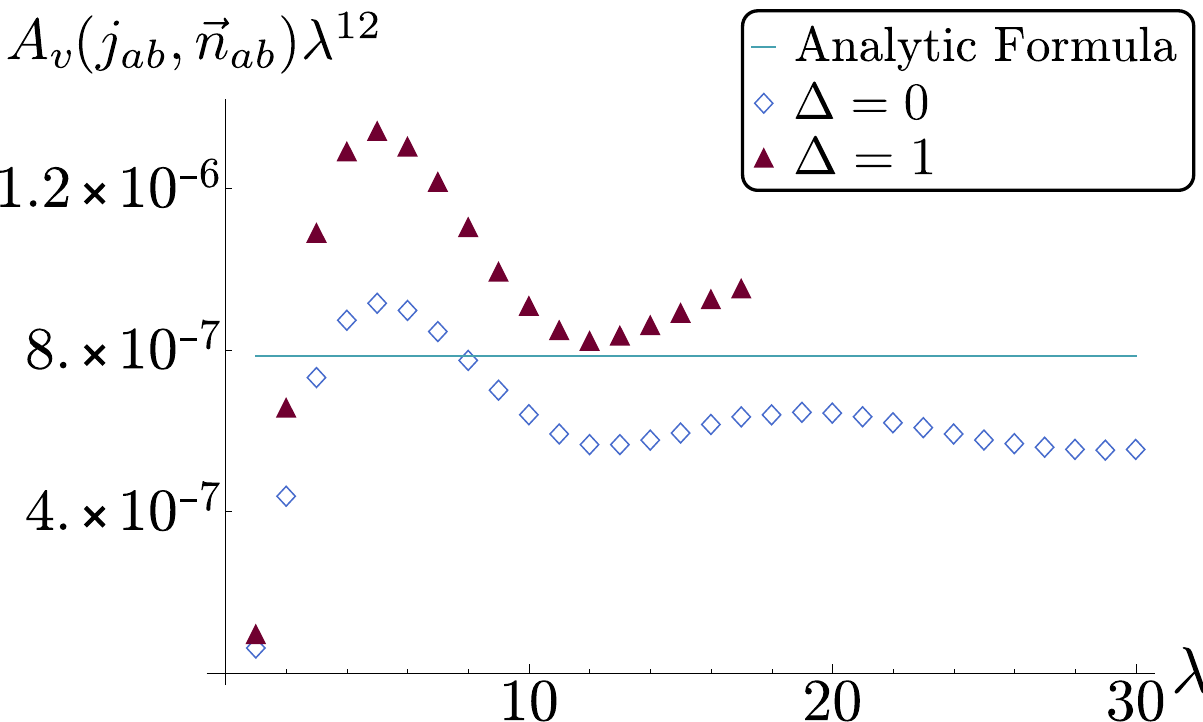}
\caption{\label{fig:Vector} \small{\emph{Asymptotics of vector geometries, $\g=1.2$.} \\ Left panel: \emph{The absolute value of the vertex amplitude with a vector geometry boundary for an homogeneous rescaling of the spins in log scale. The amplitude is following the power law \eqref{LOvec}. We plot the value of the amplitude in the simplified model ($\D=0$) with empty diamonds, and the first shelled summation ($\D=1$) with full triangles.
Here we used the cutoff on the coherent states discussed earlier, and the whole plot took roughly 30h. The line plotted is the analytical formula.
}  \\
Right panel: \emph{The absolute values rescaled by $\l^{12}$, showing a short scale nonmonotonic behavior, coming from the higher order saddle point corrections. }} } 
\end{figure}

The results are reported in Fig.~\ref{fig:Vector}, together with the analytic prediction. 
We find excellent agreement for the power law. As for the numerical coefficient, the simplified model comes out a bit short, and the first shell gives an order-one contribution approaching the predicted analytical value. Notice also the short scale nonmonotonic behavior, which is introduced by the higher order corrections to the saddle point. Presumably the first shell is already converging quite well to the analytic value \Ref{Nvg}, and the contributions from the second and higher shells will be small. This would mean that the behavior observed at $\l=1$ in Fig.~\ref{FigConv} (left panel) is maintained at higher $\l$'s. 

%----------------------------------------------------------------------------
\subsection{Euclidean Regge geometry}
\label{sec:euclidean}
%----------------------------------------------------------------------------
For Euclidean Regge data, the integral has two distinct critical points, and the asymptotic formula given in \cite{BarrettLorAsymp} reads
\begin{equation}
\label{ERLO}
A_v = 
\frac{(-1)^{\chi}}{\lambda^{12}} e^{i\l\Phi_c}
\left(N_c e^{i \l S_{\rm R}} + N_{{\scriptscriptstyle P}c} e^{-i\l S_R}\right) + O(\l^{-13}).
\end{equation}
Here $\chi$ is an integer depending on the eventual braiding of the intertwiners; $c$ and ${\scriptstyle P}c$ the two critical points, at which the action is purely imaginary and equal up to sign, up to a global phase $\Phi_c$;\footnote{The two critical points are related by a parity transformation, hence the notation ${\scr P}c$. Then one finds $S_{c}=i \l (S_{\rm R}+\Phi_c)$ and $S_{{\scr P}c}=i \l (-S_{\rm R}+\Phi_c)$, leading to Eq. \Ref{ERLO}. 
} $N_c$ contains the (inverse square root of the) Hessian determinant, but also a contribution from the integration measure at the critical point which is not trivial, and for convenience also the various factors of $2$ and $\pi$ coming from the twofold degeneracy of the critical points and the Gaussian integrations;  and most importantly,
\begin{equation}
 \label{Regge}
S_{\rm R}(j_{ab}, \vec n_{ab}) := \sum_{a<b}  j_{ab}   \th_{ab}(\vec n_{ab}),
\end{equation}
is the Regge action. Here $\th_{ab}$ are the 4D dihedral angles between tetrahedra, which can be reconstructed from the 3D normals using the spherical cosine laws. 
 
It is always worthwhile to remind the reader that the actual Regge action is a function of edge lengths, unlike \Ref{Regge}. Nonetheless, the bivector reconstruction theorem \cite{Barrett:1997gw} used in \cite{BarrettLorAsymp} to derive Eq. \Ref{ERLO} guarantees that the data span all Euclidean 4-simplices, and therefore Eq. \Ref{Regge} is equivalent to the Regge action.
 One can also give a different proof that does not rely on the bivector reconstruction theorem \cite{IoSU2asympt}. To that end, notice first that due to the rotational invariance at each node of the functions $\th_{ab}$, the independent variables of \Ref{Regge} are areas and angles between the 3D normals. It is known from \cite{DittrichSpeziale} that the area-angle action \Ref{Regge} is equivalent to the Regge action based on edge lengths provided closure and shape-matching conditions are satisfied, a result that holds for a generic triangulation and not only a single 4-simplex. In the case at hand of a single 4-simplex, closure is guaranteed by the critical point condition \Ref{clos}. And it can be shown explicitly \cite{IoSU2asympt} that the conditions for the existence of two distinct critical points is precisely the shape-matching conditions, in the form of consistency of spherical cosine laws.\footnote{More precisely, one gets \emph{angle-matching} conditions. However, for triangular faces with the areas already matching by construction, matching angles imply matching shapes. This is not the case for general polytopes with nontriangular faces, and in fact the large spin asymptotics of nonsimplicial vertices contains not just Regge geometries, but more general \emph{conformal} twisted geometries, see the discussion in \cite{IoSU2asympt}.}

If $N_c=\bar N_{{\scriptscriptstyle P}c}$, 
the asymptotics can be written as a cosine of $\l S_{\rm R}$, with a $\l$-independent phase offset given by the argument of $N_c$,
\be
A_v = 
\frac{(-1)^{\chi}}{\lambda^{12}} 2|N_c| e^{i\l\Phi_c}\cos\left( \l S_{\rm R} -\f12 \arg H_c \right) + O(\l^{-13}).
\ee 
We are not aware of an analytic proof of this property, but we verified it in all cases numerically checked. To avoid confusions between the phases $\Phi_c$ and $\arg H_c$, we will refer to the first as the global phase and to the second as the cosine phase offset.

For our numerical simulations, we considered the equilateral twisted spike configuration described in Sec.~\ref{SecER}.
Assigning the normals to the intertwiners according to the geometric picture of a 4-simplex, we have no braiding and $\chi=0$. For the Hessian, we use the explicit formulaq in Appendix~\ref{AppH}, and the Mathematica algorithm \texttt{EuclideanHessian} of \cite{code-sl2cfoam} to evaluate it at the chosen configuration. The result is real and identical at both critical points, and fixing for example $\g=1.2$, we find
\be
N_c =  N_{{\scriptscriptstyle P}c} = 2.65\times 10^{-7}.
\ee
Furthermore, using our phase convention for the coherent amplitude (which is not the same as the one in \cite{BarrettLorAsymp}, see Appendix~\ref{AppA}) and the conventional  phase convention for the coherent states \cite{Perelomov}, we numerically found that $\Phi_c=S_R$. Removing this phase (redefining for instance the phase of the coherent states), the analytic prediction is 
\be\label{EuAs}
A_v = \f{5.13\times 10^{-7}}{\l^{12}}
\cos\Big(5\l \arccos \big(-\frac14\big)\Big).
\ee
We recall that for these boundary data $\l=2j$. Notice that there is no phase offset in the argument of the cosine, unlike for SU(2) asymptotics \cite{PonzanoRegge,Esterlis:2014zoa,IoSU2asympt}. This fact will play a role below.

In the numerics we used the approximation truncating three of the Gaussians, for reference the tetrahedra 2, 3 and 4. We evaluated first the simplified model, then added the internal sums, with truncations at $\D=1$ and $\D=2$. This increases  progressively the cost of the numerics. We were able to reach $\l=40$ for the simplified model, $20$ for the first shell, and 
$18$ for the second shell. We did not push our analysis to a third shell, because we would have been forced to stop at spins too small to test the asymptotic behavior. 
The data obtained are complex, but since we identified the global phase, we can remove it a posteriori and obtain real data points.

\begin{figure}[H]
\centering
\includegraphics[width=7.5cm]{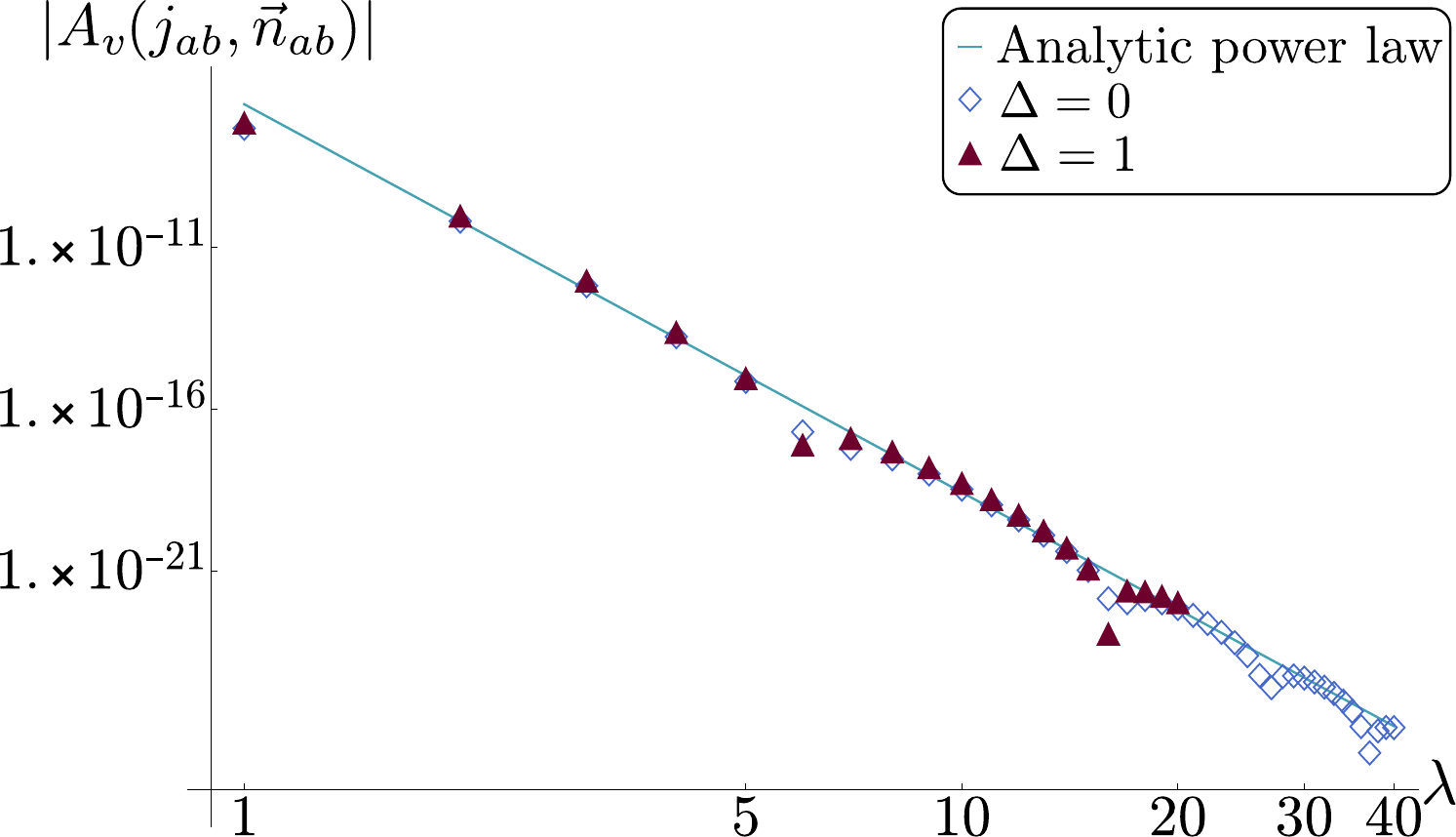}
\hspace{0.3cm}
\includegraphics[width=7.5cm]{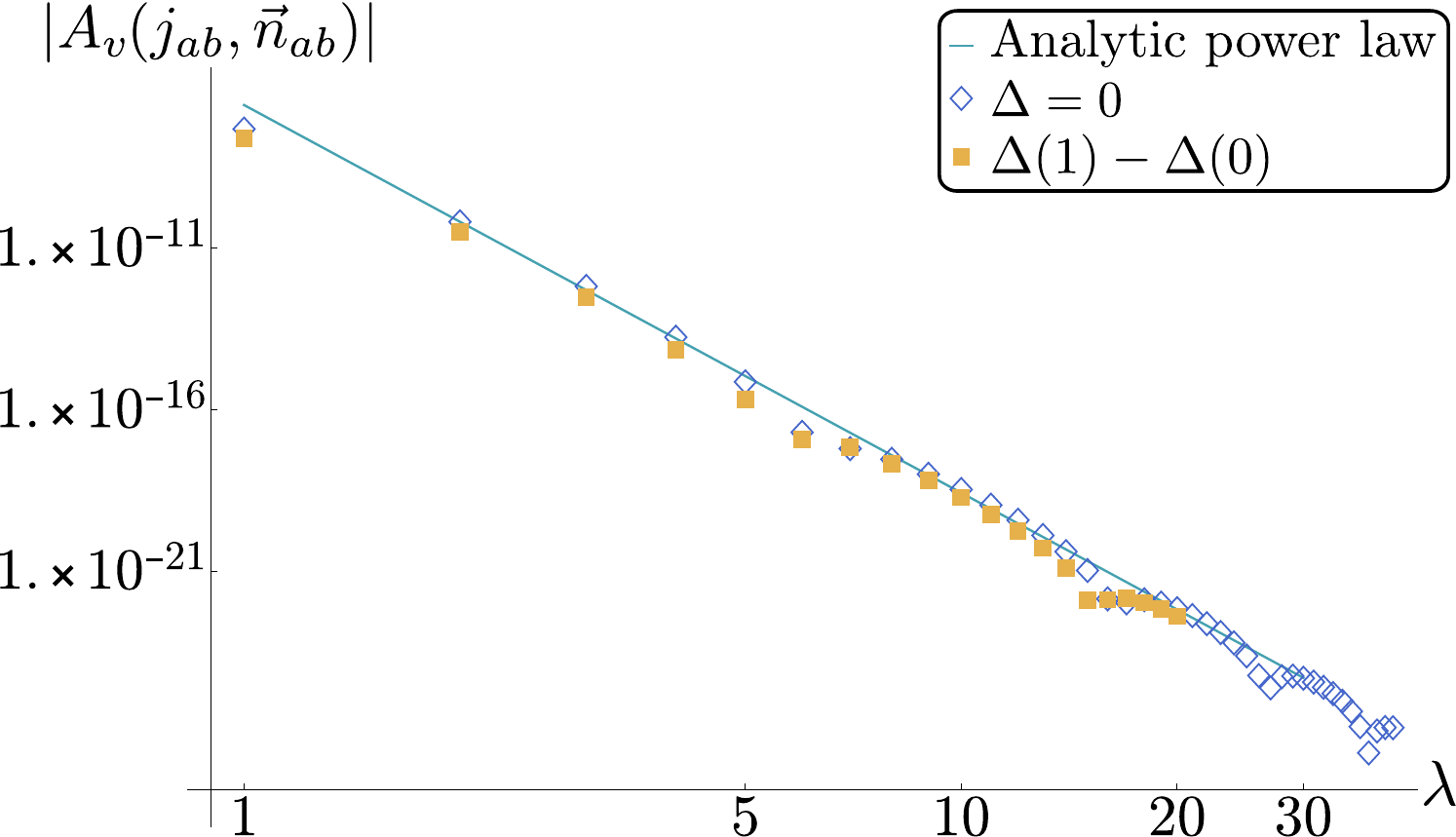} 
\caption{\label{FigEuAs1}  
\small{\emph{Numerical data versus the analytic asymptotics \Ref{EuAs}, for Euclidean Regge data, $\g=1.2$. Log-log plots, absolute values of the amplitude, versus the analytic scaling $5.13\times 10^{-7}\l^{-12}$.}
\\ Left panel: \emph{Simplified model and first shelled sum. Both scale according to the predicted power law, and feature oscillations.} \\
Right panel: \emph{The simplified model and the difference between the first shelled sum and the simplified model; namely, all contributions with one nonminimal $l$. This difference also scales with the predicted power law and features oscillations, showing that individual higher $l$ contributions see the critical behavior. 
}  } } 
\end{figure}

The first result we found is that the numerical data are in excellent agreement with the $\l^{-12}$ power law, see Fig.~\ref{FigEuAs1}.
The agreement is achieved already at small spins, and holds for both the simplified model and the first shelled sum. We plot also the difference $\D_1-\D_0$ to show that it follows the same power law: not only does the simplified model see the critical behavior, but so do (at least some of) the individual higher $l$ configurations. 
\noindent We found the same $\l^{-12}$ scaling also for the second shell $\D_2-\D_1-\D_0$, with includes individual terms with two $l=j+1$ or one $l=j+2$.
We deduce from these numerical results that the simplified model captures the right scaling, and so do individual shells. This means that the convergence of the internal sums for higher $\l$ is still qualitatively similar to the extended study that was possible to make for $\l=1$ (see Fig.~\ref{FigConv}), and also that the 
the analytic behavior \Ref{ERLO} must be the result of adding up all the relevant contributions of different $l$'s.  These are of order one initially, but we expect them to suitably decrease as guaranteed by the convergence. 

\begin{figure}[H]
\centering
\includegraphics[width=15.5cm]{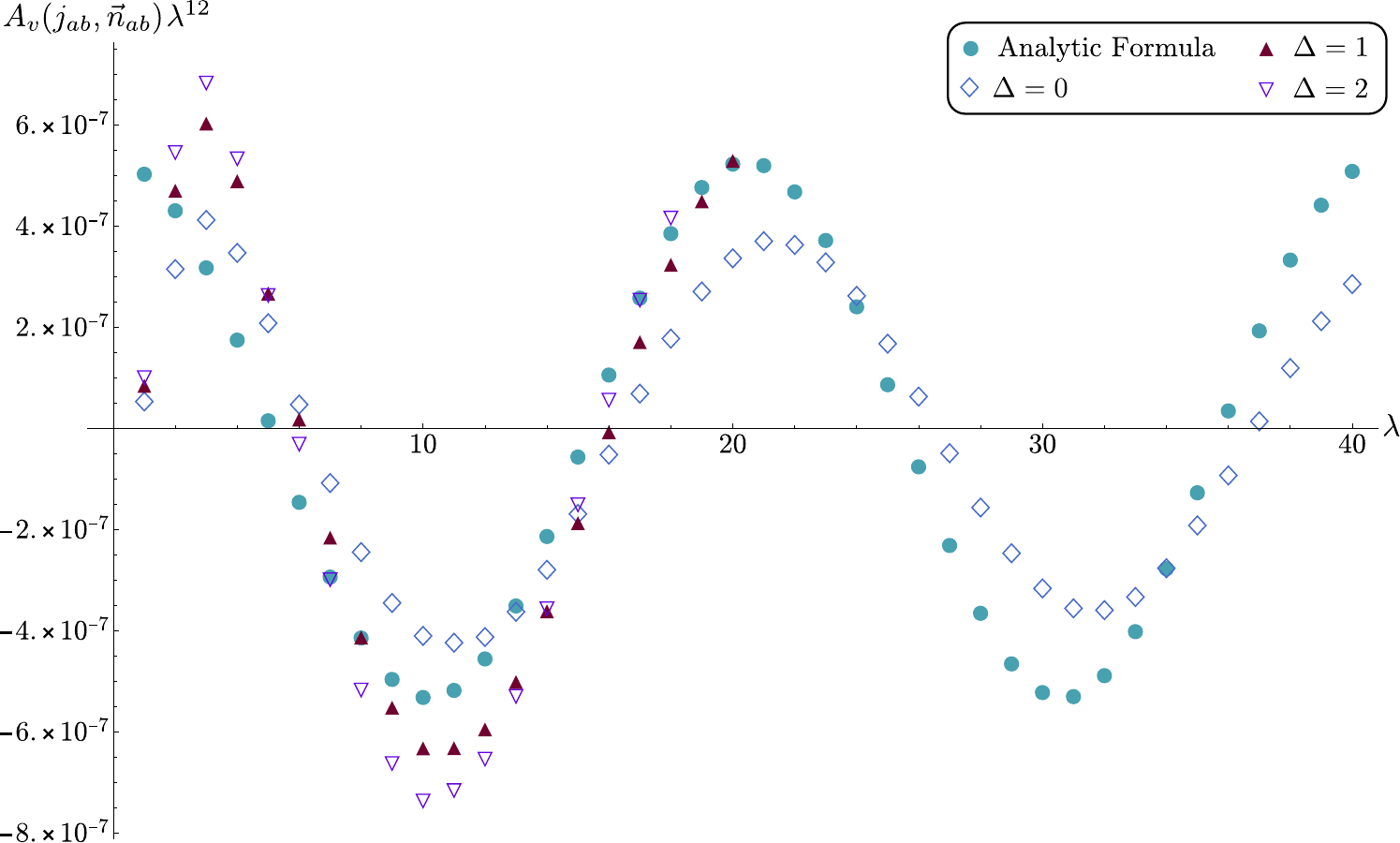}
\caption{\label{fig:EuclideanAsym}  \small{\emph{Rescaled numerical data versus the analytic asymptotics \Ref{EuAs}, for Euclidean Regge data and $\g=1.2$, for the simplified model and internal summations with cutoffs $\D=$1 and 2. The simplified model alone already sees the Regge oscillations of the analytic formula.
The two shelled sums maintain the same frequency, but give contributions of order one to the magnitude and the phase offset.}  } } 
\end{figure}

Next, let us zoom in on the palatable oscillations already visible in Fig.~\ref{FigEuAs1}. We rescale the data and asymptotic formula by $\l^{12}$ and report the result in Fig.~\ref{fig:EuclideanAsym}, the highlight plot of our paper. 
It shows a beautiful agreement between the numerical evaluation of the vertex and the analytic asymptotic formula in particular, a confirmation of the frequency of the oscillations determined by the Regge action, its most important feature.\footnote{A word of caution to avoid possible confusion: If one plots the asymptotic formula as a real function, the real frequency given by $S_{\rm R}$ is much higher than the frequency of oscillations that can be deduced interpolating the  (half-)integer sampling, a situation familiar from the study of SU(2) asymptotics. Hence when we speak of matching frequency of oscillations between the numerical data and the analytic formula, we refer to the ``effective'' frequency of the (half-)integer oscillations, which is not the Regge cation, but a function thereof.}

In more details, we observe the following situation. First, the simplified model captures already the right frequency of oscillations. What is missing is the precise magnitude and cosine phase offset. 
The first and second shells have the same frequency of oscillations, and different magnitude and cosine phase offset, which lead to a better qualitative match with the asymptotic formula.
A more precise quantitative improvement will likely require pushing to higher spins where the asymptotic formula becomes more accurate. 
As for the crude numerical values, we see that at $\l\sim20$ (which means spins $j\sim10$, recall half-integers are also being computed here) the simplified model and shelled sums agree with one another and with the analytic formula with a $10\%$ error.

From these data we deduce that individual $l$'s see the same critical behavior for Euclidean Regge data, including the simplified model, and that the asymptotic formula \Ref{ERLO} is the result of summing all the relevant contributions. However the power law and frequency of oscillations are well described by the simplified model, with the internal sums only contributing to the overall magnitude and phase offset. Given the much faster evaluation times of the simplified model, this is an interesting property, whose origin will be explained in the next section.

The data presented in the figures of this section have $\g=1.2$.
For double checking, we performed further numerical investigations with the different value $\g=0.1$, leading to similar plots. In particular, we found the same frequency of oscillations as predicted by Eq. \Ref{ERLO}.

We conclude from the numerics that the power law and frequency of the asymptotic formula can be confirmed already at spins of order 10, to a $10\%$ error. A better match requires changes in the amplitude and the phase offset, for which one needs more shells and/or higher spins.

%----------------------------------------------------------------------------
\subsection{Lorentzian Regge geometry}
%----------------------------------------------------------------------------
\label{sec:lorentzian}

For Lorentzian Regge data, we have two distinct critical points, and the oscillating power law decay \cite{BarrettLorAsymp}
\begin{equation}
\label{LRLO}
A_v = 
\frac{(-1)^{\chi+M}}{\lambda^{12}} e^{i\l\Phi_c}
\left(N_c e^{i \l S_{c}} + N_{{\scriptscriptstyle P}c} e^{-i\l S_c}\right) + O(\l^{-13}).
\end{equation}
The crucial difference is that this time the action at the critical point depends linearly on $\g$,
\begin{equation}
\label{Reggeg}
S_{c} = 
\g \sum_{a<b}  j_{ab}   \th^{\scriptscriptstyle \rm L}_{ab}(\vec n_{ab}).
\end{equation}
The Lorentzian 4D dihedral angles $\th^{\scriptscriptstyle \rm L}_{ab}$ are defined from the Lorentzian spherical cosine laws in terms of the 3D normals, or from the 4D normals as recalled in Sec.~\ref{SecLD} above.
They are boosts between the timelike normals if the latter are both future or past pointing, and in the mixed case (called the thin wedge configuration) 
one has to add a shift by a factor $i\pi$. These shifts result in the additional $(-1)^M$ phase, where
\be
M = \sum_{\rm thin\ wedges} j_{ab}.
\ee
We remark that Eq. \Ref{Reggeg} is a Lorentzian Regge action with areas are given by $\g j$, which is in agreement with the area spectrum of loop quantum gravity in the large spin limit. As for the shape-matching and equivalence of Eq. \Ref{Reggeg} to the Regge action based on lengths, the same considerations of the Euclidean case apply.

For our numerical simulations, we considered the isosceles configuration described in Sec.~\ref{SecLD}. Assigning the normals to the intertwiners according to the geometric picture of a 4-simplex, we have no braiding and $\chi=0$. Further, we have four thin wedges all associated with $j_{ab}=5\l$, thus $M=20\l$ with integer $\l$ and this phase drops out as well.
For the Hessian, we use the explicit formulas in Appendix~\ref{AppH}, and the Mathematica algorithm \texttt{LorentzianHessian} of \cite{code-sl2cfoam} to evaluate it at the chosen configuration. Taking 
%For the same value $\g=1.2$  used in the previous cases
$\g=0.1$, we find 
\be
%N_c = \overline{ N_{{\scriptscriptstyle P}c}} =2.82  \times 10^{-16} + i 5.7\times 10^{-16}, 
N_c = \overline{ N_{{\scriptscriptstyle P}c}} = 1.76 \times 10^{-13} + i 1.87 \times 10^{-14},
\ee
thus Eq. \Ref{LRLO} is a cosine up to a global phase. Reabsorbing the global phase in the definition of the coherent states and computing the corresponding Regge action, we obtain the analytic asymptotic formula
\be \label{LorAs}
A_v = \f{3.53 \times 10^{-13}}{\l^{12}}
\cos \left(0.01 \l + 0.106 \right).
\ee

In running the numerical code, we used the Gaussian truncation on only one coherent intertwiner, the equispin gauge-fixed tetrahedron 1. The data obtained are complex, and their $\l$-dependent phase this time is not simply the Regge action. We spent a considerable amount of time going through the conventions and technical details that determine this anyway irrelevant global phase, for the sake of having real data points. In the end the simplest solution was to determine the $\l$-dependent phase through a numerical fit, then remove it from the numerical data. A more brutal removal by taking absolute values would have deprived us of the numerical sign, which is important in comparing data and analytic formulas.\footnote{We shamelessly admit that we actually \emph{failed} to reproduce the observed numerical global phase from our analytic calculations -- Ref. \cite{BarrettLorAsymp} does not compute it, given its irrelevance. Appendix~\ref{AppA} collects all explicit formulas and all numerous places where a global phase arises, as well as a careful comparison between the conventions here used based on \cite{Varshalovich,Perelomov,Ruhl}, and the ones of \cite{BarrettLorAsymp}. Albeit an overlooked phase lurking there and open to inspection, we suspect that the origin of the problem lies in a mismatch in the conventions for the orientation of the links of the reducible $\{15j\}$ between the numerical code and our analytic description of the amplitude.}

The code works fine also for Lorentzian boundary data, and we were able to evaluate the amplitude for the simplified model and first shell.
The numerical results we were able to obtain are however less satisfying than for the previous configurations, in terms of comparing with the analytic asymptotic formula. The main problem is the numerical instability in the booster functions, which limits us to $\l=9$, before the $j=50$ instability of the boosters is reached. Furthermore, the numerical cost of the calculation limits us to $\D=1$. 
We can, on the other hand, take advantage of the expected linear dependence of the oscillations on $\g$, and evaluate numerically the vertex amplitude  at different values of $\g$, to try to establish different aspects of the asymptotics. We considered three different values of $\g$, chosen to have qualitatively different behaviors in the accessible $\l\in[1,9]$ range: a rather flat plateau with no oscillations, and visible oscillations but with a different frequency. The resulting asymptotic formulas needed to match the numerics in the three cases are reported in Table~\ref{Tlor}.

\bgroup
\def\arraystretch{1.3}
\begin{table}[H]
\centering
\framebox{
\begin{tabular}{c|c|c}
$\g$ & $N_c = \overline{ N}_{{\scriptscriptstyle P}c}$ & $ \l^{12} A_v$ \\ \hline
0.1 & $1.76 \times 10^{-13} + i 1.87 \times 10^{-14}$ & 
${3.53 \times 10^{-13}} \cos \left(0.963 \times 10^{-2} \l + 0.106 \right)$ \\
1.2 & $2.82  \times 10^{-16} + i 5.7\times 10^{-16}$ & 
${1.27 \times 10^{-15}} \cos \left(0.116 \l + 1.111 \right)$ \\
7 & $8.86  \times 10^{-25} - i 3.67\times 10^{-25}$ & ${1.92\times 10^{-24}}
\cos \left(0.674 \l - 0.392 \right)$ \\
8 & $1.48 \times10^{-25} - i 4.32 \times 10^{-26} $& ${3.08 \times 10^{-25}}
\cos \left(0.770 \l - 0.284 \right)$
\end{tabular}}
\caption{\label{Tlor} \small{\emph{Asymptotics for Lorentzian boundary data at different values of $\g.$}}}
\end{table}
\egroup

\begin{figure}[H]
\centering
\includegraphics[width=7.5cm]{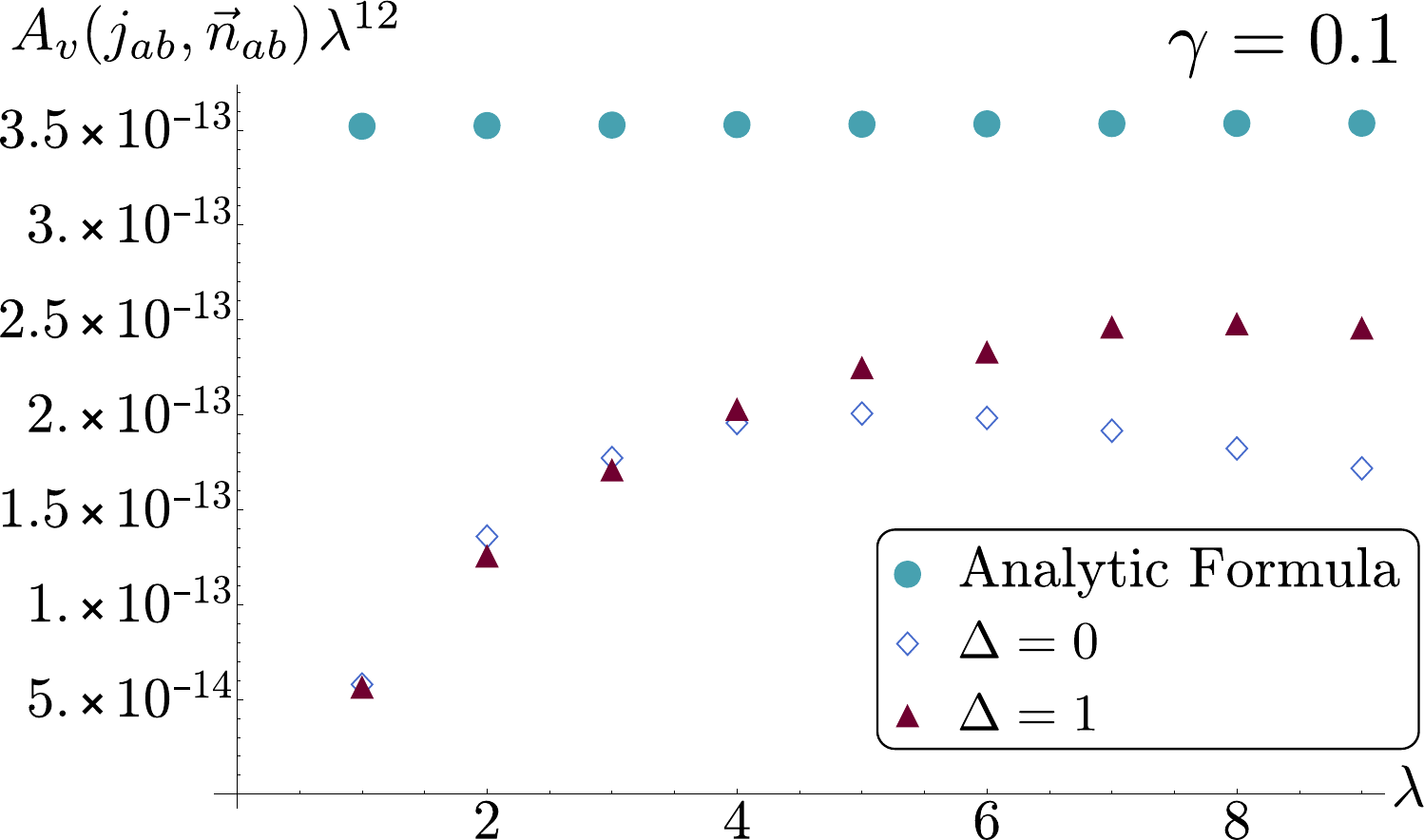} 
\caption{ \label{fig:LorentzianAsym1}  \small{\emph{Rescaled numerical data for Lorentzian Regge geometry, versus the analytic asymptotics \Ref{LRLO}, for $\g=0.1$. The bottleneck is the numerical instabilities of the booster functions at spins of order 50, which even for the optimal isosceles configuration we identified, occur already at $\l=10$. With the caveat of the few points obtained, the simplified model is decaying faster than the first shell, indicating that it lacks the critical behavior, and will likely decay exponentially. The first shell, on the other hand, is well in line with the power law decay. Its magnitude falls short of the asymptotic formula by a factor of order one: this could be completed by higher order shells, accordingly with the previous indications that convergence is slow for Lorentzian Regge geometries. } } }
\end{figure}

Not having enough points for a significative log-log plot, we present directly the data rescaled by the predicted $\l^{12}$ falloff, versus the rescaled asymptotic formula. 
The first plot shown in Fig.~\ref{fig:LorentzianAsym1} is for $\g=0.1$. The effective frequency of oscillations is low, giving a roughly constant behavior in the accessible range. The first consideration to be deduced from the data is that the simplified model appears to fall off significantly faster than the analytic power law decay of the EPRL model. We interpret this to mean that the simplified model does not have a critical behavior for Lorentzian Regge geometries. The data points of the first shell on the other hand roughly stabilize in parallel to the asymptotic formula. This indicates that the first shell is already enough to capture the critical behavior, even though with its actual magnitude being too small, we expect nearby shells to still give order-one contributions, and thus a slow convergence.

\begin{figure}[H]
\centering
\includegraphics[width=7.5cm]{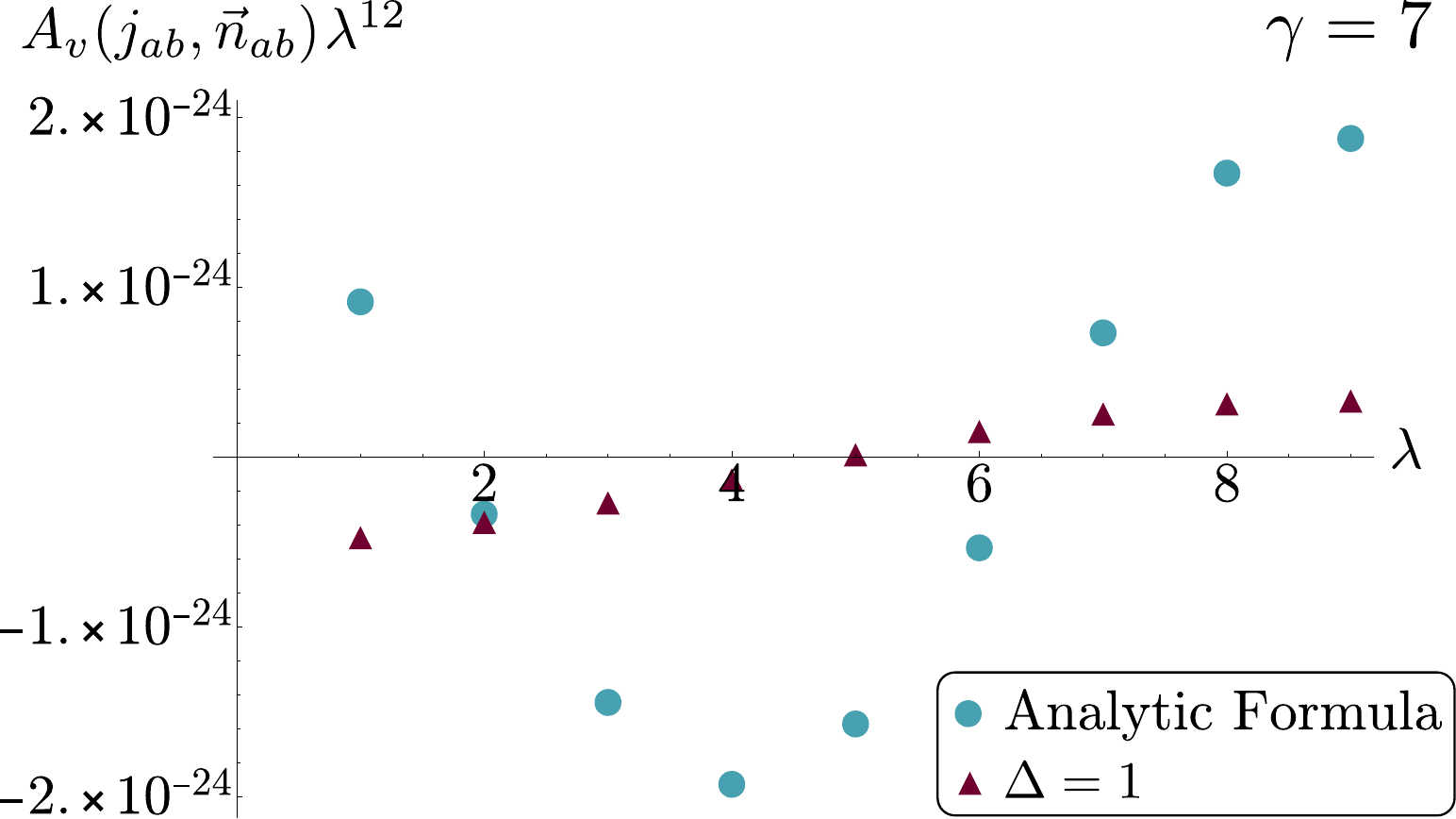}\hspace{1cm}
\includegraphics[width=7.5cm]{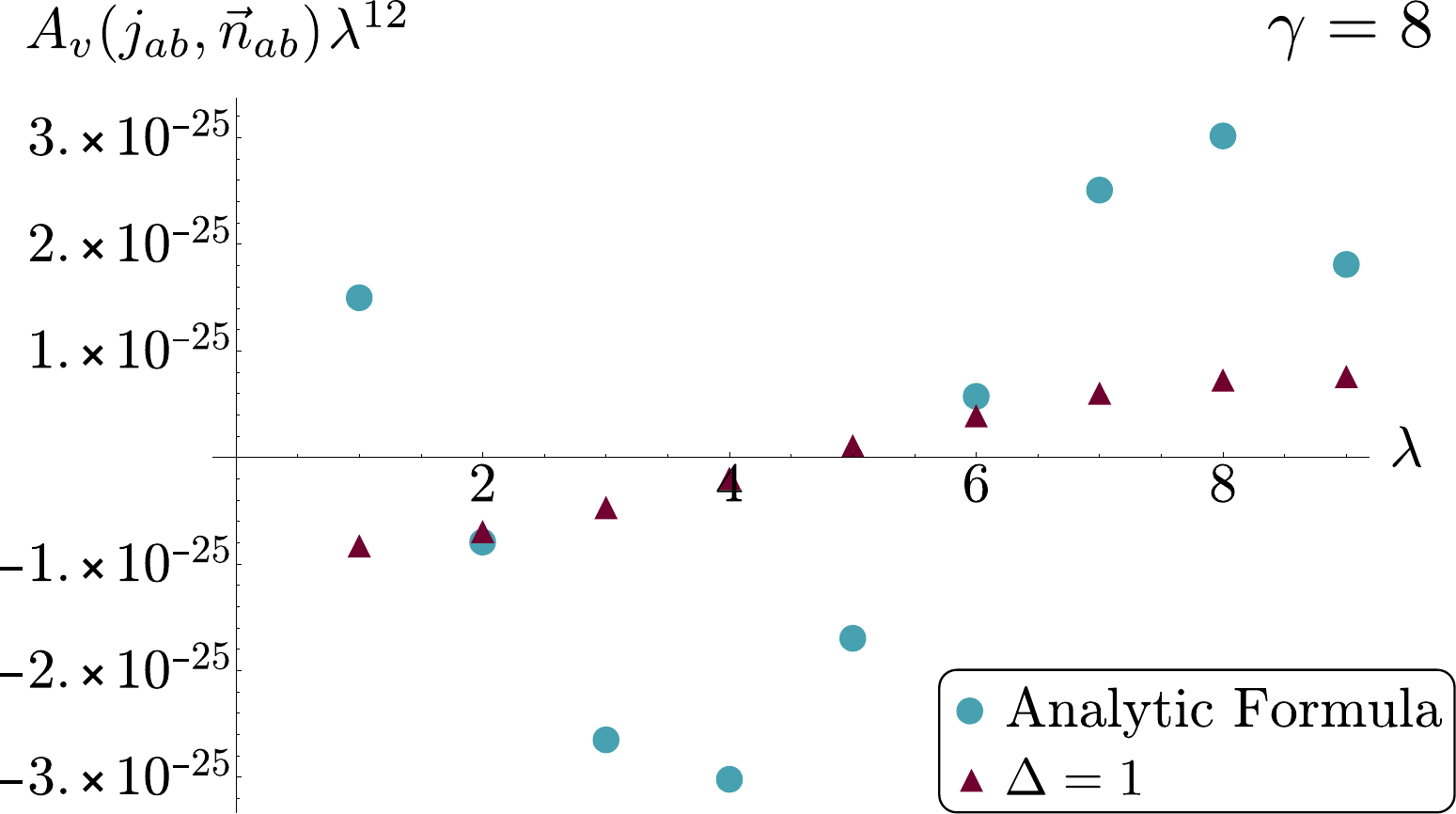}
\caption{\label{fig:LorentzianAsym}  \small{\emph{Rescaled numerical data versus the analytic asymptotics \Ref{LRLO}, for Lorentzian Regge data and $\g=7$ and 8. These values are chosen so that the analytic formula has a visible oscillation within the range numerically accessible. The faster frequency of the $\g=8$ case can be inferred comparing the eyesight interpolation of the positive slope through the $x$ axis. The data points for the first shell are well in line with the analytic power law, and show a nonmonotonic behavior qualitatively comparable with the predicted oscillations. Higher spins are needed for a better match, as well as increasing the number of shells being summed over.}  } } 
\end{figure}

Next, we increase the expected frequency looking at larger values of $\g$. Taking $\g=7$ and $\g=8$ allows us to fit roughly one full period of analytic oscillations within the accessible range, and with  different frequencies, to see whether the data offer a compatible nonmonotonic behavior.
The results are plotted in Fig.~\ref{fig:LorentzianAsym}. 
The behavior of the first shell is again compatible with the power law scaling, and we clearly observe nonmonotonic behavior and dependence on $\g$. This is nice evidence in support of the numerical evaluation of the vertex amplitude and of the validity of the asymptotic formula.
The data points at disposal are however too few and at too small $\l$ to draw any conclusions about the frequency of oscillations like we could do for the Euclidean Regge geometry. As a somewhat optimistic remark, as we increase $\g$ from 7 to 8 the (interpolated) analytic zero with positive slope moves to the left, and so does the (interpolated) zero of the numerical data.

In summary, the simplified model does not see the critical behavior for Lorentzian Regge data, but the first shell already does, and presumably all other shells. The power law decay observed confirms the asymptotic formula. There is a $\g$-dependent nonmonotonic behavior, qualitatively compatible with the oscillations predicted by the asymptotic formula. The convergence of the internal sums is  slower than with Euclidean Regge data, and higher shells/higher  spins are needed for a more quantitative confirmation of the asymptotic formula.

%----------------------------------------------------------- 
\section{Outlook}
%----------------------------------------------------------- 

\subsection{Geometric meaning of the internal sums}

The formula \Ref{Ac} and more generally the method of \cite{Boosting}, introduce a factorization of Lorentzian spin foam amplitudes so that only SU(2) objects appear at the vertices, and all Lorentzian properties are in the booster functions localized at the edges.
Since the vertices are dual to 4D structures and the edges to 3D structures, this raises the question of how the 4D Lorentzian structures are reconstructed by the internal sums.
A second question concerns the simplified model, which we have shown to capture the EPRL asymptotics for vector and Euclidean Regge geometries. 
Since the simplified model has the advantage of much faster evaluation times, it is useful to understand why this is the case. 
These two questions are closely related, and can be answered by inspecting in more detail the role of the sums over the internal spins $l_{ab}$.

Let us look again at the decomposition \Ref{Ac}, and consider the simplified model, for which the internal sums are removed.\footnote{This is achieved through a further imposition of the $Y$ map inside the product $h^{-1}_a h_b$ in Eq. \Ref{Ac}, see \cite{Boosting} for details and motivations.} The vertex amplitude reduces to the $\{15j\}$ symbol evaluated at the boundary spins, and the booster functions at their minimal configurations:
\be\label{As}
A^{\rm EPRLs}_v \left(j_{ab}, \, \vec n_{ab}\right) = \sum_{k_{a},i_{a}} \{15j\} %_{j_{ab},i_4}(l_{ab}, k_{e})
\prod_{a=2}^{5} d_{k_{a}} B^\g_{4}(j_{ab},j_{ab};i_{a}, k_{a})  \prod_{a=1}^5 c_{i_a}(\vec n_{ab}).
\ee
The internal intertwiners are still being summed over,
but in the large spin limit the booster functions are Gaussians peaked on equal intertwiners, and to lowest order \cite{Puchta:2013lza,Boosting,noiGen}
\begin{equation}\label{Bapp}
B^\g_4(j_{ab},j_{ab}; i_a,k_a) \simeq \f{b(\g,j_{ab})}{\l^{3/2} \, d_{i_a}} \delta_{i_a k_a}.
\end{equation}
The simplified vertex amplitude \Ref{As} is thus in first approximation proportional to a single coherent $\{15j\}$ symbol, which we recall is the vertex amplitude for SU(2) BF theory:
\begin{equation}\label{simpl}
A^{\rm EPRLs}_v (j_{ab},\vec{n}_{ab}) \simeq  \f1{\l^6} \left(\prod_{a=2}^{5}  b(\g,j_{ab})\right) A_v^{\scriptscriptstyle \SU(2)}(j_{ab},\vec{n}_{ab}),
\qquad 
A_v^{\scriptscriptstyle \SU(2)}(j_{ab},\vec{n}_{ab}):= \sum_{i_{a}} \{15j\}  
 \prod_{a=1}^5 c_{i_a}(\vec n_{ab}).
\end{equation}
This approximation shows that the simplified model has the same critical points of SU(2) BF theory, and $\l^{-6}$ times its scaling. 
The SU(2) coherent vertex amplitude has no critical behavior for Lorentzian Regge data and thus decays exponentially; whereas  for Euclidean Regge data it
admits two distinct critical points and oscillations with exactly the same frequency \Ref{Regge} of the EPRL model \cite{BarrettSU2},
\be\label{SU2LO}
A_v^{\scriptscriptstyle \SU(2)}(j_{ab},\vec{n}_{ab}) = \f1{\l^6}\f1{2^6\pi^2}\f{e^{i\l\Psi_c}}{|\det H^{\scr \SU(2)}_c|^{1/2}}\cos\Big(\l S_{\rm R}-\f12\arg H^{\scr \SU(2)}_c\Big) +O(\l^{-7}).
\ee
Here $H^{\scr \SU(2)}_c$ is the Hessian determinant at the critical point, and $\Psi_c$ a global phase determined by the action at the critical point, which is purely imaginary, and which depends in turn on the gauge choice and the phase of the coherent states exactly as for the EPRL model. 

From this analysis we understand why the simplified model has no critical behavior for Lorentzian Regge data, and the same critical behavior of SU(2) BF for vector and Euclidean Regge data. In particular, for Euclidean Regge data, inserting Eq. \Ref{SU2LO} into Eq. \Ref{simpl} we find that the simplified model has the same $\l^{-12}$ power law decay and the same $\g$-independent frequency of oscillations $S_{\rm R}$ of the full EPRL model. This explains why the simplified model captures the right scaling and frequency of oscillations of the EPRL asymptotics \Ref{ERLO}, as was shown in Fig.~\ref{fig:EuclideanAsym}. 

\begin{figure}[H]
\centering
\includegraphics[width=7.5cm]{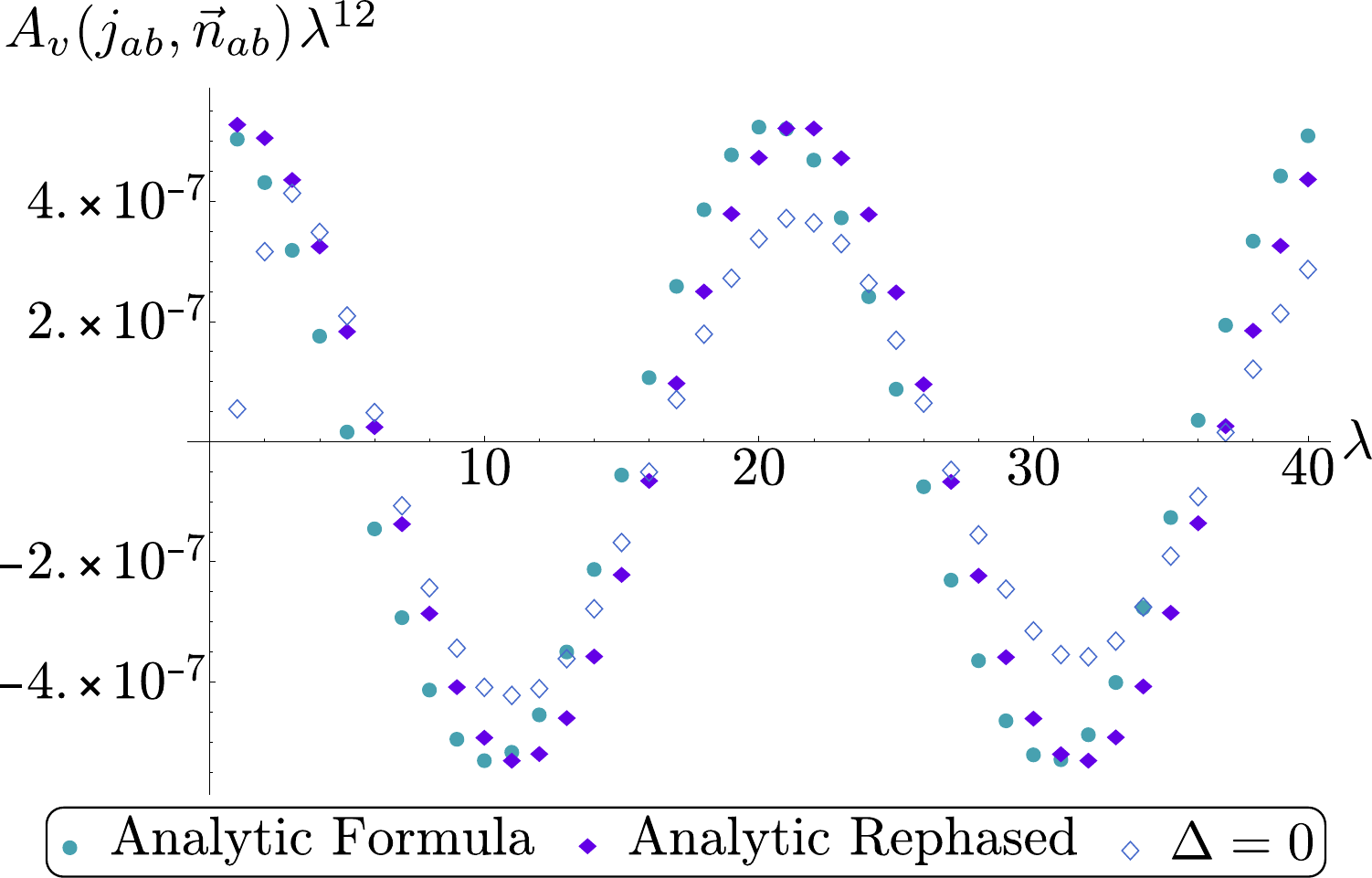}  \hspace{1cm}
\includegraphics[width=7.5cm]{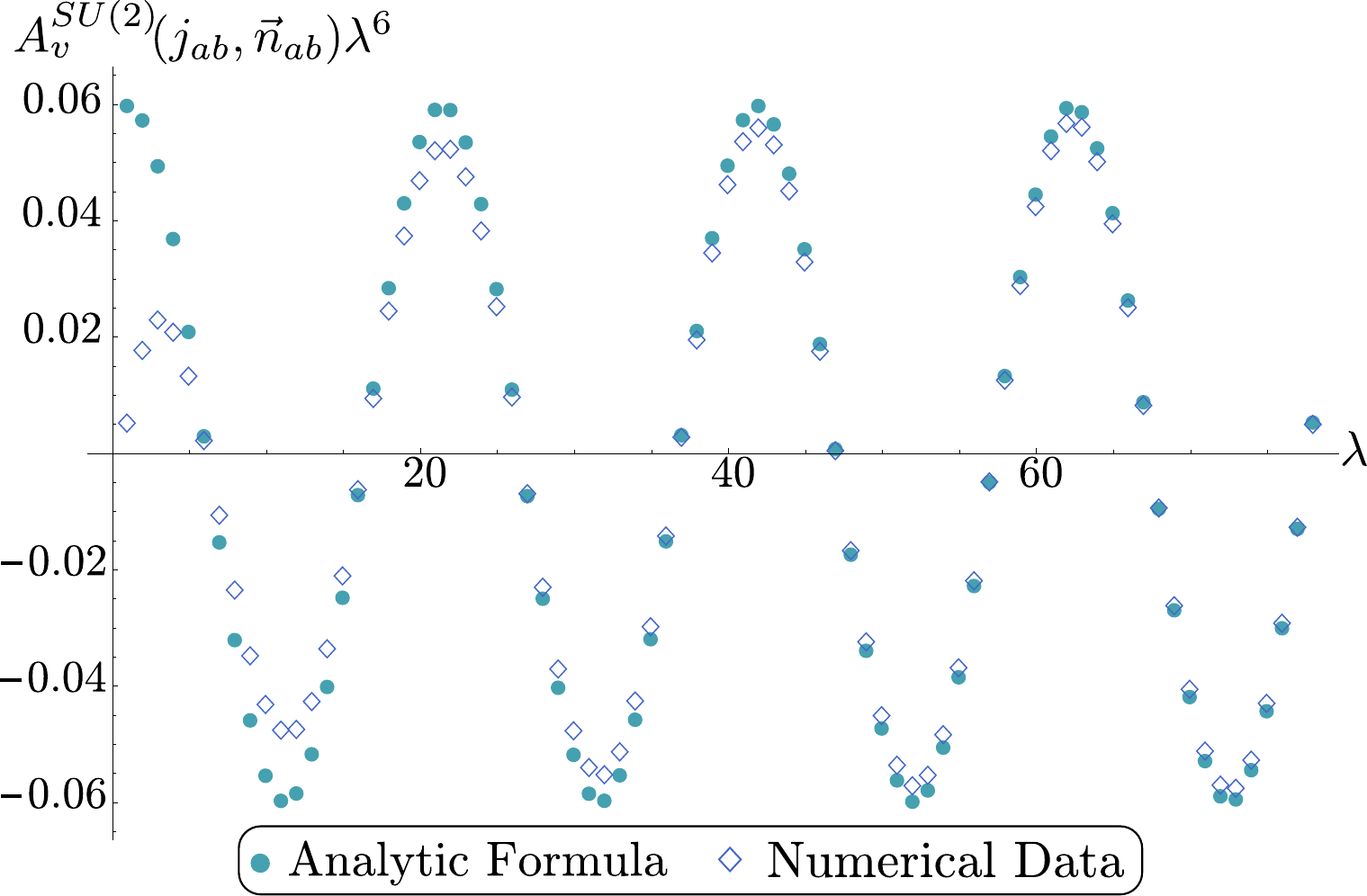}
\caption{\label{FigSU2}  \small{Left panel: 
\emph{Comparison of the simplified model with the EPRL asymptotics and the modified asymptotics with the SU(2) offset \Ref{LOs}. The improved agreement shows that the right asymptotics of the simplified model is to first approximation the same of SU(2) BF theory, and implies that the role of the internal sums in this case is to contribute to the re-phasing of the offset to match the EPRL asymptotics - as well as contributing to the magnitude.} 
\\ Right panel:
\emph{Evauation of the SU(2) BF asymptotics for an equilateral Euclidean 4-simplex, and comparison with the analytic formula \Ref{SU2LO} (It updates the plot of \cite{IoSU2asympt} which stopped at $\l=50$ - or $\l=25$ in that notation). It shows how while the frequency is captured early on, the magnitude requires higher spins.
At $\l=79$ the data for the $\{6j\}$ required weighted 230 gigabytes and saturated our server's capacity. }}}
\end{figure}

It also explains why it does \emph{not} capture the right cosine phase offset: The simplified model sees the SU(2) phase offset $-\f12\arg H^{\scr \SU(2)}_c=-0.324$ for the equilateral configuration, which is absent in the asymptotic formula for the EPRL model, since the Hessian is in that case real. Consider then a modified asymptotic formula 
\be\label{LOs}
A^s_v({\rm Euclidean \ data}) = \f{5.13\times 10^{-7}}{\l^{12}}
\cos\Big(5\l \arccos \big(- \f14\big) -0.324\Big),
\ee
where we added by hand the offset of the SU(2) BF amplitude. This formula correctly matches the cosine phase offset of the simplified model, see Fig.~\ref{FigSU2}. 
As a consequence, the internal sums pile up to the SU(2) Hessian to give the right magnitude of the EPRL amplitude, but also contribute to the phase offset of the cosine, creating an interference pattern that changes Eq. \Ref{SU2LO} to Eq. \Ref{ERLO}.

We take this opportunity to provide in the same figure an updated plot of the SU(2) BF asymptotics with respect to the one presented in \cite{IoSU2asympt}. It allows us to highlight that the frequency of the asymptotic formula is matched early on by the exact evaluation, but the magnitude only at higher spins. We expect a similar situation for the EPRL model.

\begin{figure}[H]
\centering
\includegraphics[width=7.5cm]{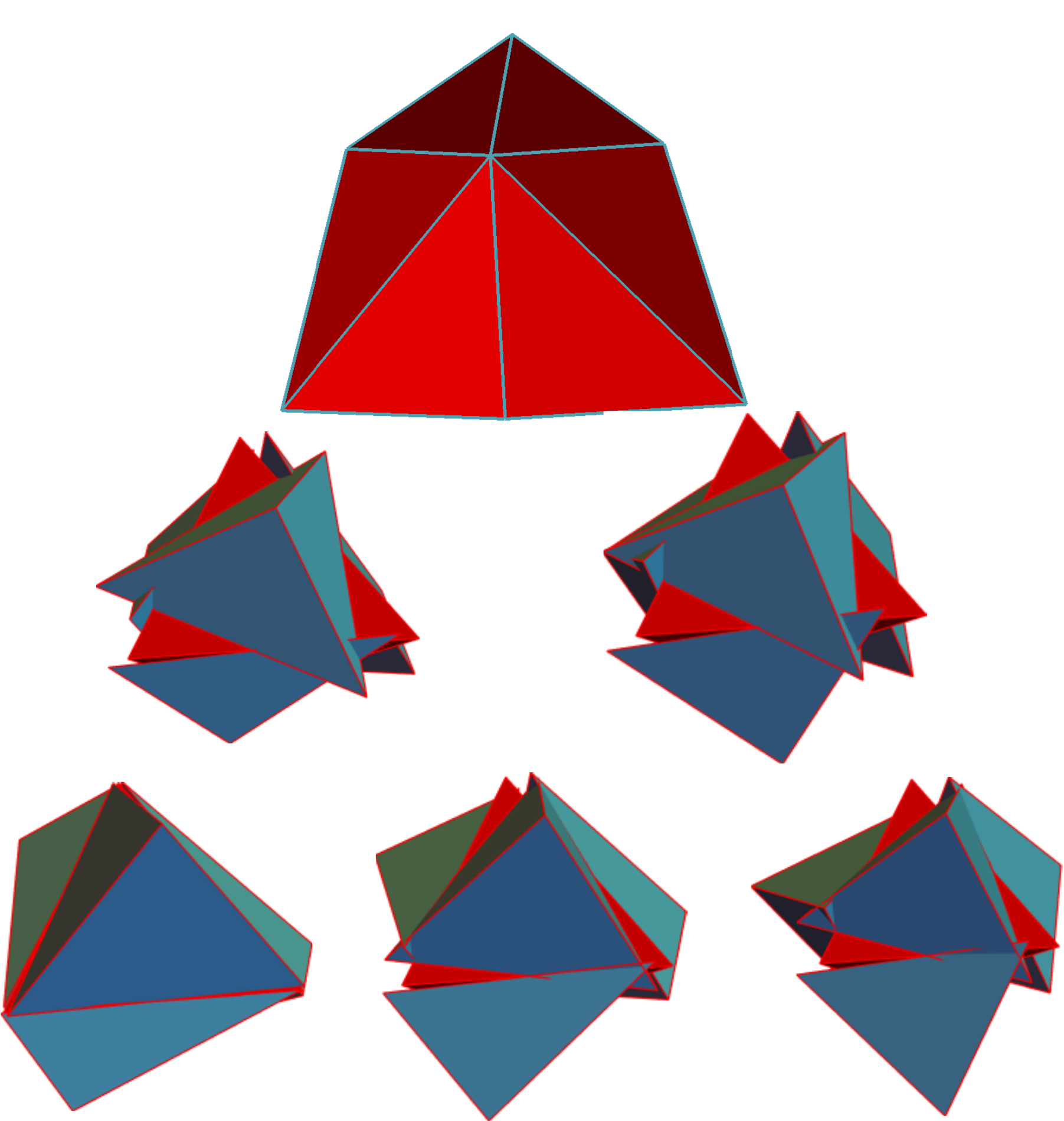} 
\caption{\label{FigCentu}  \small{Top panel: \emph{A pictorial representation of the \textit{spike} of the Lorentzian 4-simplex used in our numerical analysis. One tetrahedron is equilateral with areas 5, the other four are isosceles with base area 5 and the others 2.}\\
Middle panel: \emph{\textit{Twisted spike} for the Euclidean 4-simplices with areas compatible with the spins in the first shell. The 4-simplex on the left has one equilateral tetrahedron with areas 5, one isosceles tetrahedron with base area 5 and the other 3 and three tetrahedra with one area 5, one area 3 and the other two areas 2. The one on the right has one equilateral tetrahedron with areas 5, and four isosceles tetrahedra with base area 5 and the others 3 .}\\
Bottom panel: \emph{\textit{Twisted spike} for the Euclidean 4-simplices with areas compatible with the spins in the second shell. The 4-simplex on the left has one equilateral tetrahedron with areas 5, one isosceles tetrahedron with base area 5 and the other 4 and three tetrahedra with one area 5, one area 4 and the other two areas 2. The one in the middle is similar but with areas 5, 4, and 3. The one on the right has one equilateral tetrahedron with areas 5, and four isosceles tetrahedra with base area 5 and the others 4.}
 } } 
\end{figure}

Let us now consider Lorentzian boundary data. As already stated, the SU(2) vertex amplitude is exponentially suppressed for such data and thus also the simplified model should be, in agreement with the indication of exponential decay in Fig.~\ref{fig:LorentzianAsym1}. Only the $l$-shells can see the critical behavior. 
If we look at the factorization \Ref{Ac}, we need both booster functions and the $\{15j\}$ symbol to individually have critical behavior. The $\{15j\}$ has critical behavior only for Euclidean Regge configurations, or vector geometries. The booster functions must then admit a critical behavior precisely at those configurations for which the $j$'s correspond to a spacelike Lorentzian 4-simplex, and the $l$'s to a Euclidean one, or a vector geometry. 
A saddle point approximation of the booster function will appear in \cite{IoPierreMarco}, and shows that critical behavior appears when the two sets of spins and coherent intertwiners correspond to tetrahedra that can be boosted into one another. The action of the boosts is defined embedding each 3D normal as the electric part of a $\g$-simple bivector. 
Hence there are a priori an infinite number of admissible $l$'s contributing to the critical behavior, characterized by the existence of a certain classical boost mapping the $j$ tetrahedra to the $l$ tetrahedra, and then a critical point of the $\{15j\}$ symbol at that configuration. There are a priori infinitely many Euclidean 4-simplices for which this construction works, and we had fun exploring the ones with smallest $l$'s in a reconstruction algorithm depicted in Fig.~\ref{FigCentu}. 

According to this argument, all shells should contribute to the asymptotic formula, and Eq. \Ref{LRLO} should only be reproduced accurately when sufficiently many shells are summed over. One can reasonably hope that a limited number of shells will suffice to most applications, and in the luckiest cases, the single first shell and its qualitative matching to the asymptotic behavior. 
For situations where one needs to sum many shells, the method \cite{Boosting} here used becomes less efficient and may have to leave way for alternative approaches, like for instance adaptive Monte Carlo.

%------------------------------------------------------------
\subsection{Numerical improvements}
\label{SecOverview}
%------------------------------------------------------------

\subsubsection{Improved server}
A few of the numerics presented in this paper can be performed on a simple laptop, like computing the simplified model at spins below 10, our counting of the number of configurations compatible with the triangle inequalities. For the rest, we used a server with 32 cores at the CPT. While this enormously improves with respect to a single laptop, it is effectively very small with respect to the computing power used in other branches of theoretical physics. The CPT is in the process of acquiring a server with 128 cores. When it will come on line, it will cut our evaluation times by a factor of 4, and push forward current computing and RAM limits. 
To fully take advantage of this additional power however, one will need to improve the part of the code that computes the booster functions, to eliminate the numerical instabilities plaguing it at spins of order 50. Once this is done, a more powerful server will no doubt be able to fully test the asymptotics with Lorentzian boundary data, and show at which spins it is reached, and for which cutoff $\D$, answering the questions left open here.

A more powerful server will also allow numerical evaluations for extended triangulations. In particular for calculations not requiring the use of coherent states, computational times may be reasonable for a triangulation with a handful or more of 4-simplices.  
Beyond that, one may likely need simplification schemes, like for instance the minisuperspace models used for a numerical study of renormalization in the Euclidean EPRL \cite{Bahr:2017klw},  or approximations.
To that end, let us briefly comment on one approximation and two variations that could be considered in future work.

\subsubsection{Cutting the boosters}

Cutting the coherent boundary states is the only approximation used in some of the calculations of this paper. A second approximation that should be considered is to cut the booster functions, as it could turn out to be more efficient. 
The boosters 
are decreasing functions of the $l$'s, typically power law but in many cases exponentially decreasing. If one could identify a priori which ones, then one could cut the $l$ sums. The numerical investigations we performed do not suggest any easy way in which this can be done. We found for example cases in which most of the boosters within a shell are one to 3 orders of magnitude smaller than the few dominant ones, yet the total sum of the subdominant ones is of the same order as the sum of the dominant ones, thus forestalling any truncation. Some light over this possibility could be shed by analytic studies of the boosters' asymptotic behavior \cite{Puchta:2013lza,IoPierreMarco}.

\subsubsection{Alternative expressions for the boosters}
Our code computes the booster functions using numerical integration with the trapezoidal method, and the expression of the boost matrix elements $d^{}$ as finite sums derived in \cite{Francois}. This expression contains ratios of Euler's gamma functions with both real and complex arguments. We spent a significative amount of time optimizing this part of the code and running tests and debugs. 
The instabilities we found arising at spins of order 50 are the results of very small numbers constructed from ratios of very large numbers, 
  and it is the part of our code that we should, and we plan to, improve next. 

On the other hand, there are alternative expressions that could be considered in future work. One is the alternative formula derived in \cite{Boosting}, where the $r$ integrals are solved analytically but at the price of introducing a new integral over a virtual $\r$ label. The $\r$ integrand has a nice Gaussian-like behavior,\footnote{With a peak when the virtual $\r$ label takes value $\g j$, thus showing a certain type of simplicity off-shellness in the model.}, which can be exploited in numerical evaluations. 

Another possibility is to implement the recursion relations for the $\SL(2,\C)$ Clebsch-Gordan coefficients derived in \cite{Rashid70II}. For a simpler model with three-valent boosters and no intertwiners, this approach is extremely powerful, and one can push the internal sums to order $\D\sim 100$
\cite{PietroGiorgioToappear}.

\subsubsection{Alternative form of the coherent states}

The coherent state coefficients $c_i(\vec n_{ab})$ are a fast part of the numerical code.
However, since an enormous number of these coefficients is needed, it could be interesting to explore alternative definitions that could speed up this procedure. One possibility is the cross ratio coherent states \cite{EteraHoloQT}, and the analytic expressions studied in \cite{Bonzom:2012bn}. This basis has more economical features and proved advantageous in numerical evaluations of the coherent state volume operator defined in \cite{IoPoly}. We hope to be able to investigate its implementation in future work.

As we estimated in Sec. 4, the use of coherent states can require up to an additional $\l^5$ terms, which has a strong impact on the numerical cost of the amplitude. Coherent states are necessary for the asymptotic formula, but many spin foam calculations can be done in the more economical orthonormal basis of intertwiners. This can be an important trade-off when moving to consider amplitudes on many 4-simplices. 
We have seen that the number of terms for a single equilateral 4-simplex with coherent states scales like $\l^{10}$. This is the same number of terms needed to compute, with orthonormal intertwiners, the equispin amplitude with three 4-simplices glued together and one internal sum. This $\D_3$ amplitude is relevant for a discussion of the flatness problem \cite{Collet,HellmannFlatness}, and will be the object of future work.
More in general, avoiding coherent states and increasing the number of cores, we expect to be able to make calculations on a few vertices. Going beyond that will probably require some deeper optimization of the method, or reduction to minisuperspace models.

%----------------------------------------------------------- 
\section{Conclusions}
%----------------------------------------------------------- 

We presented for the first time a numerical evaluation of the Lorentzian EPRL vertex amplitude. 
Obtaining this result proved to be a challenging task, and the work presented in this paper took more than two years to complete. Preliminary, more qualitative results were presented at the conference Loops '17. Making them quantitatively precise was complicated by the many technical details to be worked out. 
The challenges were twofold. The main one was clearly to setup the numerical code, which includes evaluating the booster functions, the $\{15j\}$ symbols, the coherent states, and assembling them via hash tables. All making sure that there were no numerical instabilities or inconsistent conventions. A bit more surprisingly, we also faced some analytical challenges, concerning the way the boundary data were to be constructed and fed to the computer, and computing the overall phase and Hessian of the asymptotic formula.  While these quantities were well defined in principle, precise explicit formulas and reconstruction algorithms were missing. We devoted part of the paper to explaining how this can be done explicitly, and the online repository \cite{code-sl2cfoam} contains explicit Mathematica codes to answer these questions. We hope that our paper and this repository can provide a reference point for future work in this direction, so that one does not have to spend time again on this background material.

On the analytic side, we provided useful formulas and explicit configurations for the boundary data to be used in the amplitude. We shed light on the geometric meaning of the internal sums, explaining in particular why the simplified model asymptotes to the Regge action for Euclidean configurations but not for Lorentzian ones.

On the numerical side, our main conclusion is that precise numerical evaluations of the complete Lorentzian EPRL model are currently possible, and it is worthwhile to invest computational capabilities to address explicit physical questions. The method of \cite{Boosting} and its implementation developed in \cite{Dona:2018nev} have been successfully tested. Including only a few terms of the internal shells appears sufficient to most applications, avoiding the need to develop a full Monte Carlo approach of the amplitude.  The codes for the geometric reconstruction of the data and the numerical evaluation of the vertex amplitudes are publicly available at \cite{code-sl2cfoam}.
More specifically to the asymptotic large spin behavior of the vertex amplitude, our conclusions confirm the analytic formula of \cite{BarrettLorAsymp} already at spins of order 10 for vector and Euclidean Regge data. Higher spins are required for a good match in the case of Lorentzian Regge data, but the power law decays and $\g$-dependence of the oscillations are already evident at spins of order 10.

The computational resources it takes to compute the vertex amplitude in the coherent basis are considerable but at the same time affordable. 
A natural and indeed crucial question is how to extend the numerical calculations to many vertices. We performed computations using 32 parallel cores, but let us not forget that in other areas of physics, computations are done with hundreds or thousands of cores. What will be computed in the future depends on the resources available, as well as on numerical improvements and optimizations that can still be done.	
Finally, we would like to stress that while our methods and results were motivated by the EPRL model, they can be applied or adapted to other spin foam models based on or related to $\SL(2,\C)$ \cite{KKL,HellmannFlatness,Haggard:2014xoa,Engle:2015mra,Engle:2015zqa,Kaminski:2017eew,Liu:2018gfc},
their physical applications, e.g., \cite{Christodoulou:2016vny}, to applications involving many vertices through intertwiner renormalization as in \cite{Dittrich:2014mxa} or minisuperspace renormalization as in \cite{Bahr:2017klw}, as well as to calculations of tensorial and graph invariants regardless of their quantum gravity applications.

%----------------------------------------------------------- 
\subsection*{Acknowledgments}
%----------------------------------------------------------- 
We would like to thank Jean-Roch Liebgott and Lorenzo Bosi for help with the server at CPT, and Francesco Gozzini and Fran\c{c}ois Collet for  discussions and the shared work on the numerical code.
P.D. thanks Tommaso De Lorenzo and Eugenio Bianchi for useful discussions, and support from 
FQXi grants No. 2016-165616 and No. 2018-190485. P.D. is also supported by the NSF grants No. PHY-1505411 and No. PHY-1806356, and the Eberly research funds of Pennsylvania State University. We thank Muxin Han, Hongguang Liu and Dongxue Qu for exchanges on the calculation of the Hessian that led us to correct the sign of its phase. 

%----------------------------------------------------------------------------
\appendix
\setcounter{equation}{0}
\renewcommand{\theequation}{\Alph{section}.\arabic{equation}}

%----------------------------------------------------------------------------
\section{Conventions and explicit formulae}\label{AppA}
%----------------------------------------------------------------------------
In this appendix we fix our conventions and notation, and provide the explicit relation between our version \Ref{Ac} of the coherent amplitude and the one in \cite{BarrettLorAsymp}. The difference is just in the phase and a complex conjugation of the spinors, but it has multiple origins due to the many conventional choices involved. Our conventions follow those of \cite{Perelomov,Ruhl,Varshalovich}. 

%----------------------------------------------------------------------------
\subsection{SU(2) coherent states and spinors}
%----------------------------------------------------------------------------

%---------------------------------------------------------------------------
\subsubsection*{SU(2) coherent states}
%---------------------------------------------------------------------------
The uncertainty in the direction of a state $\ket{j,m}$ in the SU(2) irrep $j$ is minimized by the lowest and highest weights $m=\pm j$. Acting with a rotation $g\in\SU(2)$ on $\ket{j,\pm j}$ defines two alternative families of coherent states. Orbits of the isotropy subgroup $\U(1)\subset\SU(2)$ generate an irrelevant global phase in the coherent states, and it is sufficient to consider two-parameters group elements in the form \cite{Perelomov}
\be\label{Hs}
n(\z) := \f1{\sqrt{1+|\z|^2}}\left( \begin{array}{cc} 
1 & \z \\  -\bar\z & 1
\end{array}\right), \qquad \z = -\tan\f\Th2 e^{-i\Phi}\in\C P^1.
\ee
This is also known as the Hopf section for the $\SU(2)\simeq S^2\times S^1$ fibration, and we parametrized 
$\zeta$ as the stereographical projection of the sphere from the south pole, with $(\Th,\Phi)$ the zenith and azimuth angles. In terms of Wigner's matrices as $D(\alpha,\beta,\gamma) = e^{-i \alpha J_z} e^{-i \beta J_y} e^{-i \gamma J_z}$ \cite{Varshalovich} (caveat in reading the files in \cite{code-sl2cfoam}, Wolfram's Mathematica uses opposite sign conventions for this parametrization), 
\be\label{defDhopf}
n(\z) = D^{(1/2)}(\Phi,\Th,-\Phi).
\ee

The standard choice is to define coherent states starting from the lowest weight \cite{Perelomov}, 
\be\label{defzeta}
\ket{j,\z} := n^{(j)}(\zeta)\ket{j,-j}, \qquad \bra{j,m}j,\z \ra = D^{(1/2)}_{m,-1/2}(\Phi,\Th,-\Phi).
\ee
They satisfy 
\be
\bra{j,\z}\vec J\ket{j,\z} = -j\vec n,
\qquad \label{defn}
\vec n : = (\sin\Th\cos\Phi,\sin\Th\sin\Phi,\cos\Th),
\ee 
and the important factorization property
\be
\bra{j,\z_1}j,\z_2\ra = \bra{\tfrac12,\z_1}\tfrac12,\z_2\ra^{2j}.
\ee
The phase ambiguity associated with the U(1) subgroup, namely with choosing a different section than Hopf's, is the one that can be used to make the coherent amplitude real for any boundary data configuration, as discussed in \cite{BarrettLorAsymp} and in the main text.

It is often convenient to use also the second family, the one constructed from the highest weight. To distinguish it, we use a different ket notation,
\be
|j,\z] := n^{(j)}(\zeta)\ket{j,j}, \qquad \bra{j,m}j,\z] = D^{(1/2)}_{m,1/2}(\Phi,\Th,-\Phi).\label{defzeta+}
\ee
It satisfies
\be
[j,\z| \vec J |j,\z] = j\vec n,
\ee
namely the state points in the antipodal direction on the sphere.
The two families are in fact related by a parity transformation realized by the complex structure
\be\label{Jcs}
J^{(j)}\triangleright \ket{j,\z} :=  \eps^{(j)} \ket{j,\bar\z} = |j,\z], \qquad J^{(j)}{}^2=(-1)^{2j}.
\ee
Here $\eps^{(j)}_{mn} :=  (-1)^{j-m}\d_{m,-n} = D^{(j)}_{mn}(i\s_2)$ is the spinorial metric.
Another useful operation is the antipodal map 
\be\label{antipodal}
\vec n\mapsto -\vec n \quad  \Leftrightarrow \quad (\Th,\Phi)\mapsto (\pi-\Th,\pi+\Phi) \quad \Leftrightarrow\quad \z\mapsto\z^{\scr AP}=-\f1{\bar\z}.
\ee
We observe that
\be\label{phaseshift}
|j,\z] = e^{2ij(\pi+\Phi)}\ket{j,\z^{\scr AP}}, \qquad |j,\z^{\scr AP}] = e^{2ij\Phi}\ket{j,\z},
\ee
so the antipodal map has the effect of changing family without preserving the Hopf section because of the additional phase. This will play a role below.

\subsubsection*{Coherent intertwiners}

Group averaging the tensor product of coherent states defines the coherent intertwiners \cite{LS},
\begin{align}
\int_{\SU(2)} dg \otimes_f D^{(j_f)}(g) \ket{j_f,\z_f}= \sum_{i_{12}} d_{i_{12}} \, c_{i_{12}}(\vec n_f) \, \ket{j_1,\ldots j_4,i_{12}},
\end{align}
where ${i_{12}}$ is the standard orthonormal intertwiner basis, here taken in the 12 recoupling channel. The coherent intertwiner coefficient  $c_{i_{12}}(\vec n_f)$ is given by Eq. \Ref{eq:coeffCS} in terms of Wigner's four-legged symbol
\be\nn
\left(\begin{array}{c} j_f \\ m_f \end{array}\right)^{(i_{12})} = \Wfour{j_1}{j_2}{j_3}{j_4}{m_1}{m_2}{m_3}{m_4}{i_{12}} 
= \sum_{m_{12}} (-1)^{j_{12}-m_{12}}  \Wthree{j_1}{j_2}{j_{12}}{m_1}{m_2}{m_{12}} \Wthree{j_{12}}{j_3}{j}{-m_{12}}{m_3}{m}.
\ee
In Eq. \Ref{eq:coeffCS} we used $\vec n$ to label the argument of the Wigner matrices, instead of Eq.\Ref{defDhopf}. This was done so to reduce at most the technical details needed to read the main text.
This coherent intertwiner is directly applicable to the tetrahedron labeled 5 in Eq. \Ref{AvG}. In the other cases, one or more kets are replaced by bras, and simultaneously we put an antipodal map in the normal labeling the coherent states, like in Eq. \Ref{A1}. Where we have a bra, the intertwiner needs to be  multiplied by the epsilon matrix $\eps^{(j)}$ to take into account the link's orientation. For a tetrahedron with only one bra, say in the first position of the graphic symbol, the resulting coefficient is
\begin{align}\label{c1App}
&c^{(1)}_{i_{12}}(\vec n_f):= \Wfour{j_1}{j_2}{j_3}{j_4}{-m_1}{m_2}{m_3}{m_4}{i_{12}} (-1)^{j_1-m_1}\overline{ D^{(j_1)}_{m_1,j_1}(-\vec n_1) } \prod_{f=2}^4  D^{(j_f)}_{m_f,j_f}(\vec n_f).
\end{align}
For the numerical code, it was convenient to rewrite this expression in terms of the intertwiner with all four legs outgoing. This can be achieved by recalling that 
\be
\overline{ D^{(j)}_{m,n}(g)} = (-1)^{m-n} D^{(j)}_{-m,-n}(g),
\ee
which applied to Eq. \Ref{c1App} leads to 
\begin{align}
&c^{(1)}_{i_{12}}(\vec n_f) =  \Wfour{j_1}{j_2}{j_3}{j_4}{m_1}{m_2}{m_3}{m_4}{j_{12}}  D^{(j_1)}_{m_1,-j_1}(-\vec n) \prod_{f=2}^4  D^{(j_f)}_{m_f,j_f}(\vec n_f). 
\end{align}
These are the expressions used in \vaccachespacca.

%---------------------------------------------------------------------------
\subsubsection*{Spinors}
%---------------------------------------------------------------------------
The fundamental irrep $j=1/2$ is isomorphic to $\C^2$, and the relation of the coherent states to spinors is very useful to study properties of the $\SL(2,\C) $ unitary irreps. We denote a spinor with $z^A\in \C^2$, $A=0,1$, and use 
\be\label{defepsspinor}
\eps_{AB} = \mat{0}{1}{-1}{0} = \eps^{AB}
\ee
to raise and lower indices. We equip $\C^2$ with a complex structure $J$ and define the dual spinor
\be\label{defcomplstructure}
(Jz)^A := J\triangleright z^A := 
 \vet{\bar z^1}{-\bar z^0}, \qquad {J}^2=(-1)^{2j}.
\ee
Notice that so defined, the complex structure is consistent with Eq. \Ref{Jcs} for $j=1/2$. $\eps_{AB}$ gives an $\SL(2,\C)$-invariant bilinear coupling, and we can introduce an Hermitian norm given by complex conjugation (book kept by dotting the indices) and the identity matrix $\d_{A\dot A}$, which is nothing but a choice of canonical timelike vector in Penrose's abstract index notation.
With a Dirac-like notation for spinors \cite{IoWolfgang}, the relations between spinors and their dual can be compactly written
\be
\ket{z}=\vet{z^0}{z^1}=-\eps|\bar z]=-{J}\triangleright |z] = -|J z],\qquad |z]= {J}\triangleright\ket{z}= \ket{J z}=\eps\ket{\bar z} = \vet{\bar z^1}{-\bar z^0},
\ee
with $\eps=i\s_2$, and the Hermitian and Lorentz-invariant bilinear are
\be\label{due}
\bra{z}w\ra=\d_{A\dot A}\bar z^{\dot A}w^A = \bra{{J}w}{J} z\ra=[w|z], \qquad [z\ket{w}=\eps_{AB}w^A z^B \equiv \bra{ {J} z}w\ra.
\ee
Finally, for the norm we use
\be
\norm z^2:=\bra z z\ra = |z^0|^2+|z^1|^2.
\ee

Coming back to the SU(2) coherent states, the lowest-weight ones define spinors
\be\label{zspin}
\ket{\z}:=\ket{\tfrac12,\z} = \vet{-\sin\f\Th2 e^{-i\Phi}}{\cos\f\Th2}
\ee
of unit norm
and $\arg z^1=0$. The map $z^0 = -\sin\f\Th2 e^{-i\Phi}, z^1=\cos\f\Th2$ provides a choice of section for the 
$\C^2\simeq \C P^1\times \C P^1$ fibration. Conversely, an arbitrary spinor defines a coherent state for the fundamental irrep which is redundantly parametrized by an additional norm and phase,
\be\label{spinCS}
\ket{z} =\norm{z} e^{i\arg z^1} \ket{\tfrac12,\z}, 
\qquad 
\z=\f{z^0}{z^1}, \qquad |z^1| = \f{\norm z}{(1+|\z|^2)^{1/2}}.
\ee
The additional phase corresponds to the phase ambiguity in the coherent states previously discussed, whereas the norm can be eliminated by working always with normalized spinors. 
Spinorial SU(2) coherent states feature prominently in spin foam models and their applications, see e.g. \cite{EteraHoloEucl,IoHolo,IoWolfgang,Hnybida:2015ioa,FreidelPachner14,Bonzom:2015ova}. An immediate advantage of their use is to write the resolution of the identity as Gaussian integrals.

For the highest weight, 
\be\label{phaseshiftz}
|\z]:=|\tfrac12,\z]\equiv J\triangleright \ket{\z} = \vet{\cos\f\Th2}{\sin\f\Th2 e^{i\Phi}} = e^{i\Phi} \ket{\tfrac12,\z^{\scr AP}}.
\ee
Again, an arbitrary dual spinor defines a redundant coherent state with an additional norm and phase,
\be\label{spinCSd}
\ket{z} =\norm{z} e^{i\arg z^1} \ket{\tfrac12,\z}.
\ee
The expressions \Ref{spinCS} and \Ref{spinCSd} can be generalized to an arbitrary irrep $j$, to provide spinorial coherent states for SU(2).
To that end, we consider the matrix
\be
\label{defgz}
g(z^A) :=\f1{\norm z} \Big( |z],\ket{z}\Big) = \f1{\norm z}\mat{\bar z^1}{z^0}{-\bar z^0}{z^1} = n(\z)e^{-i\arg z^1 \s_3}, \qquad \z=\f{z^0}{z^1}, \qquad n(\z)\equiv g_{\scr R}(\z,1).
\ee
We then define the spinorial coherent states 
\begin{align}
& \ket{j,z^A} := \norm{z}^{2j} D^{({j})}\big(g_{\scr R}(z^A)\big)\ket{j,-j}= \sum_m \binom{2j}{j+m}^{1/2} (z^0)^{j+m} (z^1)^{j-m} \ket{j,m} \equiv \norm{z}^{2j}  e^{2ij\arg z^1}\ket{j,\z}, \\
& |j,z^A]:= \norm{z}^{2j} D^{({j})}\big(g_{\scr R}(z^A)\big)\ket{j,j}= \sum_m \binom{2j}{j+m}^{1/2} (-\bar z^0)^{j-m} (\bar z^1)^{j+m} \ket{j,m} \equiv \norm{z}^{2j}  e^{-2ij\arg z^1} |j,\z],
\end{align}
where in the second equality we used the following properties of the Wigner matrices:
\be
D^{(j)}_{m,-j}(g)=\binom{2j}{j+m}^{1/2} (g_{12})^{j+m} (g_{22})^{j-m}, \qquad
D^{(j)}_{mj}(g)=\binom{2j}{j-m}^{1/2} (-\bar g_{12})^{j-m} (\bar g_{22})^{j+m}.
\ee
These expressions reduce to Eqs. \Ref{spinCS} and \Ref{spinCSd} for $j=1/2$.

Finally, we note the factorization properties
\begin{align}
& [j,w\ket{j,z} =  \sum_m \binom{2j}{j+m} (w^1 z^0)^{j+m} (-w^0 z^1 )^{j-m} = [z\ket{w}^{2j}, \\
& \la j,w\ket{j,z} =  \sum_m \binom{2j}{j+m} (\bar w^0 z^0)^{j+m} (\bar w^1 z^1 )^{j-m} = \la w\ket{z}^{2j}.
\end{align}

%----------------------------------------------------------------------------
\subsection{$\SL(2,\C)$ unitary irreps of the principal series}
%----------------------------------------------------------------------------
%---------------------------------------------------------------------------
\subsubsection*{Homogeneous representation}
%---------------------------------------------------------------------------

We follow the conventions of \cite{Ruhl}, with the notational difference
\be
\r = \f12 \r_{\scr Ruhl}, \qquad k= -\f12 m_{\scr Ruhl}.
\ee
Unitary irreps of the principal series of $\SL(2,\C)$ are labeled by a pair $(\r \in \R,k\in \Z/2)$ \cite{GelfandLorentz, Naimark,Ruhl}, and can be represented on a space of homogeneous functions of two complex variables,
\be
F^{(\r,k)}(\l z^A) = \lambda^{k-1+i \rho}\bar{\lambda}^{-k-1+i \rho}\, F^{(\r,k)}(z^A), \qquad \l\in\C, \qquad A=0,1.
\ee
The scalar product is defined choosing any section of the $\C^2\simeq \C P^1\times \C P^1$ fibration,
\be\label{Fprod}
(F,F'):=\int_{\C P^1} d\m(z^A) \overline{ F(z^A) }F'(z^A), \qquad d\m(z^A):=\f i2 z_A d z^A\w \bar z_{\dot A} d \bar z^{\dot A}.
\ee
The scaling of the Lorentz-invariant measure $d\m(z^A)$ guarantees that the whole integrand is homogeneous of degree zero, and thus the integration is independent of the choice of section.

For the EPRL model with spacelike tetrahedra we pick Naimark's orthonormal basis, labeled by the eigenfunctions of $L^2$ and $L_z$ and given in the conventions of \cite{Ruhl} by 
\be\label{Fbasis}
F^{(\r,k)}_{jm}(z^A) := e^{i\Psi^\r_j}\sqrt{\f{d_j}\pi} \f1{\norm{z}^{2(1-i\r)}} D^{(j)}_{m,-k}\big(g(z^A)\big), 
\qquad j\geq|k|,\qquad j\geq m\geq-j.
\ee
Here $D^{(j)}$ are Wigner matrices, and $g(z^A)$ is the same matrix defined in Eq. \Ref{defgz}.

The function $\Psi^\r_j$ parametrizes a freedom in choosing the phase of the basis elements. This is set to zero in \cite{Ruhl} and \cite{BarrettLorAsymp}. This economical choice however leads to complex Clebsch-Gordan coefficients. 
An alternative choice is to take \cite{Boosting}
\be
e^{i\psi^\r_{j}}  = (-1)^{-\f{j}2} \f{\G(j+i\r+1)}{|\G(j+i\r+1)|},
\ee
which guarantees reality of the Clebsch-Gordan coefficients, and as a consequence also of all EPRL amplitudes $A_v(j_{ab},i_a)$ in the orthogonal intertwiner basis. 
This is the convention used in our numerical code, 
but this phase difference goes away in the expression of the coherent amplitude.

We introduce next the antilinear map $\cJ$, the infinite-dimensional analog of the complex structure $J$ for spinors, given by 
\begin{subequations}\label{calJ}\begin{align}
& {\cal J} F(z^A) := \overline{{\cal A} F(z^A)},\\
& {\cal A} F(z^A) := \f{\sqrt{\r^2+k^2}}\pi \int_{\C P^1}d\m(w^A)[w\ket{z}^{-k-1-i\r}\overline{[w\ket{z}}{}^{k-1-i\r} F(w^A).
\end{align}\end{subequations}
It satisfies
\be \label{Jproperties} {\cal J} \ :\ (\r,k)\mapsto(\r,k),\qquad ({\cal J}F,{\cal J}F')=(F',F), \qquad {\cal J}^2=(-1)^{2k}.
\ee
Using this map we can define the bilinear pairing
\be\label{defepsinfty}
\eps(F,F'):=({\cal J}F,F') =(-1)^{2k}\eps(F',F), \qquad (F,F')=\eps(F',{\cal J}F).
\ee
This is the infinite-dimensional analog of the spinorial bilinear \Ref{due}.

%---------------------------------------------------------------------------
\subsubsection*{Group elements and vertex amplitude}
%---------------------------------------------------------------------------

The $\SL(2,\C)$ group action in the homogeneous representation is given by matrix multiplication on the spinor argument, $h\triangleright F(z^A) = F(h^{\scr T} z^A)$. The matrix elements can be equally written as the Hermitian or the bilinear products,
\begin{align}\label{Dh}
D^{(\r,k)}_{jmln}(h) &:=  \bra{\r,k;j,m}h\ket{\r,k;l,n}= \left(F^{(\r,k)}_{jm}, h\triangleright F^{(\r,k)}_{ln}\right) =  \eps\left(h\triangleright F^{(\r,k)}_{ln}, {\cal J}F^{(\r,k)}_{jm}\right).
\end{align}
For the EPRL model, 
we are only interested in the ``minimal'' weights 
\be
\r=\g k, \qquad j=l=k,
\ee
and for group elements in the form $h= h_a^{-1}h_b$. Using the invariance of the integration measure, we can write these group elements as 
\begin{align}\label{Dhsimp}
D^{(\g j,j)}_{jmjn}(h_a^{-1}h_b) = \int_{\C P^1} d\m(z^A) \overline{F^{(\g j,j)}_{jm}(h_a^{\scr T}z^A)} F^{(\g j,j)}_{ln}(h_b^{\scr T}z^A).
\end{align}
The vertex amplitude in the orthonormal basis is simply the tensor associated to the complete five-valent graph,
\be
A_v(j_{ab},m_{ab}) =  \int \prod_{a=2}^5 dh_a \prod_{a<b}  D^{(\g j_{ab},j_{ab})}_{j_{ab},m_{ab}, j_{ab}, m_{ba}}(h_a^{-1}h_b).
\ee
Because of its SU(2) invariance it can be contracted for free with the four-legged Wigner's symbol to define the amplitude in the intertwiner basis
$A_v(j_{ab},i_{a}).$

\subsection{Definition of the $\SL(2,\C)$ coherent amplitude}
Since the maximal isotropy group of $\SL(2,\C)$ is the same as of SU(2), coherent states  are a direct embedding of the SU(2) ones into the unitary irreps used. For the EPRL irreps in the homogeneous representation \Ref{Fbasis}, we define the (lowest-weight) family of coherent states
\begin{align}\label{Fcs}
F^{(\g j,j)}_{j\z}(z^A) &:=\sum_m F^{(\g j,j)}_{jm}(z^A)\la j,m \ket{j,\z} 
= \sqrt{\frac{d_j}{\pi}}\, \norm{z}^{2(i\g j-1-j)} \la\bar z \ket{ \z}^{2j},
\end{align}
where the last step follows from the factorization properties of coherent states.\footnote{The complex conjugation in $\bra{\bar z}$, which could be avoided by defining the coherent states as $\sum_m F^{(\r,j)}_{jm}(z^A)\la j,\z \ket{j,m}$, is actually all right, and familiar from (holomorphic) representations generated by (unnormalized) coherent states, e.g. $f_{n}(z):=\bra{\bar z}n\ra=z^n/n!$ for the harmonic oscillator. }
It is useful to define also a dual basis associated with the highest weight \Ref{defzeta+},
\begin{align} 
F^{(\g j,j)}_{jJ \z}(z^A) &:=\sum_m F^{(\g j,j)}_{jm}(z^A) \bra{j,m} j,\z] 
= \sqrt{\frac{d_j}{\pi}}\, \norm{z}^{2(i\g j-1-j)} \bra{\bar z}\z]^{2j}. 
\end{align}
This is related to the antipodal map by the same phase shift \Ref{phaseshift} of the SU(2) case, namely
\be\label{phaseshift1}
F^{(\g j,j)}_{jJ \z}(z^A) = e^{2ij(\pi+\Phi)} F^{(\g j,j)}_{j\z^{\scr AP}}(z^A).
\ee

The map $\z\mapsto J\z$ from one family to the other can also be induced using the infinite-dimensional complex structure \Ref{calJ}.
An explicit calculation gives in fact
\begin{align}
\cJ F^{(\g j,j)}_{j\z}(z^A) 
&=e^{-i\arctan\g} \sqrt{\frac{d_j}{\pi}}\, \norm{z}^{2(i\g j-1-j)} [\z |z\ra^{2j}  
\label{phaseshift2}
= e^{-i\arctan\g}F^{(\g j,j)}_{j J\z}(z^A), \\
\cJ F^{(\g j,j)}_{jJ\z}(z^A) 
&=(-1)^{2j} e^{i\arctan\g}F^{(\g j,j)}_{j \z}(z^A), 
\end{align}
where we fixed $\g>0$ for convenience.

When defining the coherent vertex amplitude, it is convenient to have an antipodal map on the coherent state at the target of each link, so that the vectors can be interpreted consistently as outgoing (or incoming, if one prefers) normals. From the previous formulas, we know three different ways to apply the antipodal map: via $\z^{\scr AP}$, via $J\triangleright\ket \z$, or via $\cJ F$. The resulting amplitudes differ by an overall phase.
To keep the notation as simple as possible in the main text, we used the first option, so as to be able to refer only to unit vectors and bypass spinors. The coherent basis elements used in Eq. \Ref{A1} are defined as follows,
\begin{align}\label{defDcs}
D^{(\g j_{ab},j_{ab})}_{j_{ab},-\vec n_{ab}, j_{ab}, \vec n_{ba}}(h_a^{-1}h_b)&:=
D^{(\g j_{ab},j_{ab})}_{j_{ab},j_{ab}, j_{ab}, j_{ab}}\big(n^\dagger(\z^{\scr AP}_{ab})h_a^{-1}h_bn(\z_{ba})\big) 
=  e^{-2ij_{ab}\Phi_{ab}} D^{(\g j_{ab},j_{ab})}_{j_{ab}\z_{ab} j_{ab} {J\z}_{ba}}(h_a^{-1}h_b) 
\nn\\\nn
& = e^{-2ij_{ab}\Phi_{ab}}\int_{\C P^1}d\m(z_{ab}^A)  \overline{F^{(\g j_{ab},j_{ab})}_{j_{ab}\z_{ab}}(h_a^{\scr T}z_{ab}^A)} F^{(\g j_{ab},j_{ab})}_{j_{ab} J\z_{ba}}(h_b^{\scr T}z_{ab}^A)
 \\
& = e^{-2ij_{ab}\Phi_{ab}} \f{d_{j_{ab}}}\pi \int_{\C P^1} d\m(z^A_{ab}) \f{\exp S_{ab}}{\norm {h_a^{\scr T}z_{ab}}^2 \norm {h_b^{\scr T}z_{ab}}^2}, 
\end{align}
where in the last step we defined the action
\be\label{actionApp}
S_{ab} := j_{ab} \ln \frac{\bra{ h_{a}^{\scr T}z_{ab}}\bar\z_{ab}\ra^{2} [\bar{\z}_{ba}|h_{b}^{\scr T}z_{ab}\ra^{2}}{|\!| h_{a}^{\scr T}z_{ab}|\!|^2\, |\!| h_{b}^{\scr T}z_{ab}|\!|^2 }
+i\g j_{ab}\log\frac{\norm{h_{b}^{\scr T}z_{ab}}^2 }{|\!|h_{a}^{\scr T}z_{ab}|\!|^2 }.
\ee
Therefore our definition \Ref{A1} of the coherent vertex amplitude is
\begin{align}
A_v(j_{ab},\vec n_{ab}) &=  \int \prod_{a=2}^5 dh_a \prod_{a<b}  D^{(\g j_{ab},j_{ab})}_{j_{ab},-\vec n_{ab}, j_{ab}, \vec n_{ba}}(h_a^{-1}h_b)=
\nn\\ &=   e^{-2i\sum_{a<b}j_{ab}\Phi_{ab}} \left(\prod_{a<b} \f{d_{j_{ab}}}\pi\right)
\int \prod_{a=2}^5 dh_a \int_{\C P^1} \prod_{a<b}  \f{d\m(z^A_{ab})}{\norm {h_a^{\scr T}z_{ab}}^2 \norm {h_b^{\scr T}z_{ab}}^2} \, \exp S,
\end{align}
with
\be
S = \sum_{a<b}S_{ab}.
\ee
At this point we can follow the saddle point approximation developed in \cite{BarrettLorAsymp}, but not before explaining the different conventions used.

%---------------------------------------------------------------------------
\subsubsection*{Phase ambiguity and the conventions of \cite{BarrettLorAsymp}}
%---------------------------------------------------------------------------

The basis of homogeneous functions used in \cite{BarrettLorAsymp} differs from Eq. \Ref{Fbasis} in choosing $k$ instead of $-k$ and a different spinorial parametrization,
\be\label{FB}
F^{\scr B}{}^{(\r,k)}_{jm}(z^A) := \sqrt{\f{d_j}\pi} \f1{\norm{z}^{2(1-i\r)}} D^{(j)}_{m,k}\big(g_{\scr B}(z^A)\big), 
\qquad g_{\scr B}(z^A):= \f1{\norm z}\mat{z^0}{-\bar z^1}{z^1}{\bar z^0}.
\ee
In spite of this generic difference, the two choices coincide for the minimal weights used in the EPRL model, since
\be
D^{(j)}_{m,j}\big(g_{\scr B}(z^A)\big) \equiv D^{(j)}_{m,-j}\big(g(z^A)\big).
\ee
So there is no difference at this level. 
The choices however reflect that \cite{BarrettLorAsymp} takes as fundamental family of coherent states the highest weights, and furthermore the spinorial complex structure \Ref{defcomplstructure} is defined with an opposite sign. The spinorial bilinear in \cite{BarrettLorAsymp} is $[w,z]=[z\ket{w}$.

The  coherent basis is defined from the highest weights, and denoted
\be
{\cal I}\phi_{\xi}(z^A) := \sqrt{\f{d_j}\pi}\, \norm{z}^{2(i\g j-1-j)} [\xi\ket{z}^{2j} \equiv F^{(\g j,j)}_{j\overline{J\xi}}(z^A).
\ee
Then,
\be
\cJ {\cal I}\phi_\xi(z^A)= (-1)^{2j}e^{-i\arctan\g} \sqrt{\f{d_j}\pi}\, \norm{z}^{2(i\g j-1-j)} \bra\xi z\ra^{2j} \equiv (-1)^{2j}e^{-i\arctan\g}  F^{(\g j,j)}_{j\overline{\xi}}(z^A),
\ee
consistent with \Ref{phaseshift2}, since we have the same sign convention for $\cJ$.

The link ``propagator'' is defined in \cite{BarrettLorAsymp} from the  infinite-dimensional bilinear \Ref{defepsinfty} (there denoted $\b(F,F')$ but otherwise identically defined).
This is the same as a coherent unitary representation matrix up to a phase, 
\begin{align}\label{PB}
P_{ab}&:=\eps(h_a\triangleright{\cal I}\phi_{ab},h_b\triangleright{\cal I}\phi_{ba}) \equiv 
(h_a\triangleright \cJ{\cal I}\phi_{ab},h_b\triangleright {\cal I}\phi_{ba})
= (-1)^{2j_{ab}}e^{i\arctan\g} D^{(\g j_{ab},j_{ab})}_{j_{ab}\bar\xi_{ab} j_{ab} \overline{J\xi}_{ba}}(h_a^{-1}h_b)
\nn\\& =  (-1)^{2j_{ab}}e^{i\arctan\g} \int_{\C P^1} d\m(z^A)  \overline{F^{(\g j_{ab},j_{ab})}_{j_{ab}\bar{\xi}_{ab}}(h_a^{\scr T}z^A)} \, F^{(\g j_{ab},j_{ab})}_{j_{ab}\overline{J\xi}_{ba}}(h_b^{\scriptscriptstyle\rm T}z^A).
\end{align}

Hence our expression \Ref{defDcs} for the link propagator differs from the one of \cite{BarrettLorAsymp} by the overall phase 
$\exp \{i \sum_{a<b} (2j_{ab}(\pi+\Phi_{ab}) + \arctan\g)\}$
reported in footnote~\ref{footphase}, and by the identification
\be
\z_{ab}= \bar \xi_{ab}.
\ee
With this map, and the renaming $h_a\mapsto \bar h_a$, the action becomes
\be\label{actionB}
S_{ab} := j_{ab} \ln \frac{\bra{ h_{a}^{\dagger}z_{ab}}\xi_{ab}\ra^{2} \la J\xi_{ba}|h_{b}^{\dagger}z_{ab}\ra^{2}}{|\!| h_{a}^{\dagger}z_{ab}|\!|^2\, |\!| h_{b}^{\dagger}z_{ab}|\!|^2 }
+i\g j_{ab}\log\frac{\norm{h_{b}^{\dagger}z_{ab}}^2 }{|\!|h_{a}^{\dagger}z_{ab}|\!|^2 },
\ee
which coincides with \cite{BarrettLorAsymp} (with our definition of $J$ of opposite sign, but this is irrelevant because of the squares). Alternatively, we can reproduce the action of \cite{BarrettLorAsymp} keeping the same spinors $\z_{ab}=\xi_{ab}$ and the same group elements $h_a$ if we take a global complex conjugate and map $\g \to - \g$.

%-----------------------------------------------------------------------------------
\section{EPRL critical point equations and Hessian}
\label{appendixhessian}
%-----------------------------------------------------------------------------------

We can at this point recall briefly the key steps of the saddle point approximation of \cite{BarrettLorAsymp}, and complete it with the explicit calculation of the Hessian. 
We look for critical points defined by the vanishing of the gradient of Eq. \Ref{actionB} with respect to the variables integrated over, the spinors $z_{ab}$, and the group elements $h_a$. 
Vanishing of the spinor gradient gives
\begin{equation}
\ket{\xi_{ab}}=\frac{e^{i\upsilon_{ab}}}{|\!| h_{a}^{\dagger}z_{ab}|\!| }h_{a}^{\dagger}\,\ket{z_{ab}},
\qquad
\ket{J\xi_{ba}}=\frac{e^{i\upsilon_{ba}}}{|\!| h_{b}^{\dagger}z_{ab}|\!| }h_{b}^{\dagger}\,\ket{z_{ab}},\label{eq:zcrit}
\end{equation}
which can be combined to 
\begin{subequations}\label{critpoint}
\be
(h_{a}^{\dagger})^{-1}\ket{\xi_{ab}} = \frac{|\!| h_{b}^{\dagger}z_{ab}|\!|}{|\!| h_{a}^{\dagger}z_{ab}|\!| } 
e^{i(\upsilon_{ab}-\upsilon_{ba})} (h_{b}^{\dagger})^{-1}\ket{J\xi_{ba}},\label{eq:critpointeq1}
\ee
as well as
\be
h_{a}\ket{\xi_{ab}} = \frac{|\!| h_{a}^{\dagger}z_{ab}|\!|}{|\!| h_{b}^{\dagger}z_{ab}|\!| }
e^{i(\upsilon_{ab}-\upsilon_{ba})}h_{b}\ket{J\xi_{ba}}. \label{eq:critpointeq2}
\ee\end{subequations}
Here $\upsilon_{ab}$ are phases to be determined. Because of the homogeneity of the action, these are only four independent real equations per link. Two fix $z_{ab}$ up to the irrelevant choice of section, and the remaining ones fix the group element and $\upsilon_{ab}$ phases. On-shell of these equations, the vanishing of the $h_a$ gradient gives the closure conditions
\begin{equation}
\sum_{b\neq a}j_{ab}\vec{n}_{ab}=0.
\end{equation}
These are not equations for the integration variables, but directly a restriction on the boundary data.
On the other hand, also Eq. \Ref{critpoint} cannot be solved in general for $z_{ab}$ and $h_a$, but only if the boundary data satisfy special conditions, which leads us to the classification recalled in Sec.\ref{Sec2}.

In the first subset satisfying Eq. \Ref{ori}, let us further specialize the boundary data to the twisted spike configuration \Ref{btb}, where all normals are pairwise antiparallel. 
Plugging this condition into Eq. \Ref{critpoint} and looking at the norms, we find equations for the 3D representation matrices $H_a := D^{(1)}\left(h_a\right)$,
\be
H_a \vec{n}_{ab} = - H_b \vec{n}_{ba}= H_b \vec{n}_{ab}.
\ee
These have a trivial solution with $h_a=\pm \Id \ \forall a$. This corresponds in general to a vector geometry in the twisted spike gauge. However if additionally the shape-matching conditions are satisfied, it is possible to find a second, nontrivial solution, given by $h^{\scr (c)}_{1}=\mathds{1}$ and 
\be
h^{\scr (c)}_{a} = \pm\exp \left(\th_{1a} \vec{n}_{1a} \cdot \vec{\sigma}\right)
\ee
for $a\neq1$. 
This is the second critical point for Euclidean Regge data in the twisted spike. As for the spike configuration, one rotates back the tetrahedra so that all edges are aligned, and in that case the two critical points are Eq. \Ref{hER} in the main text. 
See also Fig.~8 of \cite{IoSU2asympt}.
Then to evaluate the action at the critical point, we plug the solutions back in Eq. \Ref{critpoint} and determine the $\upsilon_{ab}$ phases.

In the second subset, the solutions are not in the SU(2) subgroup, and their derivation is more involved, see \cite{BarrettLorAsymp}. Again we can simplify the analysis choosing judiciously the boundary data.
Fixing the spike configuration described in the main text, and using the freedom to gauge fix $h_1=\Id$, we can solve the system with Mathematica, finding the set of four solutions \Ref{hLR}.
Notice the presence of the $i\pi$: this is due to the inversion required by the fact that the first tetrahedron is past pointing, whereas all the others are future pointing. Plugging them back into Eq. \Ref{critpoint} we determine the $\upsilon_{ab}$ phases entering the action at the critical point.

For vector geometries in the twisted spike, $\upsilon_{ab}=0$. For the Regge geometries with two critical points, the $\upsilon_{ab}$ contribute not only to the Regge action oscillation but also to the overall phase of the amplitude. Their value for the configurations used in this paper are computed in the Mathematica files in \cite{code-sl2cfoam}. However, the overall phase is the only part of the analytic calculation that does not fit our numerical data. This could be due to an overlooked phase  in the previous appendix, or to a mismatch between the analytic definition and the numerical implementation that we were not able to identify. This is of limited harm, since the global phase is irrelevant to the model, and we determined it from a numerical fit for the sake of plotting real data points.

%-----
\subsection{Hessian}
\label{AppH}

The Hessian $H$ is the matrix of second derivatives of the action \eqref{actionB} with respect to the group elements $h_{a}$ and the spinors $z_{ab}$. It is a $44\times44$ matrix of the form

\begin{equation}\label{Hessian}
H=\left(\begin{array}{cc}
H_{hh} & H_{hz}\\
H_{zh} & H_{zz}
\end{array}\right) \ .
\end{equation}
The generic structure of the Hessian is presented in \cite{BarrettLorAsymp}, however not in a form explicit enough for numerical evaluations. Here we close this gap, and provide the explicit expressions for all the blocks. 

The top left block is a $24\times24$ block diagonal matrix 
\begin{equation}\nn
H_{h_ah_b}=\d_{ab} H_{h_a},
\end{equation}
where each block is a $6\times6$ matrix given by
\begin{equation}\nn
H_{h_{a}}=\frac{1}{2}\sum_{b\neq a}j_{ab}\left(\begin{array}{cc}
-\left(\mathds{1}-\vec{n}_{ab}\otimes\vec{n}_{ab}\right) & -i\epsilon_{ab}\left(\mathds{1}-\vec{n}_{ab}\otimes\vec{n}_{ab}\right)+(\epsilon_{ab}+i\gamma)\star\vec{n}_{ab}\\
-i\epsilon_{ab}\left(\mathds{1}-\vec{n}_{ab}\otimes\vec{n}_{ab}\right)-(\epsilon_{ab}+i\gamma)\star\vec{n}_{ab} & -\left(1+2i\epsilon_{ab}\gamma\right)\left(\mathds{1}-\vec{n}_{ab}\otimes\vec{n}_{ab}\right)
\end{array}\right)
\end{equation}
where $\epsilon_{ab}=\mathrm{sign}\left(b-a\right)$ and $\star$ is the Hodge dual in the 3D internal space.
The bottom right block of Eq. \Ref{Hessian} is a $20\times20$ block diagonal matrix
\be\nn
H_{z_{ab}z_{cd}} = \d_{ac} \d_{bd} H_{z_{ab}z_{ab}}
\ee
where each block is a $2\times2$ matrix 
\begin{align*}
H_{z_{ab}}= & j_{ab}(i\gamma+1)\left(\begin{array}{cc}
e^{i4\upsilon_{ab}}\langle \xi_{ab}|h_{a}h_{a}^{\dagger}|J\xi_{ab}\rangle ^{2} 
& |\langle J\xi_{ab}|h_{a}h_{a}^{\dagger}|J\xi_{ab}\rangle |^2-\langle J\xi_{ab}|(h_{a}h_{a}^{\dagger})^2|J\xi_{ab}\rangle \\
|\langle J\xi_{ab}|h_{a}h_{a}^{\dagger}|J\xi_{ab}\rangle |^2-\langle J\xi_{ab}|(h_{a}h_{a}^{\dagger})^2|J\xi_{ab}\rangle  
& -e^{-2i(2\upsilon_{ba}+\arctan\g)} \bra{J \xi_{ab}} h_{a}h_{a}^{\dagger} \ket{\xi_{ab}}^{2}
\end{array}\right)\\
 & +j_{ab}(i\gamma-1)\left(\begin{array}{cc}e^{i2(2\upsilon_{ba}+\arctan\g)}\langle J\xi_{ba}|h_{b}h_{b}^{\dagger}|\xi_{ba}\rangle ^{2} 
 & \langle \xi_{ba}|(h_{b}h_{b}^{\dagger})^2|\xi_{ab}\rangle+|\langle \xi_{ba}|h_{b}h_{b}^{\dagger}|\xi_{ab}\rangle|^2  \\
\langle \xi_{ba}|(h_{b}h_{b}^{\dagger})^2|\xi_{ab}\rangle+|\langle \xi_{ba}|h_{b}h_{b}^{\dagger}|\xi_{ab}\rangle|^2  
& -e^{-i4\upsilon_{ba}} \bra{\xi_{ba}} h_{b}h_{b}^{\dagger}\ket{J\xi_{ba}}^{2}
\end{array}\right).
\end{align*}
The off-diagonal block of Eq. \Ref{Hessian}  is a $24\times20$ matrix defined with sparse blocks
\begin{equation}\nn
H_{hz}=\left(\begin{array}{cccccccccc}
H_{h_{1}z_{12}} & H_{h_{1}z_{13}} & H_{h_{1}z_{14}} & H_{h_{1}z_{15}} & 0 & 0 & 0 & 0 & 0 & 0\\
H_{h_{2}z_{12}} & 0 & 0 & 0 & H_{h_{2}z_{23}} & H_{h_{2}z_{24}} & H_{h_{2}z_{25}} & 0 & 0 & 0\\
0 & H_{h_{3}z_{13}} & 0 & 0 & H_{h_{3}z_{23}} & 0 & 0 & H_{h_{3}z_{34}} & H_{h_{3}z_{35}} & 0\\
0 & 0 & H_{h_{4}z_{14}} & 0 & 0 & H_{h_{4}z_{24}} & 0 & H_{h_{4}z_{34}} & 0 & H_{h_{4}z_{45}}
\end{array}\right)
\end{equation}
where each block is the following $6\times2$ matrix,
\begin{align*}
H_{h_{c}z_{ab}}= & \delta_{cb}k_{ab}\left(\begin{array}{cc}
ie^{i2\upsilon_{ba}}\vec{v}_{ba} & 0\\
i\gamma e^{i2\upsilon_{ba}}\vec{v}_{ba} & \left(i\gamma-1\right)e^{-i2\upsilon_{ab}}\overline{\vec{v}_{ba}}
\end{array}\right)+\delta_{ca}k_{ab}\left(\begin{array}{cc}
0 & ie^{-i2\upsilon_{ab}}\vec{w}_{ab}\\
\left(i\gamma+1\right)e^{i2\upsilon_{ab}}\overline{\vec{w}_{ab}} & i\gamma e^{-i2\upsilon_{ab}}\vec{w}_{ab}
\end{array}\right)
\end{align*}
defined in terms of the 3D complex vectors 
\begin{equation}\nn
\vec{v}_{ba}=\langle J\xi_{ba}|\vec{\sigma}h_{b}h_{b}^{\dagger}|\xi_{ba}\rangle -\vec{n}_{ba}\langle J\xi_{ba}|h_{b}h_{b}^{\dagger}|\xi_{ba}\rangle \ , \quad \text{and} \quad \vec{w}_{ab}=\langle J\xi_{ab}|h_{a}h_{a}^{\dagger}\vec{\sigma}|\xi_{ab}\rangle +\vec{n}_{ab}\langle J\xi_{ab}|h_{a}h_{a}^{\dagger}|\xi_{ab}\rangle \ .
\end{equation}
By symmetry,  the block $H_{zh}$ is a $20\times24$ matrix that can be obtained by transposition of $H_{hz}$. 

The expression reported here must be evaluated at the critical point, and since we know that $h_{a}h_{a}^{\dagger}$ is a boost along $\vec n_{1a}$, everything can be written as a function of dihedral angles, 3D normals, and phases $\upsilon_{ab}$. The result is rather cumbersome and we do not write it here, but we refer to the Mathematica notebook \texttt{SL2Cderivatives} in \cite{code-sl2cfoam} that generates the symbolic expression of the Hessian starting from the action, where this final step is made. 

Finally, let us review the various numerical factors that contribute to $N_{c}$ in the asymptotic formula. This includes the inverse square root of the Hessian determinant, as well as the integration measure at the critical point.
For the spinor measure
$\Om_{ab}:=\norm{h^\dagger_a z_{ab}}^{-2}\norm{h^\dagger_b z_{ab}}^{-2}d\m(z_{ab}^A)$, one needs to make a choice of section of the tautological spinor bundle, the same made in evaluating the (inverse square root of the) Hessian at the critical point. Only their product is independent from this choice. The section we chose is $|\!| z_{ab}|\!|=1$ and $\upsilon_{ab}=0$ for $a<b$, and the result can be read from  \texttt{SL2Cderivatives}.
For the group integrals, we use the normalization of the $\SL(2,\C)$ Haar measure given in \cite{Ruhl}. This gives a factor  $1/(2^8 \pi^4)^4$  in both the numerical code and the value of the group measure at the critical point.
We then have a factor $(2 \pi)^{22}$ from the Gaussian integrations, a factor $2^4$ from the double multiplicity of the solutions $h_a=\pm h^{\scr (c)}$
for $a\neq 1$, and a factor $({2}/{\pi})^{10}$ from the normalization of the coherent states in the amplitude. 
This gives the factor $2^{36}\pi^{12}$ of \cite{BarrettLorAsymp} times the factor $2^{-32} \pi^{-16}$ from the Haar measure.
Putting everything together, 
\begin{equation}\label{eq:NC}
N_{c}=\frac{2^4}{\pi^4}\frac{1}{\sqrt{-\det H_{|\sigma}}}\prod_{a<b}{\Omega_{ab}}_{|c}.
\end{equation}

If we revert to our action \Ref{actionApp}, we can derive its Hessian from the one computed in this section by taking the complex conjugate and inverting the sign of $\gamma$. Numerical evaluations shows that the determinat of the hessian is invariant under this process. Therefore, the coefficient \Ref{eq:NC} is the same for both \Ref{actionApp} and \Ref{actionB} used in \cite{BarrettLorAsymp}.

%----------------------------------------------------------------------------
%----------------------------------------------------------------------------

\providecommand{\href}[2]{#2}\begingroup\raggedright\endgroup

%----------------------------------------------------------------------------
%----------------------------------------------------------------------------
\end{document}